\documentclass[aps,prd,superscriptaddress,preprint]{revtex4-1}

\usepackage{amsmath,amssymb,bm}
\usepackage{graphicx,graphics,color}
\usepackage[colorlinks=true,linkcolor=blue]{hyperref}
\usepackage{epsfig}
\usepackage{latexsym}
\usepackage{rotating}
\usepackage{subfigure}
\usepackage{multirow}
\usepackage{centernot}
\usepackage{longtable}
\usepackage{float}
\usepackage{mathrsfs}
\usepackage{ulem}

\begin{document}

\preprint{JLAB-THY-18-2746, ADP-18-17/T1065}

\title{Parton distributions from nonlocal chiral SU(3) effective theory.
	I. Splitting functions}

\author{Y. Salamu}
\affiliation{Institute of High Energy Physics, CAS,
	Beijing 100049, China}
\author{Chueng-Ryong Ji}
\affiliation{North Carolina State University, Raleigh,
	North Carolina 27695, USA}
\author{W. Melnitchouk}
\affiliation{Jefferson Lab, Newport News,
	Virginia 23606, USA}
\author{A. W. Thomas}
\affiliation{CoEPP and CSSM, Department of Physics,
	University of Adelaide,
	Adelaide SA 5005, Australia}
\author{P. Wang}
\affiliation{Institute of High Energy Physics, CAS,
	Beijing 100049, China}
\affiliation{Theoretical Physics Center for Science Facilities, CAS,
	Beijing 100049, China}

\begin{abstract}
We present a new formulation of pseudoscalar meson loop corrections
to nucleon parton distributions within a nonlocal covariant chiral
effective field theory, including contributions from SU(3) octet
and decuplet baryons.
The nonlocal Lagrangian, constrained by requirements of local gauge
invariance and Lorentz-invariant ultraviolet regularization,
generates additional interactions associated with gauge links.
We use these to compute the full set of proton $\to$ meson + baryon
splitting functions, which in general contain on-shell and off-shell
contributions, in addition to $\delta$-function terms at zero momentum,
along with nonlocal contributions associated with the finite size
of the proton.
We illustrate the shapes of the various local and nonlocal functions
numerically using a simple example of a dipole regulator.
\end{abstract}

\date{\today}
\maketitle

\section{Introduction}
\label{sec.introduction}

The important role played by chiral symmetry in hadron physics has
been documented for many decades.  Traditionally the purview of
low-energy hadron and nuclear physics, more recently the relevance of
chiral symmetry in QCD has become more prominent also in high-energy
reactions, in which the quark and gluon (or parton) substructure of
hadrons is manifest.
One of the most striking expressions of the chiral symmetry and its
approximate breaking is in the nonperturbative structure of the sea
quark distributions of the \mbox{nucleon~\cite{Speth:1996pz,
Kumano:1997cy}}.
In particular, the breaking of chiral SU(3) symmetry was anticipated
\cite{Thomas83} to generate unequal strange and (light) nonstrange
sea quark distributions, and, even more dramatically, an excess of
$\bar d$ antiquarks over $\bar u$.
The latter was confirmed in proton-proton and proton-deuteron
Drell-Yan experiments at CERN \cite{NA51} and Fermilab \cite{E866},
following earlier indirect indications from inclusive \cite{NMC94}
and semi-inclusive \cite{Ackerstaff:1998sr} deep-inelastic scattering
(DIS) data on proton and deuteron targets.

The observation of a large $\bar d - \bar u$ asymmetry has
also served to motivate more challenging searches for other
nonperturbative asymmetries, such as those between strange and
antistrange quarks in the proton, $s - \bar s$
\cite{Signal:1987gz, Mason:2007zz, Alekhin:2008mb}, or between
the helicity dependent light antiquark distributions,
$\Delta \bar d - \Delta \bar u$ \cite{JAM17}.
The phenomenological success in describing the $\bar d - \bar u$
asymmetry, in particular, in terms of nonperturbative models of
the nucleon in which its peripheral structure is modeled by a
pseudoscalar meson cloud suggested that signatures of chiral
symmetry breaking may also be found in other types of parton
distribution functions (PDFs)~\cite{Signal:1987gz, Schreiber91,
Melnitchouk:1995en, Steffens:1995at, Diakonov96, MM96, MM99,
Myhrer:2007cf}.

While considerable experience has been accumulated with
nonperturbative models, a challenge has been to compute the
chiral symmetry breaking effects on the PDFs in a model-independent
way from QCD.
An important step in establishing a direct connection with QCD
was made with the observation \cite{TMS00} that the leading
nonanalytic (LNA) behavior of moments of the nonsinglet PDFs,
expanded in powers of the pion mass, $m_\pi$, could be obtained
from chiral effective field theory, which encodes the same chiral
symmetry properties as present in QCD \cite{Arndt02, Chen01, Chen02}.
In addition to demonstrating how lattice QCD data on PDF moments
and other observables simulated at unphysically large pion masses
could be extrapolated to the physical point \cite{Detmold01},
the result \cite{TMS00} demonstrated unambiguously that a nonzero
component of $\bar d-\bar u$ arises as a direct consequence of the
infrared structure of QCD.

Subsequent work \cite{Burkardt12, JMT13, Moiseeva:2012zi,
Salamu:2014pka, XGWangPLB, XGWangPRD} computed the full set
of lowest order corrections to PDFs arising from pseudoscalar meson
loops, both for the PDF moments and the Bjorken-$x$ dependence.
The LNA behavior of the various contributions can be established
model-independently by considering the infrared limit; however,
the computation of the full amplitude requires specific choices
for regularizing the divergences in the loop integrals.
In the literature, regularization prescriptions such as transverse
momentum cutoffs, Pauli-Villars, dimensional regularization or
infrared regularization have been used, as well as form factors
or finite-range regulators.
The latter take into account the finite size of hadrons~\cite{Thomas03,
Thomas84}, while the others are generally more suitable for theories
that treat hadrons as pointlike.

In practice, the extended structure of the nucleon and other baryons
does become important in many traditional hadronic physics applications.
In nonrelativistic calculations, if the regulators are in
three-dimensional momentum space, such as for finite-range
regularization, charge conservation, which is related to the time
component of the current, is respected in the presence of form factors.
In relativistic calculations, on the other hand, simply replacing the
nonrelativistic regulator by a covariant one can lead to violation of
local gauge symmetry and charge conservation.

The problem of preserving gauge invariance in theories with hadronic
form factors can be formally alleviated by introducing {\it nonlocal}
interactions into the gauge invariant local Lagrangian,
which allows one to consistently generate a covariant regulator.
A method for constructing nonlocal Lagrangians with gauge fields was
described by Terning \cite{Terning91}, based on the path-ordered
exponential introduced by Wilson \cite{Wilson74} and earlier by
Bloch \cite{Bloch50}.
Variants of the method were subsequently used in phenomenological
applications to strange vector form factors and other nucleon
matrix elements by a number of authors \cite{Forkel:1994yx,
Musolf:1993fu, Wang:1996zu}.
The pion and $\sigma$ meson properties have been studied by
gauging nonlocal meson--quark interactions in relativistic
quark models \cite{Holdom:1992fn, Faessler:2003yf}.
The nonlocal Lagrangian at the hadron level was also recently
constructed and applied to electromagnetic form factors of the
nucleon \cite{Wangp, Hefc, Hefc2}.

The presence of gauge links in the nonlocal Lagrangian connecting
different spacetime coordinates generates additional diagrams
which are needed to ensure the local gauge invariance of the theory.
This guarantees that the proton and neutron charges, for example,
are unaffected by meson loops, or that contributions to the
strangeness in the nucleon from diagrams with intermediate state
kaons and hyperons sum to zero.
These basic features of the theory are not guaranteed for a local
Lagrangian with a covariant regulator, but arise automatically in the
nonlocal theory in which the Ward identities and charge conservation
are necessarily satisfied.
In fact, a nonlocal formulation may be preferable on physical grounds,
as this more naturally represents the extended structure of hadrons.

In this paper we describe how the nonlocal formulation of the chiral
SU(3) effective theory can be used to derive the contributions from
pseudoscalar meson loops to PDFs in the nucleon.  We include both
the SU(3) octet and decuplet baryons, using a covariant regulator
generated through the nonlocal Lagrangian that respects Lorentz
and gauge symmetry.
In the present paper we focus on the formalism and the derivation
of the proton $\to$ baryon $+$ meson splitting functions from the
nonlocal chiral Lagrangian; a follow-up paper \cite{nonlocal-II}
will report on the results for the nucleon PDFs, computed through
convolutions of the splitting functions and PDFs in the virtual
mesons and baryons in the loops.

We begin by reviewing in Sec.~\ref{sec.chirallag} the familiar local
effective Lagrangian in the standard chiral SU(3) effective field
theory.
The generalization of the effective Lagrangian to the nonlocal case
is described in Sec.~\ref{sec:nonlocal}, a procedure which allows
the preservation of gauge invariance in the presence of covariant
vertex functions for the nucleon--baryon--meson interaction.
The main results for the proton $\to$ meson $+$ baryon splitting
functions are derived in Sec.~\ref{sec.splitting} for the full set
of lowest order diagrams, including rainbow, bubble, tadpole and
Kroll-Ruderman contributions, as well as additional terms that arise
from the gauge links generated from the nonlocal interactions.
Here we present the model independent results for the nonanalytic
behavior of the moments of the splitting functions, and illustrate
the relative shapes and magnitudes of the various functions using
a simple example of a covariant dipole vertex form factor.
Finally, in Sec.~\ref{sec.summary} we summarize our results and
outline future applications of the new formalism.

\section{Local chiral effective Lagrangian}
\label{sec.chirallag}

In this section we review the standard local chiral effective theory
for mesons and baryons.  The lowest-order Lagrangian, consistent with
chiral SU(3)$_L\times$SU(3)$_R$ symmetry, describing the interaction of
pseudoscalar mesons ($\phi$) with octet ($B$) and decuplet ($T_\mu$)
baryons, is given by \cite{Jenkins:1991ts, Ledwig:2014rfa}
\begin{eqnarray}
\label{eq:ch8}
{\cal L}
&=&    {\rm Tr} \big[ \bar B (i\!\centernot D - M_B) B \big]
 -\frac {D}{2}\, {\rm Tr} \big[ \bar B \gamma^\mu \gamma_5 \{u_\mu, B\} \big]
 -\frac { F}{2}\, {\rm Tr} \big[ \bar B \gamma^\mu \gamma_5 [ u_\mu ,B ] \big]
							\notag\\
&+& \overline{T}_\mu^{ijk}
   (i\gamma^{\mu \nu \alpha} D_\alpha - M_T \gamma^{\mu\nu}) T_\nu^{ijk}
 - \frac { {\cal C}}{2}
   \left[ \epsilon^{ijk}\, \overline{T}_\mu^{ilm}
	  \Theta^{\mu\nu} (u_\nu)^{lj} B^{mk} + {\rm h.c.}
   \right]						\notag\\
&-&\frac {  {\cal H}}{2}\,
    \overline{T}_\mu^{ijk} \gamma^{\alpha} \gamma_5 (u_\alpha)^{kl}\,
	      T_\mu^{ijl}	
 +  \frac{f^2}{4}
    {\rm Tr} \big[ D_\mu U (D^\mu U)^\dag
	     \big],
\end{eqnarray}
where $M_B$ and $M_T$ are the octet and decuplet masses,
$D$ and $F$ are the meson--octet baryon coupling constants,
$\cal C$ and $\cal H$ are the meson--octet--decuplet and
meson--decuplet--decuplet baryon couplings, respectively,
$f=93$~MeV is the pseudoscalar decay constant,
and ``h.c.'' denotes the Hermitian conjugate.
The tensor $\epsilon^{ijk}$ is the antisymmetric tensor in
flavor space, and we define the tensors
  $\gamma^{\mu\nu}
   = \frac{1}{2} [\gamma^\mu,\gamma^\nu]$
and
  $\gamma^{\mu\nu\alpha}
   = \frac{1}{2} \{\gamma ^{\mu\nu}, \gamma^\alpha\}$
in terms of the Dirac $\gamma$-matrices.
The octet--decuplet transition tensor operator $\Theta^{\mu\nu}$
is defined as
\begin{equation}
\Theta^{\mu\nu}
= g^{\mu\nu} - \big( Z+\tfrac12 \big) \gamma^\mu \gamma^\nu,
\label{eq:Theta}
\end{equation}
where $Z$ is the decuplet off-shell parameter.
The SU(3) baryon octet fields $B^{ij}$ include the
  nucleon $N$ ($=p,n$),
  $\Lambda$,
  $\Sigma^{\pm,0}$ and
  $\Xi^{-,0}$ fields, and are given by the matrix
\begin{eqnarray}
\label{e.B}
B =
\left(
\begin{array}{ccc}
  \frac{1}{\sqrt 2} \Sigma^0 + \frac{1}{\sqrt 6} \Lambda
& \Sigma^+
& p					\\
  \Sigma^-
&-\frac{1}{\sqrt 2} \Sigma^0 + \frac{1}{\sqrt 6} \Lambda
& n					\\
  \Xi^-
& \Xi^0
&-\frac{2}{\sqrt6} \Lambda
\end{array}
\right).
\end{eqnarray}
The baryon decuplet fields $T^{ijk}_\mu$, which include the
  $\Delta$,
  $\Sigma^*$,
  $\Xi^*$ and
  $\Omega^-$ fields,
are represented by symmetric tensors with components
\begin{eqnarray}
\label{e.T}
\begin{array}{c}
T^{111} = \Delta^{++},\ \
T^{112} = \frac{1}{\sqrt 3} \Delta^+,\ \
T^{122} = \frac{1}{\sqrt 3} \Delta^0,\ \
T^{222} = \Delta^-,				\\
T^{113} = \frac{1}{\sqrt 3} \Sigma^{*+},\ \
T^{123} = \frac{1}{\sqrt 6} \Sigma^{*0},\ \
T^{223} = \frac{1}{\sqrt 3} \Sigma^{*-},	\\
T^{133} = \frac{1}{\sqrt 3} \Xi^{*0},\ \
T^{233} = \frac{1}{\sqrt 3} \Xi^{*-},		\\
T^{333} = \Omega^-.				\\
\end{array}
\end{eqnarray}
In the meson sector, the operator $U$ in Eq.~(\ref{eq:ch8}) is
defined in terms of the matrix of pseudoscalar fields $\phi$,
\begin{equation}
U = u^2,\ \ \ {\rm with}\ u = \exp\left(i \frac{\phi}{\sqrt{2}f}\right),
\end{equation}
where $\phi$ includes the $\pi$, $K$ and $\eta$ mesons,
\begin{eqnarray}
\label{e.phi}
\phi =
\left(
{\begin{array}{*{20}{c}}
  \frac{1}{\sqrt 2} \pi^0 + \frac{1}{\sqrt 6} \eta
& \pi^+
& K^+						\\
  \pi^-
& -\frac{1}{\sqrt 2} \pi^0 + \frac{1}{\sqrt 6} \eta
& K^0						\\
  K^-
& \bar K^0
& -\frac{2}{\sqrt 6} \eta
\end{array}}
\right).
\end{eqnarray}
The pseudoscalar mesons couple to the baryon fields through the
vector and axial vector combinations
\begin{eqnarray}
\Gamma_\mu
&=& \frac{1}{2}
    \left( u^\dagger \partial_\mu u + u\, \partial_\mu u^\dagger 
    \right)
 -  \frac{i}{2}
    \left( u^\dagger \lambda^a u + u\, \lambda^a u^\dagger
    \right) \upsilon_\mu^a,					\\
u_\mu
&=&\, i\,
    \left(u^\dagger \partial_\mu u - u\, \partial_\mu u^\dagger 
    \right)\,
 +\,
    \left( u^\dagger \lambda^a u - u\, \lambda^a u^\dagger 
    \right) \upsilon_\mu^a,
\label{eq:22}
\end{eqnarray}
where $\upsilon_\mu^a$ corresponds to an external vector field,
and $\lambda^a$ ($a=1, \ldots, 8$) are the Gell-Mann matrices.
The covariant derivatives of the octet and decuplet baryon fields
in the chiral Lagrangian (\ref{eq:ch8}) are defined as
\cite{Hemmert:1999mr, Hemmert:1998pi}
\begin{eqnarray}
D_\mu B
&=& \partial_\mu B
 + [\Gamma_\mu, B]
 - i \langle \lambda^0 \rangle \upsilon_\mu^0\, B,	\\
D_\mu T_\nu^{ijk}
&=& \partial_\mu T_\nu^{ijk}
 + (\Gamma_\mu, T_\nu )^{ijk}
 - i \langle \lambda^0 \rangle \upsilon_\mu^0\, T_\nu^{ijk},
\label{eq:11}
\end{eqnarray}
where $\upsilon_\mu^0$ denotes an external singlet vector field,
$\lambda^0$ is the unit matrix, and $\langle\, \cdots \rangle$
denotes a trace in flavor space.
For the covariant derivative of the decuplet field, we use the notation
\begin{equation}
(\Gamma_\mu, T_\nu)^{ijk}
= (\Gamma_\mu)_l^i\, T_\nu^{ljk}
+ (\Gamma_\mu)_l^j\, T_\nu^{ilk}
+ (\Gamma_\mu)_l^k\, T_\nu^{ijl}.
\end{equation}
For the pseudoscalar meson fields, the covariant derivarive is written
\begin{eqnarray}
D_\mu U
&=& \partial_\mu U
 + (iU \lambda^a - i \lambda^a U)\, \upsilon_\mu^a.
\end{eqnarray}
%

Expanding the Lagrangian (\ref{eq:ch8}) to leading order in the baryon
and meson fields, the relevant interaction part for a meson and baryon
coupling to a proton can be written explicitly~as
\begin{equation}
\begin{split}
{\cal L}_{\rm int} &
= \frac{(D+F)}{2f}
  \left( \bar p\, \gamma^\mu \gamma^5 p\, \partial_\mu \pi^0
       + \sqrt2\, \bar p\, \gamma^\mu \gamma^5 n\, \partial_\mu \pi^+
  \right)
- \frac{(D+3F)}{\sqrt{12} f}
  \bar p\, \gamma^\mu \gamma^5 \Lambda\, \partial_\mu K^+	\\
&
+ \frac{(D-F)}{2 f}
  \left( \sqrt2\, \bar p\, \gamma^\mu \gamma^5 \Sigma^+\, \partial_\mu K^0
       + \bar p\, \gamma^\mu \gamma^5 \Sigma^0\, \partial_\mu K^+
  \right)
- \frac{D-3F}{\sqrt{12} f}
  \bar p\, \gamma^\mu \gamma^5 p\, \partial_\mu \eta		\\
&
+ \frac{\cal C}{\sqrt{12} f}
  \left(
  - 2\, \bar p\, \Theta^{\nu\mu} \Delta_\mu^+\, \partial_\nu \pi^0
  - \sqrt2\, \bar p\, \Theta^{\nu\mu} \Delta_\mu^0\, \partial_\nu \pi^+
  + \sqrt6\, \bar p\, \Theta^{\nu\mu} \Delta_\mu^{++}\, \partial_\nu \pi^-
  \right.							\\
&
  \hspace*{1.8cm}
  \left.
  - \bar p\, \Theta^{\nu\mu} \Sigma_\mu^{*0}\, \partial_\nu K^+
  + \sqrt2\, \bar p\, \Theta^{\nu\mu} \Sigma_\mu^{*+}\, \partial_\nu K^0
  + {\rm h.c.}
  \right)							\\
&
+ \frac{i}{4f^2} \bar p\, \gamma^\mu p
  \Big[
    (\pi^+ \partial_\mu \pi^-  -  \pi^- \partial_\mu \pi^+)
    + 2 (K^+ \partial_\mu K^-  -  K^- \partial_\mu K^+)
    + (K^0 \partial_\mu \bar K^0  -  \bar K^0 \partial_\mu K^0)
  \Big].
\end{split}
\label{eq:j1}
\end{equation}
The terms involving the coupling ${\cal H}$ are not present
because of the restriction to proton initial states.
The current calculations below also do not involve the terms
with the coupling ${\cal H}$ for the proton initial states.

From the Lagrangian (\ref{eq:ch8}) one can also obtain the form
of the electromagnetic current that couples to the external field
$\upsilon_\mu^a$,
\begin{eqnarray}
J^\mu_a
&=&\frac{1}{2}{\rm Tr}
   \big[
   \bar B \gamma^\mu
   \left[ u \lambda^a u^\dagger + u^\dagger \lambda^a u, B
   \right]	
      + \frac{D}{2}{\rm Tr}
   \big[
   \bar B \gamma^\mu \gamma_5
   \left\{ u \lambda^a u^\dagger - u^\dagger \lambda^a u, B
   \right\}
   \big]						\notag\\
&+&
   \frac{F}{2}{\rm Tr}
   \big[
   \bar B \gamma^\mu \gamma_5
   \left[ u \lambda^a u^\dagger - u^\dagger \lambda^a u, B
   \right]
   \big]						\notag\\
&+&
   \frac{1}{2}\,
   \overline{T}_\nu \gamma^{\nu\alpha\mu}
   \left( u \lambda^a u^\dagger + u^\dagger \lambda^a u, T_\alpha
   \right)
   + \frac{\cal C}{2}
   \left(
   \overline{T}_\nu \Theta^{\nu\mu}
   (u \lambda^a u^\dagger - u^\dagger \lambda^a u) B
   + {\rm h.c.}
   \right)						\notag\\
&+&
   \frac{f^2}{4}{\rm Tr}
   \big[
   \partial^\mu U
   (U^\dagger i \lambda^a
    - i \lambda^a U^\dagger)
+  (U i \lambda^a
    - i \lambda^a U)
   \partial^\mu U^\dagger
   \big].
\label{eq:ch1}
\end{eqnarray}
For the SU(3) flavor singlet current coupling to the external field
$\upsilon_\mu^0$, one has
\begin{eqnarray}
J^\mu_0
&=& \langle \lambda^0 \rangle\,
    {\rm Tr}[\bar B \gamma^\mu B]
 +  \langle \lambda^0 \rangle\,
    \overline{T}_\nu \gamma^{\nu\alpha\mu}\, T_\alpha,
\label{eq:ch2}
\end{eqnarray}
where again $\lambda^0$ is the unit matrix and
$\langle\, \cdots \rangle$ denotes a trace in flavor space.

The currents for a given quark flavor are then expressed as
combinations of the SU(3) singlet and octet currents,
\begin{subequations}
\label{eq:ch3}
\begin{eqnarray}
J^\mu_u &=&\frac{1}{3} J^\mu_0
	 + \frac{1}{2} J^\mu_3
	 + \frac{1}{2\sqrt3} J^\mu_8,			\\
J^\mu_d &=&\frac{1}{3} J^\mu_0
	 - \frac{1}{2} J^\mu_3
	 + \frac{1}{2\sqrt3} J^\mu_8,			\\
J^\mu_s &=&\frac{1}{3} J^\mu_0
	 - \frac{1}{\sqrt3} J^\mu_8,
\end{eqnarray}
\end{subequations}
where $J^\mu_3$ and $J^\mu_8$ are the $a=3$ and 8 components
of the octet current, respectively.
Using Eqs.~(\ref{eq:ch1}), (\ref{eq:ch2}) and (\ref{eq:ch3}),
the currents $J^\mu_u$, $J^\mu_d$ and $J^\mu_s$ can be written
explicitly~as
\begin{subequations}
\label{eq:jq}
\begin{eqnarray}
J_u^\mu
&=& 2 \bar p \gamma^\mu p + \bar n \gamma^\mu n
+ \bar\Lambda \gamma^\mu \Lambda
+ 2 \overline{\Sigma}^+ \gamma^\mu \Sigma^+
+ {\overline\Sigma}^0 \gamma^\mu \Sigma^0
- \frac{1}{2f^2}
  \left( \bar p \gamma^\mu p\, \pi^+ \pi^-
     + 2 \bar p \gamma^\mu p\, K^+ K^-
  \right)						\notag\\
&+&
  3 \overline{\Delta}_\alpha^{++} \gamma^{\alpha\beta\mu} \Delta_\beta^{++}
+ 2 \overline{\Delta}_\alpha^+    \gamma^{\alpha\beta\mu} \Delta_\beta^+
+   \overline{\Delta}_\alpha^0    \gamma^{\alpha\beta\mu} \Delta_\beta^0
+ 2 \overline{\Sigma}_\alpha^{*+} \gamma^{\alpha\beta\mu} \Sigma_\beta^{*+}
+   \overline{\Sigma}_\alpha^{*0} \gamma^{\alpha\beta\mu} \Sigma_\beta^{*0}
							\notag\\
&+&
  i \left( \pi^- \partial^\mu\pi^+ - \pi^+ \partial^\mu\pi^- \right)
+ i \left( K^-   \partial^\mu K^+  - K^+   \partial^\mu K^-   \right)
							\notag\\
&-&
  \frac{i(D+F)}{\sqrt 2f} \bar p \gamma^\mu \gamma^5 n\, \pi^+
+ \frac{i(D+3F)}{\sqrt{12}f} \bar p \gamma^\mu \gamma^5 \Lambda\, K^+
- \frac{i(D-F)}{2f}       \bar p \gamma^\mu \gamma^5 \Sigma^0\, K^+
							\notag\\
&+&
  \frac{\cal C}{\sqrt{12}f}
  \left(
    i \sqrt{6}\, \bar p\, \Theta^{\mu\nu} \Delta_\nu^{++}\, \pi^-
  + i \sqrt{2}\, \bar p\, \Theta^{\mu\nu} \Delta_\nu^0\, \pi^+
  + i \,\bar p\, \Theta^{\mu\nu} \Sigma_\nu^{*0}\, K^+
  + {\rm h.c.}
  \right),
\label{eq:ju}
\end{eqnarray}
\begin{eqnarray}
J^\mu_d
&=& \bar p \gamma^\mu p
+ 2 \bar n \gamma^\mu n
+ 2 \overline{\Sigma}^- \gamma^\mu \Sigma^-
+ \overline{\Sigma}^0 \gamma^\mu \Sigma^0
+ \bar\Lambda \gamma^\mu \Lambda
+ \frac{1}{2f^2}
  \left( \bar p \gamma^\mu p\, \pi^+ \pi^-
       - \bar p \gamma^\mu p\, \overline{K}^0 K^0
  \right)						\notag\\
&+&
    \overline{\Delta}_\alpha^+ \gamma^{\alpha\beta\mu} \Delta_\beta^+
+ 2 \overline{\Delta}_\alpha^0 \gamma^{\alpha\beta\mu} \Delta_\beta^0
+ 3 \overline{\Delta}_\alpha^- \gamma^{\alpha\beta\mu} \Delta_\beta^-
+   \overline{\Sigma}_\alpha^{*0}  \gamma^{\alpha\beta\mu} \Sigma_\beta^{*0}
+ 2 \overline{\Sigma}_\alpha^{*0-} \gamma^{\alpha\beta\mu} \Sigma_\beta^{*-}
							\notag\\
&-&
  i (\pi^- \partial^\mu \pi^+  -  \pi^+ \partial^\mu \pi^-)
+ i (\overline{K}^0 \partial^\mu K^0  -  K^0 \partial^\mu \overline{K}^0)
							\notag\\
&+&
  \frac{i(D+F)}{\sqrt2 f}
  \bar p \gamma^\mu \gamma^5 n\, \pi^+
- \frac{i(D-F)}{\sqrt2 f}
  \bar p \gamma^\mu \gamma^5 \Sigma^+\, K^0		\notag\\
&-&
\frac{\cal C}{\sqrt6 f}
  \left(
    i \sqrt3\, \bar p\, \Theta^{\mu\nu} \Delta_\nu^{++}\, \pi^-
  + i \bar p\, \Theta^{\mu\nu} \Delta_\nu^0\, \pi^+
  + i \bar p\, \Theta^{\mu\nu} \Sigma_\nu^{*+}\, K^0
  + {\rm h.c.}
  \right),
\label{eq:jd}
\end{eqnarray}
\begin{eqnarray}
\label{eq:js}
J^\mu_s
&=&\overline{\Sigma}^+ \gamma^\mu \Sigma^+
 + \overline{\Sigma}^0 \gamma^\mu \Sigma^0
 + \bar\Lambda \gamma^\mu \Lambda
 + \frac{1}{2f^2}
   \left( 2 \bar p \gamma^\mu p\, K^+ K^-
        + \bar p \gamma^\mu p\, \overline{K}^0 K^0
   \right)						\notag\\
&+&
   \overline{\Sigma}_\alpha^{*+} \gamma^{\alpha\beta\mu} \Sigma_\beta^{*+}
 + \overline{\Sigma}_\alpha^{*0} \gamma^{\alpha\beta\mu} \Sigma_\beta^{*0}
 - i (K^- \partial^\mu K^+  -  K^+ \partial^\mu K^-)
 - i (\overline{K}^0 \partial^\mu K^0
     - K^0 \partial^\mu \overline{K}^0)			\notag\\
&+&
   \frac{i(D-F)}{\sqrt2 f}
   \bar p \gamma^\mu \gamma^5 \Sigma^+\, K^0
 + \frac{i(D-F)}{2f}
   \bar p \gamma^\mu \gamma^5 \Sigma^0\, K^+
 - \frac{i(D+3F)}{\sqrt{12} f}
   \bar p \gamma^\mu \gamma^5 \Lambda\, K^+		\notag\\
&+&
   \frac{\cal C}{\sqrt{12} f}
   \left(
   - i \bar p\, \Theta^{\mu\nu} \Sigma_\nu^{*0}\, K^+
   + i \sqrt2\, \bar p\, \Theta^{\mu\nu} \Sigma_\nu^{*+}\, K^0
   + {\rm h.c.}
   \right),
\end{eqnarray}
\end{subequations}
where the terms involving the doubly-strange baryons $\Xi^{0,-}$
and $\Xi^{*0,-}$ and the triply-strange $\Omega^-$ are not present
because they cannot couple to the proton initial states.

\section{Nonlocal chiral Lagrangian}
\label{sec:nonlocal}

In this section we describe the generation of the nonlocal Lagrangian
from the local meson--baryon Lagrangian in Sec.~\ref{sec.chirallag}.
Evaluating the traces in Eq.~(\ref{eq:ch8}) and introducing the
minimal substitution for the electromagnetic field $\mathscr{A_\mu}$,
the local Lagrangian density can be rewritten more explicitly in the
form
\begin{eqnarray}
{\cal L}^{\rm (local)}(x)
&=&\bar B(x)(i \gamma^\mu \mathscr{D}_{\mu,x} - M_B) B(x)
 + \frac{C_{B\phi}}{f}
   \left[ \bar{p}(x) \gamma^\mu \gamma^5 B(x)\,
	  \mathscr{D}_{\mu,x} \phi(x) + {\rm h.c.}
   \right]						\notag\\
&+&\overline{T}_\mu(x)
   (i \gamma^{\mu\nu\alpha} \mathscr{D}_{\alpha,x} - M_T \gamma^{\mu\nu})\,
   T_\nu(x)
 + \frac{C_{T\phi}}{f}
   \left[\, \bar{p}(x) \Theta^{\mu\nu} T_\nu(x)\,
	 \mathscr{D}_{\mu,x} \phi(x) + {\rm h.c.}
   \right]						\notag\\
&+&\frac{iC_{\phi\phi^\dag}}{2 f^2}
   \bar p(x) \gamma^\mu p(x)
   \left[ \phi(x) (\mathscr{D}_{\mu,x} \phi)^\dag(x)
	- \mathscr{D}_{\mu,x} \phi(x) \phi^\dag(x)
   \right]						\notag\\
&+& \mathscr{D}_{\mu,x} \phi(x)
    (\mathscr{D}_{\mu,x} \phi)^\dag(x)\, + \cdots ,
\label{eq:Llocal}
\end{eqnarray}
where for the interaction part we show only those terms that
contribute to a meson--baryon coupling to a proton, and we keep
the dependence on the space-time coordinate $x$ explicitly.
The covariant derivatives here are written so as to indicate
the coordinate with respect to which the derivative is taken,
\begin{subequations}
\begin{eqnarray}
\mathscr{D}_{\mu,x} B(x)
&=& \left[ \partial_\mu - i e^q_B\, \mathscr{A_\mu}(x)
    \right] B(x),					\\
\mathscr{D}_{\mu,x} T^\nu(x)
&=& \left[ \partial_\mu - i e^q_T\, \mathscr{A_\mu}(x)
    \right] T^\nu(x),					\\
\mathscr{D}_{\mu,x} \phi(x)
&=& \left[ \partial_\mu - i e^q_\phi\, \mathscr{A_\mu}(x)
    \right] \phi(x),
\end{eqnarray}
\end{subequations}
where $e^q_B$, $e^q_T$ and $e^q_\phi$ are the quark flavor charges of
the octet baryon $B$, decuplet baryon $T$ and meson $\phi$,
respectively.
For example, for the proton one has the charges
$e^u_p = 2 e^d_p = 2$,
$e^s_p = 0$,
while for the $\Sigma^+$ hyperon
$e^u_{\Sigma^+} = 2 e^s_{\Sigma^+} = 2$,
$e^d_{\Sigma^+} = 0$,
and so forth.
For the mesons, the flavor charges for the $\pi^+$ are
$e^u_{\pi^+} = -e^d_{\pi^+} = 1$
but $e^q_{\pi^0} = 0$ for all $q$,
and for the $K^+$ these are
$e^u_{K^+} = -e^s_{K^+} = 1$,
$e^d_{K^+} = 0$,
and similarly for the charge conjugate states.
These flavor charges may be read off from the currents given
in Eqs.~(\ref{eq:ju})--(\ref{eq:js}).
The coefficients $C_{B\phi}$ in Eq.~(\ref{eq:Llocal}) depend on the
coupling constants $D$, $F$ and ${\cal C}$, and are given explicitly
in Table~\ref{tab:C} for the processes discussed in this work.

\begin{table}[t]
\begin{center}
\caption{Coupling constants $C_{B\phi}$, $C_{T\phi}$ and
	$C_{\phi\phi^\dag}$ for the $p B \phi$, $p T \phi$
	and $p p \phi \phi^\dag$ interactions, respectively,
	for the various	allowed flavor channels.\\}
\begin{tabular}{c|ccccc}
\hspace*{0.5cm}$(B\phi)$\hspace*{0.5cm}
& \hspace*{0.5cm}$(p \pi^0)$\hspace*{0.4cm}
& \hspace*{0.4cm}$(n \pi^+)$\hspace*{0.4cm}
& \hspace*{0.4cm}$(\Sigma^+ K^0)$\hspace*{0.4cm}
& \hspace*{0.4cm}$(\Sigma^0 K^+)$\hspace*{0.4cm}
& \hspace*{0.4cm}$(\Lambda  K^+)$\hspace*{0.4cm}	\\
\hspace*{0.5cm}$C_{B\phi}$\hspace*{0.5cm}
& \hspace*{0.4cm}$\frac12 (D+F)$\hspace*{0.4cm}
& \hspace*{0.4cm}$\frac{1}{\sqrt2} (D+F)$\hspace*{0.4cm}
& \hspace*{0.4cm}$\frac{1}{\sqrt2} (D-F)$\hspace*{0.4cm}
& \hspace*{0.4cm}$\frac12 (D-F)$\hspace*{0.4cm}
& \hspace*{0.4cm}$-\frac{1}{\sqrt{12}} (D+3F)$\hspace*{0.4cm} \\ \hline
($T\phi$)
& $(\Delta^0 \pi^+)$
& $(\Delta^+ \pi^0)$
& $(\Delta^{++} \pi^-)$
& $(\Sigma^{*+} K^0)$
& $(\Sigma^{*0} K^+)$				\\
$C_{T\phi}$
& $-\frac{1}{\sqrt 6}\, {\cal C}$
& $-\frac{1}{\sqrt 3}\, {\cal C}$
& $\frac{1}{\sqrt 2}\, {\cal C}$
& $\frac{1}{\sqrt 6}\, {\cal C}$
& $-\frac{1}{\sqrt{12}}\, {\cal C}$		\\ \hline
$(\phi\phi^\dag)$
& $(\pi^+\pi^-)$
& $(K^0 \overline{K}^0)$
& $(K^+ K^-)$
&
&						\\
$C_{\phi\phi^\dag}$
& $\dfrac{1}{2}$
& $\dfrac{1}{2}$
& 1
&
&						\\
\end{tabular}
\label{tab:C}
\end{center}
\end{table}

Using the methods described in Refs.~\cite{Terning91, Holdom:1992fn,
Faessler:2003yf, Wang:1996zu, Wangp, Hefc, Hefc2}, the nonlocal version
of the local Lagrangian (\ref{eq:Llocal}) can be written as
\begin{eqnarray}
{\cal L}^{\rm (nonloc)}(x)
&=& \bar B(x) (i\gamma^\mu \mathscr{D}_{\mu,x} - M_B) B(x)
+ \overline{T}_\mu(x)
  (i\gamma^{\mu\nu\alpha} \mathscr{D}_{\alpha,x} - M_T \gamma^{\mu\nu})
  T_\nu(x)						\notag\\
&+&
  \bar{p}(x)
  \left[
    \frac{C_{B\phi}}{f} \gamma^\mu \gamma^5 B(x)\,
  + \frac{C_{T\phi}}{f} \Theta^{\mu\nu} T_\nu(x)
  \right]						\notag\\
& & \hspace*{2cm} \times\
  \int d^4a\, {\cal G}_\phi^q(x,x+a) F(a)\
  \mathscr{D}_{\mu,x+a} \phi(x+a) + {\rm h.c.}		\notag\\
&+&
  \frac{i C_{\phi\phi^\dag}}{2 f^2}
  \bar{p}(x) \gamma^\mu p(x)				
  \int d^4a\!\int d^4b\ {\cal G}_\phi^q(x+b,x+a) F(a) F(b)\  \notag\\
& & \hspace*{2cm}
\times
  \left[ \phi(x+a) (\mathscr{D}_{\mu,x+b} \phi)^\dag(x+b)
       - \mathscr{D}_{\mu,x+a} \phi(x+a) \phi^\dag(x+b)
  \right]						\notag\\
&+&
  \mathscr{D}_{\mu,x} \phi(x) (\mathscr{D}_{\mu,x} \phi)^\dag(x)
 + \cdots,
\label{eq:j4}
\end{eqnarray}
where the gauge link ${\cal G}_\phi^q$ is introduced to preserve
local gauge invariance,
\begin{equation}
{\cal G}_\phi^q(x,y)
= \exp\left[ -i e^q_\phi \int_x^y dz^\mu \mathscr{A}_\mu(z) \right],
\label{eq:link}
\end{equation}
and the function $F(a)$ is the meson--baryon vertex form factor
in coordinate space.
One can verify that the nonlocal Lagrangian in Eq.~(\ref{eq:j4}),
as well as local Lagrangian in Eq.~(\ref{eq:Llocal}), are invariant
under the gauge transformations
\begin{subequations}
\begin{eqnarray}
B(x)     &\to&
  B'(x)     = B(x)     \exp\left[i e^q_B\, \theta(x)\right],	\\
T_\mu(x) &\to&
  T_\mu'(x) = T_\mu(x) \exp\left[i e^q_T\, \theta(x)\right],	\\
\phi(x)  &\to&
  \phi'(x)  = \phi(x)  \exp\left[i e^q_\phi\, \theta(x)\right],
\end{eqnarray}
for the matter fields, and
\begin{equation}
{\mathscr{A}^\mu}(x) \to
\mathscr{A}'^\mu(x) = {\mathscr{A}^\mu}(x)
		    + {\partial^\mu} \theta(x)
\end{equation}
\end{subequations}
for the electromagnetic field, where $\theta(x)$ is an arbitrary
function of the space-time coordinate~$x^\mu$.

The nonlocal Lagrangian density in Eq.~(\ref{eq:j4}) can be further
decomposed by expanding the gauge link (\ref{eq:link}) in powers of
the charge $e_\phi^q$,
\begin{eqnarray}
{\cal G}^q_\phi(x+b,x+a)
&=& \exp \Big[ -i e^q_\phi\, (a-b)^\mu
	      \int_0^1 dt\, \mathscr{A}_\mu\big(x+at+b(1-t)\big)
	 \Big]						\notag\\
&=& 1\ +\ \delta {\cal G}^q_\phi\
       +\ \cdots,
\label{eq:linkexpand}
\end{eqnarray}
where the ${\cal O}(e^q_\phi)$ term is
\begin{eqnarray}
\delta {\cal G}^q_\phi
&=& -\ i e^q_\phi\, (a-b)^\mu
       \int_0^1 dt\, \mathscr{A}_\mu\big(x+at+b(1-t)\big)
\label{eq:deltaG}
\end{eqnarray}
and we have used a change of variables
	$z^\mu \to x^\mu + a^\mu\, t + b^\mu\, (1-t)$.
This allows the Lagrangian ${\cal L}^{\rm (nonloc)}$ to be written
as a sum of free and interacting parts,
where to lowest order the latter consists of
purely hadronic (${\cal L}^{\rm (nonloc)}_{\rm had}$),
electromagnetic (${\cal L}^{\rm (nonloc)}_{\rm em}$), and
gauge link (${\cal L}^{\rm (nonloc)}_{\rm link}$) components.
The higher order terms in Eq.~(\ref{eq:linkexpand}) contribute to
higher order electromagnetic corrections, which are in practice
negligible.  The higher order terms can also be related to other
processes, such as those involving two or more photons emitted
in the final state.

The hadronic and electromagnetic interaction parts of the nonlocal
Lagrangian arise from the
${\cal O}(e^q_\phi)$ term in Eq.~(\ref{eq:linkexpand}),
and are given by
\begin{eqnarray}
{\cal L}^{\rm (nonloc)}_{\rm had}(x)
&=& \bar{p}(x)
    \left[ \frac{C_{B\phi}}{f}\, \gamma^\mu \gamma^5 B(x)
	 + \frac{C_{T\phi}}{f}\, \Theta^{\mu\nu} T_\nu(x)
    \right]
    \!\int\! d^4a\, F(a)\, \partial_\mu \phi(x+a)
	+ {\rm h.c.}					\notag\\
&+& \frac{iC_{\phi\phi^\dag}}{2f^2}
    \bar{p}(x) \gamma^\mu p(x)
    \int\!d^4a\!\int\!d^4b\ F(a) F(b)			\notag\\
& & \hspace*{3cm} \times
    \left[ \phi(x+a) \partial_\mu \phi^\dag(x+b)
	 - \partial_\mu \phi(x+a) \phi^\dag(x+b)
    \right],
\label{eq:Lnonloc_had}
\end{eqnarray}
and
\begin{eqnarray}
{\cal L}^{\rm (nonloc)}_{\rm em}(x)
&=& e^q_B\, \bar{B}(x) \gamma^\mu B(x)\, \mathscr{A}_\mu(x)\
 +\ e^q_T\, \overline{T}_\mu(x) \gamma^{\mu\nu\alpha} T_\nu(x)\,
	    \mathscr{A}_\alpha(x)				\notag\\
&+& i e^q_\phi \left[ \partial^\mu \phi(x) \phi^\dag(x)
		    - \phi(x) \partial^\mu \phi^\dag(x)
	       \right] \mathscr{A}_\mu(x)			\notag\\
&-& i e^q_\phi\, \bar{p}(x)
    \left[
      \frac{C_{B\phi}}{f}\, \gamma^\mu \gamma^5 B(x)
    + \frac{C_{T\phi}}{f}\, \Theta^{\mu\nu} T_\nu(x)
    \right]							\notag\\
& & \hspace*{3cm} \times
    \int\!d^4a\, F(a)\,
    \phi(x+a) \mathscr{A}^\mu(x+a) + {\rm h.c.}			\notag\\
&-& \frac{e^q_\phi C_{\phi\phi^\dag}}{2f^2}\,
    \bar{p}(x) \gamma^\mu p(x)
    \int\!d^4a\, F(a)
    \int\!d^4b\, F(b)						\notag\\
& & \hspace*{3cm} \times\
   \phi(x+a) \phi^\dag(x+b)
   \left[ \mathscr{A}^\mu(x+a) + \mathscr{A}^\mu(x+b) \right],
\label{eq:Lnonloc_em}
\end{eqnarray}
respectively.
For the $\delta {\cal G}^q_\phi$ term in Eq.~(\ref{eq:deltaG}),
which explicitly depends on the gauge link, the nonlocal interaction
with the external gauge field yields the additional contribution to
the Lagrangian density,
\begin{eqnarray}
{\cal L}^{\rm (nonloc)}_{\rm link}(x)
&=& -i e^q_\phi\, \bar{p}(x)
    \left[ \frac{C_{B\phi}}{f}\, \gamma^\rho \gamma^5 B(x)
	 + \frac{C_{T\phi}}{f}\, \Theta^{\rho\nu} T_\nu(x)
    \right]							\notag\\
& & \hspace*{2cm} \times
    \int_0^1 dt \int\!d^4a\, F(a)\, a^\mu\,
    \partial_\rho \phi(x+a) \mathscr{A}_\mu(x+at)
    + {\rm h.c.}						\notag\\
& & \hspace*{-1cm}
 +\ \frac{e^q_\phi C_{\phi\phi^\dag}}{2f^2}\,
    \bar{p}(x) \gamma^\rho p(x)
    \int_0^1 dt \int\!d^4a\! \int\!d^4b\,
	F(a)\, F(b)\, (a-b)^\mu\,				\notag\\
& & \hspace*{-0.7cm} \times
    \left[ \phi(x+a) \partial_\rho \phi^\dag(x+b)
	 - \partial_\rho \phi(x+a) \phi^\dag(x+b)
    \right] \mathscr{A}_\mu\big(x+at+b(1-t)\big).
\label{eq:Lnonloc_link}
\end{eqnarray}
For the nonlocal theory the quark current has two contributions:
the usual electromagnetic current, $J_{\rm em}^{\mu, q}$, obtained
with minimal substitution from Eq.~(\ref{eq:Lnonloc_em}),
\begin{eqnarray}
J_{q, \rm em}^\mu(x)
&\equiv& \frac{\delta \int\!d^4y\,
		{\cal L}_{\rm em}^{\rm (nonloc)}(y)}
	      {\delta \mathscr{A_\mu}(x)}			\notag\\
&=& e^q_B\, \bar{B}(x) \gamma^\mu B(x)
 +  e^q_T\, \overline{T}_\alpha(x) \gamma^{\alpha\nu\mu} T_\nu(x)
 +  i e^q_\phi \left[ \partial^\mu \phi(x) \phi^\dag(x)
		    - \phi(x) \partial^\mu \phi^\dag(x)
	       \right]						\notag\\
&-& i e^q_\phi\,
    \int\!d^4a\, F(a)\, \bar{p}(x-a)
    \left[ \frac{C_{B\phi}}{f}\, \gamma^\mu \gamma^5 B(x-a)\
	+\ \frac{C_{T\phi}}{f}\, \Theta^{\mu\nu} T_\nu(x-a)
    \right]
    \phi(x)\ +\ {\rm h.c.}					\notag\\
&-& \frac{e^q_\phi C_{\phi\phi^\dag}}{2f^2}
    \int\!d^4a\, F(a)
    \int\!d^4b\, F(b)
    \Big[ \bar{p}(x-a) \gamma^\mu p(x-a)\, \phi(x) \phi^\dag(x+b-a)
								\notag\\
& & \hspace*{5cm}
      +\, \bar{p}(x-b) \gamma^\mu p(x-b)\, \phi(x+a-b) \phi^\dag(x)
    \Big],
\label{eq:Jem}
\end{eqnarray}
and an additional term obtained from the gauge link,
\begin{eqnarray}
\delta J_q^\mu(x)
&\equiv& \frac{\delta \int\!d^4y\, {\cal L}^{\rm (nonloc)}_{\rm link}(y)}
              {\delta \mathscr{A_\mu}(x)}                       \notag\\
& & \hspace*{-2cm}
 =\ -i e^q_\phi
    \int_0^1 dt \int\!d^4a\, F(a)\, a^\mu\,
    \bar{p}(x-at)
    \left[ \frac{C_{B\phi}}{f}\, \gamma^\rho \gamma^5 B(x-at)\,
       +\, \frac{C_{T\phi}}{f}\, \Theta^{\rho\nu} T_\nu(x-at)
    \right]							\notag\\
& & \hspace*{3cm} \times\
    \partial_\rho \phi(x+a(1-t)) + {\rm h.c.}			\notag\\
& & \hspace*{-2cm}
 +\ \frac{e^q_\phi C_{\phi\phi^\dag}}{2f^2}
    \int_0^1 dt \int\!d^4a\, F(a) \int\!d^4b\, F(b)\, (a-b)^\mu\,
    \bar{p}\big(x-at-b(1-t)\big) \gamma^\rho p\big(x-at-b(1-t)\big)
								\notag\\
& & \hspace*{3cm} \times\
    \Big[ \phi\big(x+(a-b)(1-t)\big)
	  \partial_\rho \phi^\dag\big(x-(a-b)t\big)		\notag\\
& & \hspace*{3.4cm}
        - \partial_\rho \phi\big(x+(a-b)(1-t)\big)
	  \phi^\dag\big(x-(a-b)t\big)
    \Big],
\label{eq:Jlink}
\end{eqnarray}
respectively.  Compared with Eqs.~(\ref{eq:j1}) and (\ref{eq:jq}),
the nonlocal interaction Lagrangian and currents in
Eqs.~(\ref{eq:Lnonloc_had})--(\ref{eq:Jlink}) include the
extra regulator function $F(a)$.
The local limit can be obtained by taking $F(a)$ to be a
$\delta$-function, $F(a) \to \delta^{(4)}(a)$, which is equivalent
to taking the form factor in momentum space to be unity.
Since the Fourier transform of the $\delta$-function in position
space is a plane wave in momentum space, the value of the plane
wave at the origin is unity.

Note that compared with traditional power counting schemes in
chiral perturbation theory that use dimensional regularization
\cite{Manohar:1996cq}, the introduction of the regulator
function $F(a)$ in the nonlocal interactions
(\ref{eq:Lnonloc_had})--(\ref{eq:Lnonloc_link})
leads to the generation of higher order terms in $m_\phi$ with
coefficients that in general will depend on the regulator mass,
such as the large momentum cutoff parameter $\Lambda$
(see Sec.~\ref{ssec.dipole} below).
This is analogous to a resummation of the standard chiral
perturbation theory, which goes beyond the usual power counting
regime, at the expense of introducing model dependence into
the calculation.
An advantage of this resummed approach is that one can obtain
better convergence in $m_\phi$ in regions where the usual
power counting schemes would not be applicable
(see Refs.~\cite{Young2003, Hall2010}).

\section{Splitting functions}
\label{sec.splitting}

With the nonlocal interaction and current derived in
Sec.~\ref{sec:nonlocal}, in this section we will discuss the
splitting functions describing the interaction of the external
field with the proton dressed by the pseudoscalar fields.
We will derive the general expressions for the proton $\to$
pseudoscalar meson $+$ baryon splitting functions for the full
set of SU(3) octet and decuplet states.
After giving the general results for an arbitrary
regulating function $F(a)$, we derive explicit expressions
for a specific choice of regulator in which the momentum
dependence is given by a dipole shape.

\subsection{Model independent results}

\begin{figure}[t]
\centering
\includegraphics[width=0.85\textwidth]{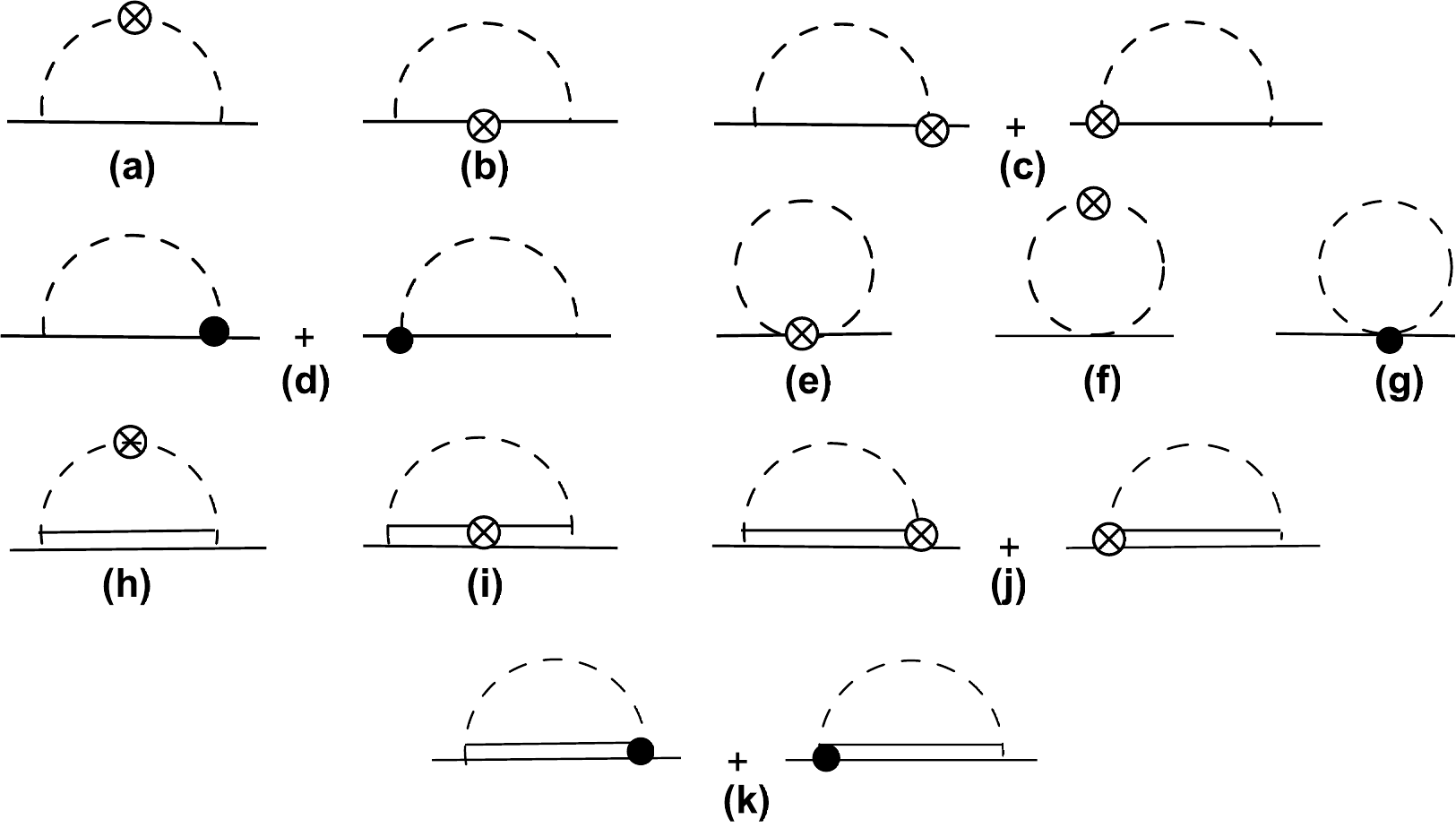}
\caption{Diagrams representing the interaction of an external current
	(denoted by the crossed circles) with the proton involving
	SU(3) octet [(a)--(g)] and decuplet [(h)--(j)] states:
	(a) and (h) are for meson coupling rainbow diagrams;
	(b) and (i) are for octet and decuplet baryon coupling
	    rainbow diagrams;
	(c) and (k) are for Kroll-Ruderman;
	(d) and (j) are for Kroll-Ruderman type diagrams generated
	    by the gauge link (denoted by the filled circle);
	(e) is for meson tadpole;
	(f) is for meson bubble; and
	(g) is for meson tadpole diagram generated by the gauge link.}
\label{fig:loop8}
\end{figure}

The interaction of an external probe with a proton dressed
by pseudovector mesons at leading order is given in
Fig.~\ref{fig:loop8} for octet intermediate states and in
Fig.~\ref{fig:loop8} for decuplet intermediate states.
The diagrams in
Figs.~\ref{fig:loop8}(a)--(c), (e), (f), (h)--(j) correspond
to those in the local effective theory, while those in
Figs.~\ref{fig:loop8}(d), (g) and (k) arise from the new
interactions in the nonlocal theory given by
Eqs.~(\ref{eq:Lnonloc_had})--(\ref{eq:Lnonloc_link}).
The resulting amplitudes will be expressed in terms of specific
meson--baryon splitting functions convoluted with corresponding
PDFs in the bare or undressed mesons and baryons.
These will be used to compute the contributions from meson loops to
PDFs in the nucleon, the most direct predictions for which will be
for nonsinglet PDF combinations in which perturbative QCD effects
largely cancel.  Examples include the light-antiquark flavor
asymmetry $\bar{d}-\bar{u}$ and the strange asymmetry $s-\bar{s}$.
In the valence approximation for the undressed hadrons, the former
will only receive contributions from the direct meson coupling
diagrams in Figs.~\ref{fig:loop8}(a), (f) and (h), while all the
diagrams in Fig.~\ref{fig:loop8} will be relevant for the
$s - \bar s$ asymmetry.

\subsubsection{SU(3) octet intermediate states}

Beginning with the meson rainbow diagram in Fig.~\ref{fig:loop8}(a),
the vertex function for the nonlocal theory can be written as
\cite{Tegen:1983gg}
\begin{eqnarray}
\boldsymbol{\Gamma}_{\phi B}^\mu\, (2\pi)^4\delta^{(4)}(p-p)
&=& \big\langle p\big|
    i^2\!\int\!d^4x\, d^4y\, d^4z\,
    {\cal L}^{\rm (nonloc)}_{{\rm had} (B)}(x)\,
    J_{q, \rm em}^\mu(y)\,
    {\cal L}^{\rm (nonloc)}_{{\rm had} (B)}(z)
    \big| p \big\rangle					\notag\\
& & \hspace*{-3cm}
 =\ \frac{i^2 C_{B\phi}^2}{f^2}
    \big\langle p \big|
    \int\!d^4x\, d^4y\, d^4z\!
    \int\!d^4a\, F(a) \int\!d^4b\, F(b)\,
    \bar p(x)\, \gamma^\nu \gamma^5 B(x)\,
    \partial_\nu \phi(x+a)				\notag\\
& & \hspace*{-2cm} \times\,
    \left( -i [ \phi(y) \partial^\mu \phi^\dag(y)
	      - \phi^\dag(y) \partial^\mu \phi(y) ]
    \right)
    \bar B(z) \gamma^\rho \gamma^5 p(z)\,
    \partial_\rho \phi^\dag(z+b)\,
    \big| p \big\rangle,
\label{eq:25}
\end{eqnarray}
where ${\cal L}^{\rm (nonloc)}_{{\rm had} (B)}$ is the part of
the hadronic nonlocal Lagrangian (\ref{eq:Lnonloc_had}) that
depends on the octet baryon fields $B$.
(Note also that we defined the vertex such that the quark flavor
charge $e^q_\phi$ is included explicitly in the bare meson and
baryon PDFs discussed in the next section.)
Integrating over the space-time coordinates $x^\mu$, $y^\mu$
and $z^\mu$, one has
\begin{eqnarray}
\boldsymbol{\Gamma}_{\phi B}^\mu
&=& \frac{C_{B\phi}^2}{f^2}\,
    \bar u(p) \int\!\frac{d^4k}{(2\pi)^4}
    \int\!d^4a\, F(a) \int\!d^4b\, F(b)\,
    (\centernot k \gamma^5)
    \frac{i \left[ (\centernot p - \centernot k) + M_B \right]}{D_B}
    (\gamma^5 \centernot k)				\notag\\
& & \hspace*{2cm} \times\
    \frac{i}{D_\phi} 2k^\mu \frac{i}{D_\phi} u(p)\,
    \exp[-ik\cdot (a-b)],
\label{eq:23}
\end{eqnarray}
where the Dirac spinor $u$ is normalized such that $\bar u u = 1$,
and $D_\phi$ and $D_B$ denote the propagator factors for the
intermediate baryon and meson, respectively,
\begin{subequations}
\begin{eqnarray}
D_\phi &=& k^2 - m_\phi^2 + i\varepsilon,
\label{eq:Dphi}					\\
D_B &=& (p-k)^2 - M_B^2 + i\varepsilon,	
\label{eq:DB}
\end{eqnarray}
\end{subequations}
where $m_\phi$ and $M_B$ are for the meson and octet baryon masses.
Defining the regulator in momentum space as
\begin{equation}
\widetilde{F}(k) \equiv \int d^4a\, \exp[-ia\cdot k]\, F(a),
\end{equation}
the vertex operator becomes
\begin{eqnarray}
\boldsymbol{\Gamma}_{\phi B}^\mu
&=& \frac{C_{B\phi}^2}{f^2}
    \bar u(p) \int\!\frac{d^4k}{(2\pi)^4}
    (\centernot k \gamma^5)\,
    \widetilde{F}(k)\,
    \frac{i[(\centernot p - \centernot k) + M_B]}{D_B}
    \frac{i}{D_\phi} 2k^\mu \frac{i}{D_\phi}
    (\gamma^5 \centernot k)\,
    \widetilde{F}(-k)\, u(p)			\notag\\
&\equiv& \int\!\frac{d^4k}{(2\pi)^4}
    \boldsymbol{\widetilde\Gamma}_{\phi B}^\mu.
\label{eq:j8}
\end{eqnarray}
Taking the $\mu = +$ component of the integrand
$\boldsymbol{\widetilde\Gamma}_{\phi B}^\mu$, we define the
splitting function $f^{\rm (rbw)}_{\phi B}(y)$ in terms of the
light-cone projection of $\boldsymbol{\widetilde\Gamma}_{\phi B}^\mu$,
\begin{eqnarray}
f^{\rm (rbw)}_{\phi B}(y)
&=& \frac{M}{p^+} \int\!\frac{d^4k}{(2\pi)^4}\,
    \boldsymbol{\widetilde\Gamma}^+_{\phi B}\,
    \delta\left(y - \frac{k^+}{p^+}\right),
\label{eq:hg}
\end{eqnarray}
where $k^+ = k^0 + k^z$ and $M$ is the nucleon mass.
From Eq.~(\ref{eq:j8}) the splitting function for the meson
rainbow diagram is then given by
\begin{eqnarray}
f^{\rm (rbw)}_{\phi B}(y)
&=& \frac{C_{B\phi}^2}{f^2}
    \int\!\frac{d^4k}{(2\pi)^4}\, \bar u(p)
    (\centernot k \gamma^5)
    \frac{i[(\centernot p - \centernot k) + M_B]}{D_B}
    \frac{i}{D_\phi} (2 k^+) \frac{i}{D_\phi}
    (\gamma^5 \centernot k)			\notag\\
& & \times\,
    \widetilde{F}^2(k)\, u(p)\,
    \frac{M}{p^+}
    \delta\left(y - \frac{k^+}{p^+}\right).
\label{eq:f-rbw-def}
\end{eqnarray}

Similarly, the splitting functions for the baryon rainbow diagram
of Fig.~\ref{fig:loop8}(b) and the Kroll-Ruderman (KR) diagram of
Fig.\ref{fig:loop8}(c) can be expressed as
\begin{eqnarray}
f^{\rm (rbw)}_{B\phi}(y)
&=& \frac{C_{B\phi}^2}{f^2}
    \int\!\frac{d^4k}{(2\pi)^4}\, \bar u(p)
    (\centernot k \gamma^5)
    \frac{i[(\centernot p - \centernot k) + M_B]}{D_B}
    \gamma^+
    \frac{i[(\centernot p - \centernot k) + M_B]}{D_B}
    (\gamma^5 \centernot k)			\notag\\
& & \hspace*{2cm} \times\
    \frac{i}{D_\phi} \widetilde{F}^2(k)\, u(p)\,
    \frac{M}{p^+} \delta\left(y - \frac{k^+}{p^+}\right)
\label{eq:bm}
\end{eqnarray}
and
\begin{eqnarray}
f^{\rm (KR)}_{B}(y)
&=& \frac{C_{B\phi}^2}{f^2}
    \int\!\frac{d^4k}{(2\pi)^4}\, \bar u(p)
    \left\{
    (i \gamma^+ \gamma^5)
    \frac{i[(\centernot p - \centernot k) + M_B]}{D_B}
    (\gamma^5 \centernot k)
    \right.					\notag\\
& & \left. \hspace*{3.5cm}
 +\ (\centernot k \gamma^5)
    \frac{i[(\centernot p - \centernot k) + M_B]}{D_B}
    (i \gamma^5 \gamma^+)
    \right\}					\notag\\
& & \hspace*{3cm} \times\
    \frac{i}{D_\phi}
    \widetilde{F}^2(k)\, u(p)\,
    \frac{M}{p^+} \delta\left(y - \frac{k^+}{p^+}\right),
\label{eq:kr}
\end{eqnarray}
respectively.

As discussed in Sec.~\ref{sec:nonlocal}, the current generated by
the gauge link in Eq.~(\ref{eq:Jlink}) produces the additional
diagrams in Fig.~\ref{fig:loop8}(d), \ref{fig:loop8}(g) and
\ref{fig:loop8}(k).  The amplitude for the Kroll-Ruderman
additional diagram in Fig.~\ref{fig:loop8}(d) can be written as
\begin{eqnarray}
\delta\boldsymbol{\Gamma}_{B}^\mu\, (2\pi)^4\delta^{(4)}(p-p)
&=& \big\langle p \big|
    i\!\int\!d^4y\, d^4z
    \left( {\cal L}^{\rm (nonloc)}_{{\rm had}(B)}(y)\,
	   \delta J_q^\mu(z)\,
        +\ \delta J_q^\mu(y)\,
	   {\cal L}^{\rm (nonloc)}_{{\rm had}(B)}(z)
    \right)						\notag\\
& & \hspace*{-4.9cm}
 =\ \frac{i C_{B\phi}^2}{f^2}
    \int_0^1 dt\, \big\langle p \big|
    \int\!d^4y\, d^4z
    \int\!d^4a\, F(a) \int\!d^4b\, F(b)			\notag\\
& & \hspace*{-4.3cm} \times\
    \Big[   -i\, b^\mu\, \bar p(y) \gamma^\nu \gamma^5 B(y)\,
	     \partial_\nu \phi(y+a)\,
	     \bar B(z-bt) \gamma^\rho \gamma^5 p(z-bt)\,
	     \partial_\rho \phi^\dag(z+b(1-t))		\notag\\
& & \hspace*{-3.5cm}
	 +\, i\, a^\mu\, \bar p(y-at) \gamma^\nu \gamma^5\, B(y-at)\,
	     \partial_\nu \phi(y+t(1-a))\,
	     \bar B(z) \gamma^\rho \gamma^5 p(z)\,
	     \partial_\rho \phi^\dag(z+b)
    \Big]
    \big| p \big\rangle,
\label{eq:a6}
\end{eqnarray}
which after Wick contraction and integration over $x^\mu$,
$y^\mu$ and $z^\mu$, becomes
\begin{eqnarray}
\delta\boldsymbol{\Gamma}_{B}^\mu
&=& \frac{i C_{B\phi}^2}{f^2}\,
    \bar u(p) \int\!d^4a\, F(a) \int\!d^4b\, F(b)
    \int\!\frac{d^4k}{(2\pi)^4}				\notag\\
& & \hspace*{0cm} \times\
    \left\{
      - i b^{\mu}\, (\centernot k \gamma^5)
      \frac{i[(\centernot p - \centernot k) + M_B]}{D_B}
      (\centernot k \gamma ^5)
      \frac{i}{D_\phi}\
      +\ i a^\mu\, (\centernot k \gamma^5)
      \frac{i}{D_\phi}
      \frac{i[(\centernot p - \centernot k) + M_B]}{D_B}
      (\centernot k \gamma^5)
    \right\}						\notag\\
& & \hspace*{0cm} \times\
    u(p)\, \exp[-i k \cdot (a-b)].
\label{eq:a10}
\end{eqnarray}
Performing the integrations over the space-time coordinates
$a^\mu$ and $b^\mu$, the vertex can be further simplified to
\begin{eqnarray}
\hspace*{-1cm}
\delta\boldsymbol{\Gamma}_{B}^\mu
&=& \frac{i C_{B\phi}^2}{f^2}\,
    \bar u(p) \int\!\frac{d^4k}{(2\pi)^4}
    \left\{
    - \frac{\partial \widetilde{F}(-k)}{\partial k^\mu} \widetilde{F}(k)
      (\centernot k \gamma^5)
      \frac{i [(\centernot p - \centernot k) + M_B]}{D_B}
      (\centernot k \gamma^5)
    \right.						\notag\\
& & \hspace*{3.7cm}
    \left.
    - \frac{\partial \widetilde{F}(k)}{\partial k^\mu}
      \widetilde{F}(-k)
      (\centernot k \gamma^5)
      \frac{i [(\centernot p - \centernot k) + M_B]}{D_B}
      (\centernot k \gamma^5)
    \right\}
    \frac{i}{D_\phi}\ u(p).
\label{eq:a9}
\end{eqnarray}
In analogy with the definition of the splitting function in
Eq.~(\ref{eq:j8}), the splitting function for the nonlocal
Kroll-Ruderman diagram in Fig.~\ref{fig:loop8}(d) induced
by the gauge link can be written as
\begin{eqnarray}
\delta f^{(\rm KR)}_{B}(y)
&=& \frac{2C_{B\phi}^2}{f^2}
    \int\!\frac{d^4k}{(2\pi)^4}\, \bar u(p)
      (i\centernot k \gamma^5)
      \frac{i [(\centernot p - \centernot k) + M_B]}{D_B}
      \frac{i}{D_\phi} (-\centernot k \gamma^5) u(p)	\notag\\
& & \hspace*{3cm} \times\
    \frac{\partial \widetilde{F}^2(k)}{\partial k^-}
    \frac{M}{p^+}\,
    \delta\left(y-\frac{k^+}{p^+}\right).
\label{eq:adoc}
\end{eqnarray}
The main additional feature here compared with the splitting
functions in the local theory is the dependence on the derivative
of the hadronic form factor $\widetilde{F}$ on $k^-$.

For the remaining meson tadpole and bubble diagrams in
Fig.~\ref{fig:loop8}(e) and \ref{fig:loop8}(f), the splitting
functions are given by
\begin{eqnarray}
f^{\rm (tad)}_\phi(y)
&=& \frac{C_{\phi\phi^\dag}}{f^2}
    \int\!\frac{d^4k}{(2\pi)^4}\,
    \bar u(p)\, \gamma^+\, \frac{i}{D_\phi}\, u(p)\,
    \widetilde{F}^2(k)\, \frac{M}{p^+}\,
    \delta\left(y-\frac{k^+}{p^+}\right),
\label{eq:tad}
\end{eqnarray}
and
\begin{eqnarray}
f^{\rm (bub)}_\phi(y)
&=&-\frac{i C_{\phi\phi^\dag}}{f^2}
    \int\!\frac{d^4k}{(2\pi)^4}\,
    \bar u(p)\, 2 \centernot k k^+\, \left( \frac{i}{D_\phi} \right)^2\,
    u(p)\,
    \widetilde{F}^2(k)\, \frac{M}{p^+}\,
    \delta\left(y-\frac{k^+}{p^+}\right),
\label{eq:bub}
\end{eqnarray}
where the coupling constant $C_{\phi\phi^\dag}$ is listed in
Table~\ref{tab:C}.

Finally, the vertex associated with the nonlocal tadpole diagram in
Fig.~\ref{fig:loop8}(g), generated by the gauge link, is defined by
\begin{eqnarray}
\delta\boldsymbol{\Gamma}_\phi^\mu\,
(2\pi)^4 \delta^{(4)}(p'-p)
&=& \big\langle p' \big|
    \int\!d^4x\, \delta J_q^\mu(x)
    \big| p \big\rangle,
\label{eq:a10b}
\end{eqnarray}
and can be reduced to
\begin{eqnarray}
\delta\boldsymbol{\Gamma}_\phi^\mu
&=& \frac{C_{\phi\phi^\dag}}{f^2}
    \int\!\frac{d^4k}{(2\pi)^4}\,
    \bar u(p)\, \centernot k \frac{i}{D_\phi}\, u(p)\,
    \left[
      \widetilde{F}(-k) \frac{\partial\widetilde{F}(k)}{\partial k^\mu}
    + \widetilde{F}(k) \frac{\partial\widetilde{F}(-k)}{\partial k^\mu}
    \right].
\label{eq:a20}
\end{eqnarray}
The splitting function for the nonlocal tadpole diagram is
then given by
\begin{eqnarray}
\delta f^{\rm (tad)}_\phi(y)
&=& \frac{C_{\phi\phi^\dag}}{f^2}
    \int\!\frac{d^4k}{(2\pi)^4}\, \bar u(p)\,
    \centernot k \frac{i}{D_\phi}\, u(p)\,
    \frac{2 \partial \widetilde{F}^2(k)}{\partial k^-}
    \frac{M}{p^+}\, \delta\left(y-\frac{k^+}{p^+}\right).
\label{eq:adtad}
\end{eqnarray}

\subsubsection{Decuplet intermediate states}

For the splitting functions associated with the decuplet
intermediate states in Fig.~\ref{fig:loop8}, the diagrams
in Figs.~\ref{fig:loop8}(h), \ref{fig:loop8}(i) and
\ref{fig:loop8}(j) arising from the local Lagangian are
supplemented by the additional nonlocal Kroll-Ruderman diagram
in Fig.~\ref{fig:loop8}(k) induced by the gauge link in the
nonlocal theory.
Similarly to the meson rainbow contribution in Eq.~(\ref{eq:25}),
the vertex function for the meson rainbow diagram in
Fig.~\ref{fig:loop8}(h) with an intermediate decuplet baryon $T$
can be written
\begin{eqnarray}
\boldsymbol{\Gamma}_{\phi T}^\mu\,
(2\pi)^4\, \delta^{(4)}(p-p)
&=& \big\langle p \big|
    i^2\!\int\!d^4x\, d^4y\, d^4z\,
    {\cal L}^{\rm (nonloc)}_{{\rm had} (T)}(x)\,
    J_{q, {\rm em}}^\mu(y)\,
    {\cal L}^{\rm (nonloc)}_{{\rm had} (T)}(z)
    \big| p \big\rangle				\notag\\
& & \hspace*{-2cm}
 =\ \frac{i^2 C_{T\phi}^2}{f^2}
    \big\langle p \big|
    \int\!d^4x\, d^4y\, d^4z\!\int\!d^4a\, F(a) \int\!d^4b\, F(b)\,
    \bar p(x) \Theta^{\alpha\beta} T_\beta(x)\,
    \partial_\alpha\phi(x+a)			\notag\\
& & \hspace*{-1.5cm} \times\
    \left\{ - i[ \phi(y) \partial^\mu \phi^\dag(y)
	       - \phi^\dag(y) \partial^\mu \phi(y) ]
    \right\}
    \overline{T}_\rho(z) \Theta^{\rho\sigma} p(z)\,
    \partial_\sigma\phi^\dag(z+b)
    \big| p \big\rangle,
\end{eqnarray}
where ${\cal L}^{\rm (nonloc)}_{{\rm had} (T)}$ is the part of the
hadronic nonlocal Lagrangian (\ref{eq:Lnonloc_had}) that depends
on the decuplet baryon fields $T$, and the operator
$\Theta^{\alpha\beta}$ is given in Eq.~(\ref{eq:Theta}).
Integrating over the space-time coordinates, one finds
\begin{eqnarray}
\boldsymbol{\Gamma}_{\phi T}^\mu
&=& \frac{i^2 C_{T\phi}^2}{f^2}\,
    \bar u(p) \int\!\frac{d^4k}{(2\pi)^4}
    \int\!d^4b\, F(b) \int\!d^4a\, F(a)\,
    k_\alpha \Theta^{\alpha\beta}\,
    \frac{-i [(\centernot p - \centernot k) + M_T]\, P_{\beta\rho}(p-k)}
	 {D_T}					\notag\\
& & \hspace*{2cm} \times\
    \frac{i}{D_\phi} 2k^\mu \frac{i}{D_\phi}
    \Theta^{\rho\sigma} k_\sigma\,
    u(p)\, \exp[-i k \cdot (a-b)],
\label{klkl}
\end{eqnarray}
where the decuplet baryon propagator $D_T$ is the same as $D_B$ in
Eq.~(\ref{eq:DB}), but with $M_B$ replaced by decuplet baryon mass $M_T$.
The spin-3/2 projection operator $P_{\alpha\beta}$, like the
octet--decuplet vertex function $\Theta_{\alpha\beta}$, depends
on the off-shell parameter $Z$, defined in Eq.~(\ref{eq:Theta}).
However, as physical quantities do not depend on $Z$, it makes sense
to simplify the form of the spin-3/2 propagator, and hence in our
calculation we choose $Z = 1/2$, following Refs.~\cite{Hacker:2005fh,
Nath:1971wp}, in which case the projector $P_{\alpha\beta}$ is written
\begin{equation}
P_{\alpha\beta}(p)
= g_{\alpha\beta}
- \frac13 \gamma_\alpha \gamma_\beta
- \frac{\gamma_\alpha p_\beta - \gamma_\beta p_\alpha}{3 M_T}
- \frac{2\, p_\alpha p_\beta}{3 M_T^2}.
\label{eq:projector}
\end{equation}
Note that for this choice one then has the operator
$\Theta^{\alpha\beta} = g^{\mu\nu} - \gamma^\mu \gamma^\nu$.
Performing the integrations over the space-time coordinates
$a^\mu$ and $b^\mu$ then gives
\begin{eqnarray}
\boldsymbol{\Gamma}_{\phi T}^\mu
&=& \frac{i^2 C_{T\phi}^2}{f^2}\,
    \bar u(p)\!\int\!\frac{d^4k}{(2\pi)^4}\,
    k_\alpha \Theta^{\alpha\beta}\, \widetilde F(k)
    \frac{-i [(\centernot p - \centernot k) + M_T]\, P_{\beta\rho}(p-k)}
	 {D_T}						\notag\\
& & \hspace*{2cm} \times\
    \frac{i}{D_\phi} 2k^\mu \frac{i}{D_\phi}
    \Theta^{\rho\sigma} k_\sigma\, \widetilde F(-k)\, u(p).
\label{eq:j88}
\end{eqnarray}
The splitting function for the meson rainbow diagram with decuplet
intermediate state is therefore given by
\begin{eqnarray}
f^{\rm (rbw)}_{\phi T}(y)
&=& \frac{C_{T\phi}^2}{f^2}
    \int\!\frac{d^4k}{(2\pi)^4}\, \bar u(p)\,
    k_\alpha \Theta^{\alpha\beta}\,
    \frac{-i [(\centernot p - \centernot k) + M_T]\, P_{\beta\rho}(p-k)}
	 {D_T}						\notag\\
& & \hspace*{2cm} \times\
    \frac{i}{D_\phi} 2k^+ \frac{i}{D_\phi}
    (-\Theta^{\rho\sigma} k_\sigma)\, u(p)\,
    \widetilde{F}^2(k)\, \frac{M}{p^+}\,
    \delta\left(y-\frac{k^+}{p^+}\right).
\label{eq:dmb}
\end{eqnarray}

Following similar procedures as for the octet baryon case, the
splitting functions for the decuplet baryon rainbow diagram in
Fig.~\ref{fig:loop8}(i) and the decuplet Kroll-Ruderman diagram
in Fig.~\ref{fig:loop8}(j) can be written as
\begin{eqnarray}
f^{\rm (rbw)}_{T\phi}(y)
&=& \frac{C_{T\phi}^2}{f^2}
    \int\!\frac{d^4k}{(2\pi)^4}\,
    \bar u(p)\, k_\mu \Theta^{\mu\nu}\,
    \frac{-i [(\centernot p - \centernot k) + M_T]\, P_{\nu\alpha}(p-k)}
	 {D_T}\, \gamma^{\alpha\beta +}			\notag\\
& & \hspace*{2cm} \times\
    \frac{-i [(\centernot p - \centernot k) + M_T]\, P_{\beta\rho}(p-k)}
	 {D_T}
    \frac{i}{D_\phi} (-\Theta^{\rho\sigma} k_\sigma)\, u(p)	\notag\\
& & \hspace*{2cm} \times\
    \widetilde{F}^2(k)\, \frac{M}{p^+}\,
    \delta\left(y-\frac{k^+}{p^+}\right)
\label{eq:dbm}
\end{eqnarray}
and
\begin{eqnarray}
f^{\rm (KR)}_{T}(y)
&=& \frac{C_{T\phi}^2}{f^2}
    \int\!\frac{d^4k}{(2\pi)^4}\, \bar u(p)\,
    \left\{
    \frac{i}{D_\phi} (i \Theta^{+ \nu})\,
    \frac{-i [(\centernot p - \centernot k) + M_T] P_{\nu\alpha}(p-k)}
	 {D_T}
    (-\Theta^{\alpha\sigma} k_\sigma)
    \right.					\notag\\
& & \left. \hspace*{3.5cm}
 +\ k_\mu \Theta^{\mu\nu}\,
    \frac{-i [(\centernot p - \centernot k) + M_T] P_{\nu\alpha}(p-k)}
	 {D_T}\,
    (-i\Theta^{\alpha +})
    \frac{i}{D_\phi}
    \right\}\, u(p)				\notag\\
& & \hspace*{3cm} \times\
    \widetilde{F}^2(k)\, \frac{M}{p^+}\,
    \delta\left(y-\frac{k^+}{p^+}\right),
\label{eq:dkr}
\end{eqnarray}
respectively.
Finally, the splitting function for the nonlocal Kroll-Ruderman
decuplet diagram in Fig.~\ref{fig:loop8}(k) induced by the gauge
link is
\begin{eqnarray}
\delta f^{\rm (KR)}_{T}(y)
&=& \frac{2 C_{T\phi}^2}{f^2}
    \int\!\frac{d^4k}{(2\pi)^4}\, \bar u(p)\,
    (i k_\sigma \Theta^{\sigma\nu})\,
    \frac{-i [(\centernot p - \centernot k) + M_T] P_{\nu\alpha}(p-k)}
	 {D_T}
    \frac{i}{D_\phi} (-\Theta^{\alpha\sigma} k_\sigma) u(p)	\notag\\
& & \hspace*{2cm} \times
    \frac{\partial\widetilde{F}^2(k)}{\partial k^-}\, \frac{M}{p^+}\,
    \delta\left(y-\frac{k^+}{p^+}\right).
\label{eq:dad}
\end{eqnarray}
The set of functions
$\big\{$
$f^{\rm (rbw)}_{\phi B}$, $f^{\rm (rbw)}_{B\phi}$,
$f^{\rm (KR)}_{B}$, $\delta f^{\rm (KR)}_{B}$,
$f^{\rm (bub)}_\phi$, $f^{\rm (tad)}_\phi$, $\delta f^{\rm (tad)}_\phi$
$\big\}$ for the octet baryons,
and
$\big\{$
$f^{\rm (rbw)}_{\phi T}$, $f^{\rm (rbw)}_{T\phi}$,
$f^{\rm (KR)}_{T}$, $\delta f^{\rm (KR)}_{T}$
$\big\}$
for the decuplet baryons,
then represent the complete set of functions that describe the
dressing at one loop of the interaction of an external current
with the proton in the nonlocal meson--baryon field theory.

\subsection{Covariant dipole form factor}
\label{ssec.dipole}

To evaluate the splitting functions derived in the previous section
requires a specific choice for the meson--baryon vertex form factor
$\widetilde F(k)$.  Consistency with Lorentz invariance restricts
the form factor to in general be a function of the meson virtuality
$k^2$ and the baryon virtuality $(p-k)^2$.
For illustration, we choose the regulator to have a simple dipole
shape in $k^2$ with a cutoff parameter $\Lambda$ \cite{Forkel:1994yx,
Musolf:1993fu}, independent of the details of the baryon state,
\begin{equation}
\widetilde{F}(k)
= \frac{\overline\Lambda^4}{D_\Lambda^2},
\label{eq:re}
\end{equation}
where $D_\Lambda = k^2 - \Lambda^2 + i\varepsilon$ and we define
$\overline\Lambda^2 \equiv \Lambda^2 - m_\phi^2$.
Other forms, such as Guassian, monopole or sharp cutoff, have also been
used in the literature~\cite{Young2003, Hall2010}, and, with appropriate
choices of regulator mass for the different regulators, give rise to
qualitatively similar results.
An advantage of the dipole form (\ref{eq:re}) is that it allows a more
direct comparison with previous literature~\cite{Forkel:1994yx,
Musolf:1993fu, MTS91} that has used the same functional form.

\subsubsection{Octet splitting functions}

With the dipole regulator in Eq.~(\ref{eq:re}), after reduction of the
$\gamma$ matrices in Eq.~(\ref{eq:f-rbw-def}) the splitting function for
the meson rainbow diagram in Fig.~\ref{fig:loop8}(a) can be written as
\begin{eqnarray}
f^{\rm (rbw)}_{\phi B}(y)
&=& \frac{i C_{B\phi}^2 \overline\Lambda^8}{f^2}
    \int\!\frac{d^4k}{(2\pi)^4}
    \left[ \frac{y \overline{M}^2 (\Delta ^2 - m_\phi^2)}
		{D_\phi^2\, D_B\, D_\Lambda^4}
	 - \frac{y \overline{M}^2}
		{D_\phi\, D_B\, D_\Lambda^4}
	 + \frac{y (\overline{M} \Delta - 2 p \cdot k)}
		{D_\phi^2\, D_\Lambda^4}
    \right]					\notag\\
& & \hspace*{3cm} \times\
    \delta\left(y-\frac{k^+}{p^+}\right),
\label{eq:31}
\end{eqnarray}
where the average mass $\overline{M}$ and mass difference $\Delta$
are defined as
\begin{equation}
\overline{M} = M + M_B, ~~~~~~ \Delta = M_B - M.
\label{eq:mass8}
\end{equation}
It will be convenient to perform the $d^4k$ integration in terms of
light-cone momentum components $k^\pm = k^0 \pm k^z$ and transverse
momentum ${\bm k}_\perp$.
The first two terms in Eq.~(\ref{eq:31}) have poles both on the
upper and lower half-plane, so the integration over $k^-$ can be
obtained using the residue of $D_B$ or $D_\phi$.
For the third term, proportional to $1/D_\phi^2$, when $k^+ \neq 0$
both $D_\phi$ and $D_\Lambda$ have poles on same half-plane, so the
integral vanishes.
On the other hand, when $k^+ = 0$ the integral becomes divergent.
We can simplify this term using
\begin{eqnarray}
\int\!d^4k\, \frac{2y\, p\cdot k}{D_\phi^2\, D_\Lambda^4}
&=& \frac{\partial^4}{6\, \partial\Omega^4}
    \int_0^1 dz \int\!d^4k\,
    \frac{2p\cdot k\ y(1-z) z^3}
	 {(k^2 - \Omega + i\varepsilon)^2}	\notag\\
&=& \frac{\partial^4}{6\, \partial\Omega^4}
    \int_0^1 dz \int d^4k\,
    \frac{(1-z) z^3}{(k^2 - \Omega + i\varepsilon)},
\label{eq:identity}
\end{eqnarray}
where we define
\begin{equation}
\Omega\ \equiv\ (1-z) m_\phi^2 + z \Lambda^2.
\end{equation}
The integration over $k^-$ in Eq.~(\ref{eq:identity}) can be
written as \cite{Burkardt:2001iy, Burkardt12}
\begin{equation}
\int_{-\infty}^{\infty} dk^- \frac{1}{k^2 - \Omega + i\varepsilon}
= 2\pi i\, \log\left(\frac{k_\perp^2+\Omega}{\mu^2}\right) \delta(k^+),
\end{equation}
where $\mu$ is a momentum independent constant.
After the $k^-$ integration, the splitting function for the meson
rainbow diagram can be expressed as a sum of an on-shell term,
$f^{\rm (on)}_B$, and $\delta$-function terms,
$f^{(\delta)}_\phi$ and $\delta f^{(\delta)}_\phi$,
generated by the contact interaction,
\begin{eqnarray}
f^{\rm (rbw)}_{\phi B}(y)
&=& \frac{C_{B\phi}^2 \overline{M}^2 }{(4\pi f)^2}
    \Big[ f^{\rm (on)}_B(y)
	+ f^{(\delta)}_\phi(y)
	- \delta f^{(\delta)}_\phi(y)
    \Big].
\label{eq:phiB-rbw}
\end{eqnarray}
The on-shell function is given by
\begin{eqnarray}
f^{\rm (on)}_B(y)
&=& \overline\Lambda^8
    \int\!dk_\bot^2\,
    \frac{y\, \big[ k_\bot^2 + (y M + \Delta)^2 \big]}
	 {\bar{y}^2\, D_{\phi B}^2\, D_{\Lambda B}^4},
\label{eq:phiB-rbw-on}
\end{eqnarray}
where $\bar{y} = 1 - y$, and we employed the shorthand notations
\cite{XGWangPRD}
\begin{subequations}
\label{eq:LL}
\begin{eqnarray}
D_{\phi B}
&=& -\frac{k_\bot^2 + y M_B^2 - y\, \bar{y}\, M^2 + \bar{y}\, m_\phi^2}
	  {\bar{y}},			\\
D_{\Lambda B}
&=& -\frac{k_\bot^2 + y M_B^2 - y\, \bar{y}\, M^2 + \bar{y}\, \Lambda^2}
	  {\bar{y}}.
\end{eqnarray}
\end{subequations}
The $\delta$-function contributions are nonzero only at $y=0$,
and arise from the local and nonlocal interactions.
The local $\delta$-function term is given by
\begin{eqnarray}
f^{(\delta)}_\phi(y)
&=& -\frac {\overline\Lambda^8}{\overline{M}^2}
    \int dk_\bot^2 \int_0^1 dz\,
    \frac{z^3}{(k_\perp^2+\Omega)^4}\,
    \delta(y)				\notag\\
&=& \frac{1}{\overline{M}^2}
    \int dk_\bot^2
    \left[
      \log\frac{\Omega_\phi}{\Omega_\Lambda}
    + \frac{\overline\Lambda^2 (11\, \Omega_\Lambda^2
			 	- 7\, \Omega_\Lambda \Omega_\phi
				+ 2\, \Omega_\phi^2)}
	   {6\Omega_\Lambda^3}
    \right]
    \delta(y),
\label{eq:phi-delta}
\end{eqnarray}
with
\begin{equation}
\Omega_\phi\    =\ k_\bot^2 + m_\phi^2\ , ~~~~~~~~
\Omega_\Lambda\ =\ k_\bot^2 + \Lambda^2.
\label{eq:Omega_phi_L}
\end{equation}
The $\log\Omega_\phi$ term in Eq.~(\ref{eq:phi-delta}) gives rise to
the leading nonanalytic contribution, which is independent of the
regularization method, as we have verified using various methods,
including Pauli-Villars, dimensional regularization or a hadronic
form factor.
In the limit when $\Lambda \to \infty$, the second term in
Eq.~(\ref{eq:phi-delta}) $\sim \overline\Lambda^2/\Omega_\Lambda$
becomes a constant.
Within dimensional regularization, the integral of a constant
is defined to be zero, in which case the result coincides
with that in Ref.~\cite{XGWangPLB},
\begin{eqnarray}
f^{(\delta)}_\phi(y)
&\ \underset{\Lambda\to\infty}{\longrightarrow}\ \ &
\frac{1}{\overline{M}^2}
\int dk_\bot^2\,
\log\frac{\Omega_\phi}{\Omega_\Lambda}\, \delta(y).
\end{eqnarray}

The nonlocal $\delta$-function contribution, $\delta f^{(\delta)}_B$,
in Eq.~(\ref{eq:phiB-rbw}) is given by
\begin{eqnarray}
\delta f^{(\delta)}_\phi(y)
&=&-\frac{\overline\Lambda^8}{\overline{M}^2}
    \int dk_\bot^2 \int_0^1 dz\,
    \frac{z^4}{(k_\perp^2+\Omega)^4}\,
    \delta(y)					\nonumber\\
&=& \frac{1}{\overline{M}^2}
    \int dk_\bot^2
    \left[ - 4 \frac{\Omega_\phi}{\overline\Lambda^2}
             \log\frac{\Omega_\phi}{\Omega_\Lambda}
	   - \frac{3 \Omega_\Lambda^3
		   + 13 \Omega_\Lambda^2 \Omega_\phi
		   -  5 \Omega_\Lambda \Omega_\phi^2
		   +    \Omega_\phi^3}
		  {3 \Omega_\Lambda^3}
    \right]\, \delta(y).
\label{eq:phi-deltalink}
\end{eqnarray}
In the $\Lambda \to \infty$ limit the first term in the integrand of
$\delta f^{(\delta)}_\phi$ vanishes, while the second term becomes
a constant, independent of $k_\bot$.  In dimensional regularization
the latter can again be taken to be zero.
The local function $f^{(\delta)}_\phi$, on the other hand, retains
a dependence on $k_\bot$ through the $\log\Omega_\phi$ term,
so that the splitting function for the rainbow diagram in
Eq.~(\ref{eq:phiB-rbw}) will reduce in this limit to the local
splitting function.
In the same limit, for the case $\phi=\pi$ and $B=N$, the integrand
of Eq.~(\ref{eq:phiB-rbw-on}) reduces to the familiar on-shell form
found in the literature \cite{Speth:1996pz, Holtmann:1996be, MST98},
\begin{eqnarray}
f_{\pi^+ n}^{\rm (on)}(y)
&\longrightarrow&
    \int\!dk_\bot^2
    \frac{y\, \big( k_\bot^2 + y^2 M^2 \big)}
	 {\big[ k_\bot^2 + y^2 M^2 + \bar{y}\, m_\pi^2 \big]^2}
\end{eqnarray}
for the specific dissociation $p \to \pi^+ n$.

For the baryon coupling rainbow diagram, Fig.~\ref{fig:loop8}(b),
the splitting function in Eq.~(\ref{eq:bm}) can be reduced to
\begin{eqnarray}
f^{\rm (rbw)}_{B\phi}(y)
&=& \frac{i C_{B\phi}^2 \overline\Lambda^8}{f^2}
    \int\!\frac{d^4k}{(2\pi)^4}
    \left[ \frac{\bar{y}\, \overline{M}^2 (\Delta^2 - m_\phi^2)}
		{D_B^2 D_\phi D_\Lambda^4}
	 - \frac{\bar{y} \overline{M}^2}{D_B^2 D_\Lambda^4}
	 + \frac{(2-y)\, \overline{M}\Delta}{D_B D_\phi D_\Lambda^4}
	 + \frac{1}{D_\phi D_\Lambda^4}
    \right]					\notag\\
& & \hspace*{3cm} \times\
    \delta\left(y-\frac{k^+}{p^+}\right).
\label{eq:Bphi-rbw}
\end{eqnarray}
Performing the $k^-$ integral, this can then be expressed as a sum
of on-shell, local and nonlocal off-shell, and $\delta$-function terms,
\begin{eqnarray}
f^{\rm (rbw)}_{B\phi}(y)
&=& \frac{C_{B\phi}^2 \overline{M}^2}{(4\pi f)^2}
    \Big[ f^{\rm (on)}_B(y)
	+ f^{\rm (off)}_B(y)
	+ 4\, \delta f^{\rm (off)}_B(y)
	- f^{(\delta)}_\phi(y)
    \Big].
\label{eq:f-rbw-Bphi}
\end{eqnarray}
Note that the on-shell splitting functions for the baryon and meson
couplings are equivalent, while the $\delta$-function contribution
$f^{(\delta)}_\phi$ is as in Eq.~(\ref{eq:phi-delta}).
The off-shell contributions in Eq.~(\ref{eq:f-rbw-Bphi}) include
local and nonlocal terms.  The local off-shell contribution,
\begin{eqnarray}
f^{\rm (off)}_B(y)
&=& \frac{2\overline\Lambda^8}{\overline{M}}
    \int dk_\bot^2
    \frac{ (yM + \Delta)}
	 {\bar y\, D_{\phi B}\, D_{\Lambda B}^4},
\label{eq:f-off-Bphi}
\end{eqnarray}
is similar to that derived in Refs.~\cite{JMT13, XGWangPLB},
while the nonlocal off-shell term is given by
\begin{eqnarray}
\delta f^{\rm (off)}_B(y)
&=& \overline\Lambda^8
    \int dk_\bot^2
    \frac{y \big[k_\bot^2 + (y M + \Delta)^2 \big]}
	 {\bar y^2\, D_{\phi B}\, D_{\Lambda B}^5}.
\label{eq:f-offlink-Bphi}
\end{eqnarray}
In the $\Lambda \to \infty$ limit, the nonlocal term behaves
as $\overline\Lambda^8/D_{\Lambda B}^5 \sim 1/\Lambda^2$,
so vanishes, as expected.

For the Kroll-Ruderman diagram in Fig.~\ref{fig:loop8}(c), the
splitting function in Eq.~(\ref{eq:kr}) for the dipole regulator
becomes
\begin{eqnarray}
f^{\rm (KR)}_B(y)
&=& -\frac{2i C_{B\phi}^2 \overline\Lambda^8}{f^2}
    \int\!\frac{d^4k}{(2\pi)^4}\,
    \bigg[ \frac{(y M + \Delta) \overline{M}}{D_\phi D_B D_\Lambda^4}
	 + \frac{1}{D_\phi D_\Lambda^4}
    \bigg]
    \delta\left(y-\frac{k^+}{p^+}\right),
\end{eqnarray}
which after the $k^-$ integration can be written in terms of the
off-shell and $\delta$-function terms,
\begin{eqnarray}
f^{\rm (KR)}_B(y)
&=& \frac{C_{B\phi}^2 \overline{M}^2}{(4\pi f)^2}
    \Big[- f^{\rm (off)}_B(y)
	 + 2 f^{(\delta)}_\phi(y)
    \Big],
\label{eq:f-KR-Bphi}
\end{eqnarray}
as given in Eqs.~(\ref{eq:phi-delta}) and (\ref{eq:f-off-Bphi}).
(Note that the notation used here differs slightly from that of
Ref.~\cite{XGWangPRD}, where for strange octet baryons coupled to
kaons the Kroll-Ruderman function was labelled by $f^{\rm (KR)}_{YK}$;
here we drop the meson label, as for a proton target the choice of
baryon intermediate state uniquely specifies the meson, and we also
label the $\delta$-function contribution by the baryon involved
rather than the meson.)
For the nonlocal gauge link contribution in Fig.~\ref{fig:loop8}(d),
reduction of the Dirac matrices with the dipole form factor allows
the corresponding splitting function $\delta f^{\rm (KR)}_B$ to be
rearranged as
\begin{eqnarray}
\delta f^{\rm (KR)}_B(y)
&=& \frac{i C_{B\phi}^2 \overline\Lambda^8}{f^2}
    \int\!\frac{d^4k}{(2\pi)^4}
    \left[ - \frac{4 y \overline{M}^2 (\Delta^2 - m_\phi^2)}
		  {D_\phi D_B D_\Lambda ^5}
	   + \frac{4 y \overline{M}^2}{D_B D_\Lambda^5}
	   - \frac{4 y (\overline{M} \Delta - 2 p \cdot k)}
		  {D_\phi D_\Lambda^5}
    \right]				\notag\\
& & \hspace*{3cm} \times\
    \delta\left(y-\frac{k^+}{p^+}\right).
\end{eqnarray}
After the $k^-$ integration, this reduces to a sum of the
nonlocal off-shell and $\delta$-function contributions,
\begin{eqnarray}
\delta f^{\rm (KR)}_B(y)
&=& \frac{C_{B\phi}^2 \overline{M}^2}{(4\pi f)^2}
    \left[-4\, \delta f^{\rm (off)}_B(y)\,
	  -\,  \delta f^{(\delta)}_\phi(y)
    \right],
\label{eq:f-link-Bphi}
\end{eqnarray}
as given in Eqs.~(\ref{eq:phi-deltalink}) and (\ref{eq:f-offlink-Bphi}),
respectively.
From Eqs.~(\ref{eq:phiB-rbw}), (\ref{eq:f-rbw-Bphi}),
(\ref{eq:f-KR-Bphi}) and (\ref{eq:f-link-Bphi}) one can verify
that the splitting functions satisfy the relation
\begin{equation}
  f^{\rm (rbw)}_{\phi B}(y)
= f^{\rm (rbw)}_{B\phi}(y)
+ f^{\rm (KR)}_B(y)
+ \delta f^{\rm (KR)}_B(y),
\label{eq:g0}
\end{equation}
which generalizes the result in Ref.~\cite{XGWangPLB}
to the nonlocal theory.
Note that the local and nonlocal off-shell contributions
$f^{\rm (off)}_B$ and $\delta f^{\rm (off)}_B$ cancel between
the three terms on the right hand side of Eq.~(\ref{eq:g0}).
As noted above, in the $\Lambda \to \infty$ limit each of the
functions induced by the nonlocal gauge link,
	$\delta f^{\rm (off)}_B$ and
	$\delta f^{(\delta)}_\phi$,
vanishes, reproducing the local result from Ref.~\cite{JMT13} that
does not include the gauge link function $\delta f^{\rm (KR)}_B$.
Remarkably, the nonlocal generalization (\ref{eq:g0}) means that
gauge invariance is satisfied even in the presence of a finite
form factor cutoff $\Lambda$!

A similar analysis can be applied to the tadpole and bubble diagrams in
Fig.~\ref{fig:loop8}(e)--(g) in the presence of a hadronic form factor.
From Eq.~(\ref{eq:tad}), the splitting function for the tadpole
contribution with the dipole form factor can be written as
\begin{eqnarray}
f^{\rm (tad)}_\phi(y)
&=&-\frac{C_{\phi\phi^\dag} \overline{M}^2}{(4\pi f)^2}
    f^{(\delta)}_\phi(y),
\label{eq:f-tad}
\end{eqnarray}
where $f^{(\delta)}_\phi$ is given in Eq.~(\ref{eq:phi-delta}).
For the bubble diagram in Eq.~(\ref{eq:bub}) the corresponding
splitting function is given by
\begin{eqnarray}
f^{\rm (bub)}_\phi(y)
&=&-\frac{C_{\phi\phi^\dag} \overline{M}^2}{(4\pi f)^2}
    \Big[ f^{(\delta)}_\phi(y)
	- \delta f^{(\delta)}_\phi(y)
    \Big],
\label{eq:f-bub}
\end{eqnarray}
where the nonlocal function $\delta f^{(\delta)}_\phi$ is given by
Eq.~(\ref{eq:phi-deltalink}).
Finally, the splitting function for the nonlocal tadpole gauge link
diagram in Fig.~\ref{fig:loop8}(g) from Eq.~(\ref{eq:adtad}) with
a dipole regulator is
\begin{eqnarray}
\delta f^{\rm (tad)}_\phi(y)
&=& \frac{C_{\phi\phi^\dag} \overline{M}^2}{(4\pi f)^2}
    \delta f^{(\delta)}_\phi(y).
\label{eq:ad45}
\end{eqnarray}
Combining Eqs.~(\ref{eq:f-tad})--(\ref{eq:ad45}), one finds that
the tadpole and bubble diagrams satisfy the generalized relation
\begin{eqnarray}
f^{\rm (bub)}_\phi(y)
&=& f^{\rm (tad)}_\phi(y)
 +  \delta f^{\rm (tad)}_\phi(y),
\label{eq:g1}
\end{eqnarray}
which confirms the gauge invariance of the nonlocal theory.

\subsubsection{Decuplet splitting functions}

Turning now to the splitting functions for the decuplet baryon
intermediate states in Fig.~\ref{fig:loop8}(h)--\ref{fig:loop8}(k),
the contribution from the rainbow diagram with coupling to the
pseudoscalar meson in Eq.~(\ref{eq:dmb}) for the covariant dipole
form factor (\ref{eq:re}) is given by
\begin{eqnarray}
f^{\rm (rbw)}_{\phi T}(y)
&=& \frac{i C_{T\phi}^2\, \overline\Lambda^8}{6 M_T^2 f^2}
    \int\!\frac{d^4k}{(2\pi)^4}
\Bigg[
  \frac{y (\overline{M}_T^2 - m_\phi^2)^2 (\Delta_T^2 - m_\phi^2)}
       {D_\phi^2\, D_T\, D_\Lambda^4}			\notag\\
&-&
  \frac{y (\overline{M}_T^2 - m_\phi^2)
	  (\overline{M}_T^2 + 2 \Delta_T^2 - 3 m_\phi^2)}
       {D_\phi\, D_T\, D_\Lambda^4}
+ \frac{y (2 \overline{M}_T^2 + \Delta_T^2 - k^2 - 2 m_\phi^2)}
       {D_T\, D_\Lambda^4}				\notag\\
&+&
  \frac{y}{D_\phi^2\, D_\Lambda^4}
  \Big( 4 (p \cdot k)^2
      - 2 (\overline{M}_T^2 - k^2)\, p \cdot k
      + (M_T^2 - k^2)^2					\notag\\
& & \hspace*{1cm}
      + M (2 M_T^3 - M^3 - 2 M^2 M_T)
      - 2 M k^2 (2 M + M_T)
  \Big)
\Bigg] \delta\left(y-\frac{k^+}{p^+}\right),
\label{eq:80}
\end{eqnarray}
where the coupling constants $C_{T\phi}$ for the decuplet intermediate
states are listed in Table~\ref{tab:C}, and the masses $\overline{M}_T$
and $\Delta_T$ here are defined in analogy with Eq.~(\ref{eq:mass8}),
\begin{equation}
\overline{M}_T = M + M_T, ~~~~~~ \Delta_T = M_T - M.
\label{eq:mass10}
\end{equation}
After performing the $k^-$ integration, the splitting function can be
decomposed in terms of on-shell decuplet, end point, and local and
nonlocal $\delta$-function terms,
\begin{eqnarray}
f^{\rm (rbw)}_{\phi T}(y)
&=& \frac{C_{T\phi}^2\overline{M}_T^2 }{(4\pi f)^2}
\Bigg[ f^{\rm (on)}_T(y)
     + f^{\rm (on\, end)}_T(y)
     - \frac{1}{18} f^{(\delta)}_T(y)		\notag\\
& & \hspace*{1.5cm}
   +\, \frac{\overline{M}^2 (\overline{M}_T^2 - m_\phi^2)}
	    {6 M_T^2\, \overline{M}_T^2}
       \Big( f^{(\delta)}_\phi(y) - \delta f^{(\delta)}_\phi(y) \Big)
\Bigg].
\label{eq:dkaon}
\end{eqnarray}
As for the octet case, the first term in Eq.~(\ref{eq:dkaon})
is the on-shell splitting function for the meson rainbow with
a decuplet spectator,
\begin{eqnarray}
f^{(\rm on)}_T(y)
&=& \frac{\overline\Lambda^8}{6 M_T^2\, \overline{M}_T^2}
\int\!dk_\bot^2
\frac{y\, (\overline{M}_T^2 - m_\phi^2)}{\bar y}
\Bigg[
  \frac{(\overline{M}_T^2 - m_\phi^2) (\Delta_T^2 - m_\phi^2)}
       {D_{\phi T}^2\, D_{\Lambda T}^4}
- \frac{3 (\Delta_T^2 - m_\phi^2) + 4 M M_T}
       {D_{\phi T}\, D_{\Lambda T}^4}
\Bigg],						\notag\\
& &
\label{eq:Ton}
\end{eqnarray}
where $D_{\phi T}$ and $D_{\Lambda T}$ are defined analogously
to Eqs.~(\ref{eq:LL}),
\begin{subequations}
\begin{eqnarray}
D_{\phi T}
&=& -\frac{k_\bot^2 + y M_T^2 - y\, \bar{y}\, M^2 + \bar{y}\, m_\phi^2}
	  {\bar{y}},			\\
D_{\Lambda T}
&=& -\frac{k_\bot^2 + y M_T^2 - y\, \bar{y}\, M^2 + \bar{y}\, \Lambda^2}
	  {\bar{y}}.
\label{eq:DLT}
\end{eqnarray}
\end{subequations}
Since $\overline\Lambda^8/D_{\Lambda T}^4 \to 1$ in the
$\Lambda \to \infty$ limit, the decuplet on-shell function
(\ref{eq:Ton}) reduces to the pointlike result found in
Ref.~\cite{Salamu:2014pka}.

The function $f^{(\rm on\, end)}_T$ in Eq.~(\ref{eq:dkaon})
is finite for finite values of $\Lambda$,
\begin{eqnarray}
f^{(\rm on\, end)}_T(y)
&=& \frac{\overline\Lambda^8}{6 M_T^2\, \overline{M}_T^2}
\int\!dk_\bot^2\,
\frac{y}{\bar{y}^2\, D_{\Lambda T}^4}		\notag\\
& & \times
\left[ k_\bot^2 + y^2 M^2
     - 2 y (\overline{M}_T^2 - M \Delta_T)
     - 2 {\bar y}\, m_\phi^2
     + 3 \overline{M}_T^2 - 4 M M_T
\right],
\label{eq:Tend}
\end{eqnarray}
but in the $\Lambda \to \infty$ limit corresponds to the end point
function in Ref.~\cite{Salamu:2014pka}, with a singularity at $y=1$.
To see this, first note that $D_{\Lambda T}$ in Eq.~(\ref{eq:DLT})
can be written in the form
  $\bar{y} D_{\Lambda T} = -(X_T+ \bar{y}\, \Omega_\Lambda)$,
where
  $X_T = y\Omega_T - y\bar{y} M^2$
and
  $\Omega_T = k_\bot^2 + M_T^2$.
In the $\Lambda \to \infty$ limit, one can then write the factor
\begin{eqnarray}
\frac{\overline\Lambda^8}{\bar{y}^4 D_{\Lambda T}^4}
&\ \underset{\Lambda\to\infty}{\longrightarrow}\ \ &
  \mathop{\lim}\limits_{\Omega_0 \to \infty}
  \left.
  \int_{\Omega_0}^{\Omega_T} dt
  \frac{-4 y \overline\Lambda^8}
       {(y\, t - y\, \bar{y} M^2 + \bar{y}\, \Omega_\Lambda)^5}
  \right|_{\Lambda\to\infty}			\notag\\
&=& \frac{\overline\Lambda^6}{\bar{y}^3 \Omega_\Lambda^3}
  \mathop{\lim}\limits_{\Omega_0 \to \infty}
  \bigg(
    \frac{\overline\Lambda^2 \bar{y}^3 \Omega_\Lambda^3}
	 {\bar{y}^4 D_{\Lambda T}^4}
   -\frac{\overline\Lambda^2 \bar{y}^3 \Omega_\Lambda^3}
	 {\bar{y}^4 D_0^4}
  \bigg)_{\Lambda\to\infty},
\label{eq:ghj}
\end{eqnarray}
where
  $\bar{y} D_0 = -(X_0+ \bar{y}\, \Omega_\Lambda)$,
with
  $X_0 = y\Omega_0 - y\bar{y} M^2$
and
  $\Omega_0$ is a $\Lambda$-independent constant.
At finite $\Lambda$, the term involving $D_0$ vanishes; however,
care must be taken when evaluating this for $\Lambda \to \infty$.
Replacing $\bar{y} \Omega_\Lambda$ in the first and second terms
in Eq.~(\ref{eq:ghj}) by
  $(-\bar{y} D_{\Lambda T} - X_T)$
and
  $(-\bar{y} D_0 - X_0)$,
respectively, one obtains
\begin{eqnarray}
\frac{\overline\Lambda^8}{\bar{y}^4 D_{\Lambda T}^4}
&\ \underset{\Lambda\to\infty}{\longrightarrow}\ \ &
- \frac{\overline\Lambda^6}{\bar{y}^3 \Omega_\Lambda^3}
  \mathop{\lim}\limits_{\Omega_0 \to \infty}
  \Bigg[
    \bigg(
      \frac{\overline\Lambda^2}{\bar{y} D_{\Lambda T}}
     -\frac{\overline\Lambda^2}{\bar{y} D_0}
    \bigg)\,
+\ 3\bigg(
      \frac{\overline\Lambda^2 X_T}{\bar{y}^2 D_{\Lambda T}^2}
     -\frac{\overline\Lambda^2 X_0}{\bar{y}^2 D_0^2}
    \bigg)							\notag\\
& & \hspace*{2cm}
+\ 3\bigg(
      \frac{\overline\Lambda^2 X_T^2}{{\bar y}^3 D_{\Lambda T}^3}
     -\frac{\overline\Lambda^2 X_0^2}{{\bar y}^3 D_0^3}
   \bigg)\,
+\ \bigg(
     \frac{\overline\Lambda^2 X_T^3}{{\bar y}^4 D_{\Lambda T}^4}
    -\frac{\overline\Lambda^2 X_0^3}{{\bar y}^4 D_0^4}
   \bigg)
\Bigg]_{\Lambda\to\infty}.
\label{eq:interim}
\end{eqnarray}
Since in the $\Lambda \to \infty$ limit one has
  $\bar{y} D_{\Lambda T} \to
   -\overline{\Lambda}^2 (\bar{y} + X_T/\Omega_\Lambda)$,
the first term in parentheses in Eq.~(\ref{eq:interim}) can be written
\begin{eqnarray}
\Bigg(
  \frac{\overline\Lambda^2}{\bar{y} D_{\Lambda T}}
 -\frac{\overline\Lambda^2}{\bar{y} D_0}
\Bigg)_{\Lambda\to\infty}
&=&
-\Bigg(
  \frac{1}{\bar y + X_T/\Omega_\Lambda}
 -\frac{1}{\bar y + X_0/\Omega_\Lambda}
\Bigg)_{\Lambda\to\infty},
\label{eq:HGHG}
\end{eqnarray}
where we have taken $\Omega_0 \ll \Lambda^2$.
The right hand side of Eq.~(\ref{eq:HGHG}) has the properties that
it vanishes if $\bar{y} \not= 0$, is divergent if $\bar{y} = 0$,
and becomes $\log(X_T/X_0)$ when integrated over $\bar{y}$, so that
it can be represented by a $\delta$ function,
\begin{eqnarray}
\Bigg(
  \frac{\overline\Lambda^2}{\bar{y} D_{\Lambda T}}
 -\frac{\overline\Lambda^2}{\bar{y} D_0}
\Bigg)_{\Lambda\to\infty}
&=& \delta(\bar y) \log\frac{X_T}{X_0}.
\end{eqnarray}
Similarly, for the $1/(\bar{y} D_{\Lambda T})^n$ terms in
Eq.~(\ref{eq:ghj}) with $n \geq 2$, one can write in the
$\Lambda \to \infty$ limit
\begin{eqnarray}
\frac{\overline\Lambda^2 X_T^{n-1}}
     {(-\bar{y})^n\, D_{\Lambda T}^n}
\Bigg|_{\Lambda\to\infty}\
&=&
\frac{(X_T/\Lambda^2)^{n-1}}
     {(\bar{y} + X_T/\Omega_\Lambda)^n}
\Bigg|_{\Lambda\to\infty}\
=\ \frac{\delta(\bar{y})}{n-1},\ \ \ \ \ \ n \geq 2.
\label{eq:gy}
\end{eqnarray}
Since the same result is obtained when $X_T$ is replaced by $X_0$,
the $1/(\overline y D_{\Lambda T})^n$ and $1/ (\overline y D_0)^n$
terms cancel for $n \geq 2$, and one obtains
\begin{eqnarray}
\frac{\overline\Lambda^8}{\bar{y}^4 D_{\Lambda T}^4}
&\ \underset{\Lambda\to\infty}{\longrightarrow}\ \ &
   -\frac{1}{\bar{y}^3}
    \log\frac{\Omega_T}{\Omega_0}\, \delta(\bar y)\
=\ -\frac{1}{\bar{y}^3}
    \left( \log\frac{\Omega_T}{\mu^2}-1 \right) \delta(\bar{y}),
\label{eq:RTRTR}
\end{eqnarray}
where $\mu$ is defined such that
  $\log(\Omega_T/\mu^2) = \log(\Omega_T/\Omega_0) + 1$.
With this result, one can finally write the end point splitting
function in the $\Lambda \to \infty$ limit as
\begin{eqnarray}
f^{\rm (on\, end)}_T(y)
& \underset{\Lambda\to\infty}{\longrightarrow}\ &
\frac{1}{6 M_T^2\, \overline{M}_T^2}
\int\!dk_\bot^2
\Big\{
  \Big[ \Omega_T - 2 (\Delta_T^2 - m_\phi^2) - 6 M M_T \Big]
  \log\frac{\Omega_T}{\mu^2}
- \Omega_T					\notag\\
& & \hspace*{4cm}
+\, 2 (\Delta_T^2 - m_\phi^2) + 6 M M_T
\Big\}
\delta(\bar{y}).
\label{eq:933}
\end{eqnarray}
This expression is identical to that for the end point term in
Ref.~\cite{Salamu:2014pka}, except for the $k_\bot$-independent
terms in (\ref{eq:933}).  For dimensional regularization, however,
these are again defined to be zero, so that the result does indeed
match that in \cite{Salamu:2014pka}.

Note also that at finite values of $\Lambda$ the sum of the on-shell
function $f^{(\rm on)}_T$ in Eq.~(\ref{eq:Ton}) and the on-shell
end point function $f^{(\rm on\, end)}_T$ in Eq.~(\ref{eq:Tend})
gives the usual result found in the literature by taking the pole
contribution alone \cite{Speth:1996pz, Kumano:1997cy, Salamu:2014pka,
MTS91, MST98},
\begin{eqnarray}
\hspace*{-0.3cm}
f^{(\rm on)}_T(y) + f^{(\rm on\, end)}_T(y)
&=& \frac{\overline\Lambda^8}{6 M_T^2\, \overline{M}_T^2}
\int\!dk_\bot^2\,
\frac{y \left[ k_\bot^2 + (\Delta_T + y M)^2 \right]
	\left[ k_\bot^2 + (\overline{M}_T - y M)^2 \right]^2}
     {\bar{y}^4 D_{\phi T}^2\, D_{\Lambda T}^4},
\label{eq:Tsull}
\end{eqnarray}
for $0 < y < 1$.
Separately, however, the on-shell and end point functions are not
guaranteed to be positive definite for large values of $m_\phi$,
since the individual functions do not correspond to physical
processes~\cite{Xuangong-pc}.
The combined contribution in Eq.~(\ref{eq:Tsull}) is, however,
positive for any combination of masses and kinematics.

For the $\delta$-function contributions at $y=0$, there are
three distinct terms in the decuplet rainbow function
$f^{\rm (rbw)}_{\phi T}$.  The new decuplet $\delta$-function
term in Eq.~(\ref{eq:dkaon}) for the nonlocal case is given by
\begin{eqnarray}
\hspace*{-0.8cm}
f^{(\delta)}_T(y)
&=& \frac{\overline\Lambda^8}{M_T^2\, \overline{M}_T^2}
    \int\!dk_\bot^2 \int_0^1 dz\,
    \frac{z^3}{(k_\perp^2+\Omega)^3}\,
    \delta(y)						\notag\\
&=& \frac{1}{M_T^2\, \overline{M}_T^2}
    \int dk_\bot^2
    \frac{1}{2 \Omega_\Lambda^2}
    \bigg[ 6 \Omega_\Lambda^2 \Omega_\phi
	   \log\frac{\Omega_\phi}{\Omega_\Lambda}
  + (\Omega_\phi-\Omega_\Lambda)
    (\Omega_\phi^2 - 5 \Omega_\phi \Omega_\Lambda - 2\Omega_\Lambda^2)
    \bigg]
    \delta(y),
\label{eq:fTdelta}
\end{eqnarray}
where $\Omega_\phi$ and $\Omega_\Lambda$ are as in
Eq.~(\ref{eq:Omega_phi_L}).
In the $\Lambda \to \infty$ limit, only the first term in the
integrand of Eq.~(\ref{eq:fTdelta}) survives, so that the local
limit of the function $f^{(\delta)}_T$ is
\begin{eqnarray}
f^{(\delta)}_T(y)
&\ \underset{\Lambda\to\infty}{\longrightarrow}\ &
\frac{3}{M_T^2\, \overline{M}_T^2}
\int dk_\bot^2
\bigg[ \Omega_\phi \log\frac{\Omega_\phi}{\mu^2}
     - \Omega_\phi
\bigg] \delta(y),
\label{ghg}
\end{eqnarray}
where the constant $\mu$ here is defined by
$ \log(\Omega_\phi/\mu^2) = \log(\Omega_\phi/\Omega_\Lambda) + 17/6$.

The remaining $\delta$-function terms in Eq.~(\ref{eq:dkaon}),
namely, the local $f^{(\delta)}_\phi$ and nonlocal
$\delta f^{(\delta)}_\phi$ functions, are given in
Eqs.~(\ref{eq:phi-delta}) and (\ref{eq:phi-deltalink}), respectively.
The combined contribution of the $\delta$-function terms to
$f^{\rm (rbw)}_{\phi T}$ in the local limit is then
\begin{eqnarray}
\frac{1}{18}
\left[
  \frac{3 \overline{M}^2 (\overline{M}_T^2 - m_\phi^2)}
       {M_T^2\, \overline{M}_T^2} f_\phi^{(\delta)}
- f_T^{(\delta)}
\right]
& \underset{\Lambda\to\infty}{\longrightarrow}\ &
\frac{1}{6 M_T^2 \overline{M}_T^2}
\int\!dk_\perp^2
\left[
  \Omega_\phi
+ \big(  \overline{M}_T^2-m_\phi^2
	- \Omega_\phi
  \big) \log\frac{\Omega_\phi}{\mu^2}
\right]
\delta(y).				\nonumber\\
& &
\end{eqnarray}\\
Note that this expression differs from the total local $\delta(y)$
contribution in Ref.~\cite{Salamu:2014pka}, which was computed using
the projector $P_{\alpha\beta}$ in Eq.~(\ref{eq:projector}) but with
$Z=-1/2$ for the interaction $\Theta^{\mu\nu}$ in Eq.~(\ref{eq:Theta}).
As discussed in Ref.~\cite{Salamu:2014pka}, for values of the off-shell
parameter $Z \neq -1/2$, the additional interaction term
$\sim \gamma_\mu \gamma_\nu$ in $\Theta^{\mu\nu}$ contributes only
to the $\delta(y)$ contribution.
The result here supercedes that in Ref.~\cite{Salamu:2014pka}.

For the decuplet baryon coupling rainbow diagram in
Fig.~\ref{fig:loop8}(i), reduction of the $\gamma$-matrices in
Eq.~(\ref{eq:dbm}) yields
\begin{eqnarray}
f^{\rm (rbw)}_{T \phi}(y)
&=&
\frac{i C_{T\phi}^2\, \overline\Lambda^8}{6 M_T^2 f^2}
\int\!\!\frac{d^4k}{(2\pi)^4}
\Bigg[
   \frac{\bar{y} (\overline{M}_T^2-m_\phi^2)^2 (\Delta_T^2 - m_\phi^2)}
	{D_\phi\, D_T^2\, D_\Lambda^4}			\notag\\
& &
 + \frac{\bar{y}
	 \big[ (k^2+m_\phi^2)\, (2\overline{M}_T^2+\Delta_T^2-m_\phi^2)
	     - \overline{M}_T^4 - 2\overline{M}_T^2 \Delta_T^2 - k^4
         \big]}
	{D_T^2\, D_\Lambda^4}				\notag\\
& &
 - \frac{(\overline{M}_T^2-m_\phi^2)
	 \left[ (y-2) M_T^2 - 2y\, M M_T + (y+2) (M^2 - m_\phi^2)
	 \right]}
	{D_T\ D_\phi\, D_\Lambda^4}			\notag\\
& &
 + \frac{(y+2) (2 M^2 - m_\phi^2 - k^2) + (y M_T + 2M) 2M_T}
	{D_T\, D_\Lambda^4}				\notag\\
& &
 + \frac{\overline{M}_T^2 + 2y\, \overline{M}_T M
	 - 2y\, p \cdot k + 3 k^2}
	{D_\phi\, D_\Lambda^4}
\Bigg]
\delta\left(y-\frac{k^+}{p^+}\right).
\label{eq:fTphi-rbw}
\end{eqnarray}
Integrating over $k^-$, the splitting function for the decuplet
coupling rainbow diagram can be written analogously to the
function $f^{\rm (rbw)}_{\phi T}$ in Eq.~(\ref{eq:dkaon}),
\begin{eqnarray}
f^{\rm (rbw)}_{T\phi}(y)
&=& \frac{C_{T\phi}^2 \overline{M}_T^2}{(4\pi f)^2}
\Bigg[
  f^{\rm (on)}_T(y)
+ f^{\rm (on\, end)}_T(y)
- 2\, \Big( f^{\rm (off)}_T(y)
	  + f^{\rm (off\, end)}_T(y)
	  - 2\, \delta f^{\rm (off)}_T(y)
      \Big)						\notag\\
& & \hspace*{1.5cm}
+\ \frac{1}{18}
   \Big( f^{(\delta)}_T(y) - 3\, \delta f^{(\delta)}_T(y)
   \Big)
-  \frac{\overline{M}^2 (\overline{M}_T^2 + 3 m_\phi^2)}
        {6 M_T^2\, \overline{M}_T^2}\, f^{(\delta)}_\phi(y)
\Bigg].
\label{eq:dsigma}
\end{eqnarray}
The first term in Eq.~(\ref{eq:dsigma}) is the on-shell splitting
function for the decuplet baryon rainbow, and is identical to
that for the meson coupling rainbow in Eq.~(\ref{eq:dkaon}).
The second term is the same as the end point function contribution
in Eq.~(\ref{eq:Tend}).

The off-shell decuplet contributions to $f^{\rm (rbw)}_{T\phi}$
appear as three individual terms --- a local off-shell piece,
$f^{\rm (on)}_T$, an off-shell end point contribution,
$f^{\rm (off\, end)}_T$, and a purely nonlocal term,
$\delta f^{\rm (off)}_T$.
The local off-shell function is given by
\begin{eqnarray}
f^{\rm (off)}_T(y)
&=& \frac{\overline\Lambda^8}{6 M_T^2 \overline{M}_T^2}
\int\!dk_\bot^2
\frac{(\overline M_T^2-m_\phi^2)\,
      \left[ \bar{y}\, (M^2 - m_\phi^2) - (1+y) M_T^2 \right]}
     {\bar{y}\, D_{\phi T}\, D_{\Lambda T}^4},
\label{eq:Tphi-delta1}
\end{eqnarray}
which in the $\Lambda \to \infty$ limit reduces to
\begin{eqnarray}
f^{\rm (off)}_T(y)
& \underset{\Lambda\to\infty}{\longrightarrow}\ &
\frac{1}{6 M_T^2 \overline{M}_T^2}
\int\!dk_\bot^2
\frac{(\overline M_T^2-m_\phi^2)\,
      \left[ \bar{y}\, (M^2 - m_\phi^2) - (1+y) M_T^2 \right]}
     {\bar{y}\, D_{\phi T}}.
\label{eq:Tphi-delta1_Linf}
\end{eqnarray}

In addition to the end point function for the on-shell contribution
in Eq.~(\ref{eq:Tend}), a separate end point contribution exists
for the off-shell case, $f^{\rm (off\, end)}_T$, and is given by
\begin{eqnarray}
f^{\rm (off\, end)}_T(y)
&=& -\frac{\overline\Lambda^8}{6 M_T^2 \overline{M}_T^2}
\int\!dk_\bot^2
\frac{\left[ k_\bot^2
	   + \bar{y}^2\, M^2
	   + \bar{y}\, (\overline{M}_T^2 - m_\phi^2)
	   - M_T^2
      \right]}
     {\bar{y}\, D_{\Lambda T}^4}.
\label{eq:Tphi-offend}
\end{eqnarray}
Using the relation in Eq.~(\ref{eq:RTRTR}), one can show that in the
$\Lambda \to \infty$ limit this term is proportional to a $\delta$
function at $y=1$,
\begin{eqnarray}
f^{\rm (off\, end)}_T(y)
& \underset{\Lambda\to\infty}{\longrightarrow}\ &
\frac{1}{6 M_T^2 \overline{M}_T^2}
\int\!dk_\bot^2
\Big\{
  \big[ \Omega_T - 2 M_T^2 \big]
  \log\frac{\Omega_T}{\mu^2}
- \Omega_T
\Big\}\,
\delta(\bar{y}).
\label{eq:Tphi-delta2}
\end{eqnarray}
As for the octet case in Eq.~(\ref{eq:f-offlink-Bphi}), the decuplet
splitting function also includes a nonlocal decuplet off-shell term,
given by
\begin{eqnarray}
\delta f^{\rm (off)}_T(y)
&=& \frac{\overline\Lambda^8}{6 M_T^2 \overline{M}_T^2}
\int\!dk_\bot^2
\frac{y\, \big[ k_\bot^2 + (y M - \overline{M_T})^2 \big]^2
	  \big[ k_\bot^2 + (y M + \Delta_T)^2 \big]}
     {\bar{y}^4\, D_{\phi T}\, D_{\Lambda T}^5},
\label{eq:T-offlink-delta}
\end{eqnarray}
The presence of the $1/D_{\Lambda T}^5$ in the integrand of
(\ref{eq:T-offlink-delta}) ensures that in the $\Lambda \to \infty$
limit the nonlocal function vanishes, $\delta f^{\rm (off)}_T \to 0$.

For the $\delta$-function contributions at $y=0$, the local terms
$f^{(\delta)}_\phi$ and $f^{(\delta)}_T$ in Eq.~(\ref{eq:dsigma}) are
given above in Eqs.~(\ref{eq:phi-delta}) and (\ref{eq:phi-deltalink}),
respectively, while the new nonlocal $\delta$-function term,
$\delta f^{(\delta)}_T$, is given by
\begin{eqnarray}
\delta f^{(\delta)}_T(y)
&=& \frac{\overline\Lambda^8}{M_T^2 \overline{M}_T^2}
\int\!dk_\bot^2
\frac{1}{\Omega_\Lambda^3}\, \delta(y).
\label{eq:Omega-delta}
\end{eqnarray}
As with the other nonlocal contributions, this term also vanishes
in the $\Lambda \to \infty$ limit.

The final diagram in Fig.~\ref{fig:loop8} is that for the
Kroll-Ruderman contribution with a decuplet intermediate state,
Fig.~\ref{fig:loop8}(j).  The splitting function corresponding
to this diagram, after reducing the $\gamma$-matrices in
Eq.~(\ref{eq:dkr}), can be written
\begin{eqnarray}
f^{(\rm KR)}_T(y)
&=& -i \frac{C_{T\phi}^2 \overline\Lambda^8}{3 M_T^2\, f^2}
\int\!\frac{d^4k}{(2\pi)^4}
\Bigg[
   \frac{(\overline{M}_T^2 - m_\phi^2)^2
	 \big[ (1+y) M_T^2 - \bar{y} (M^2 - m_\phi^2) \big]}
	{D_\phi\, D_T\, D_\Lambda^4}			\notag\\
& & \hspace*{1.5cm}
+\ \frac{(1-y) k^2 - 2(1+y)\, p \cdot k
	 + y (2 M^2+\overline{M}_T^2) + \overline{M}_T^2}
	{D_\phi\, D_\Lambda^4}				\notag\\
& & \hspace*{1.5cm}
-\ \frac{2 y M_T^2 + \bar{y} (k^2 + m_\phi^2 - 2 M \overline{M}_T)}
	{D_T\, D_\Lambda^4}				
\Bigg]
\delta\left(y-\frac{k^+}{p^+}\right).
\label{eq:93}
\end{eqnarray}
After integrating over $k^-$, the splitting function for the
decuplet KR diagram can be expressed in terms of local and nonlocal
off-shell and $\delta$-function terms,
\begin{eqnarray}
f^{(\rm KR)}_T(y)
&=& \frac{C_{T\phi}^2 \overline{M}_T^2}{(4\pi f)^2}
\Bigg[
2 \left( f^{\rm (off)}_T(y) + f^{\rm (off\, end)}_T(y) \right)
- \frac{1}{9}
  \left( f^{(\delta)}_T(y) - \delta f^{(\delta)}_T(y) \right)
						\notag\\
& & \hspace*{1.7cm}
+\ \frac{\overline{M}^2 (\overline{M}_T^2 + m_\phi^2)}
        {3 M_T^2\, \overline{M}_T^2}\, f^{(\delta)}_\phi(y)
\Bigg],
\label{eq:fTKR}
\end{eqnarray}
each of which has been defined previously.
Finally, the splitting function for the additional decuplet diagram
induced by the gauge link, Fig.~\ref{fig:loop8}(k), is obtained from
Eq.~(\ref{eq:dad}),
\begin{eqnarray}
\delta f^{\rm (KR)}_T(y)
&=& -2i \frac{C_{T\phi}^2 \overline\Lambda^8}{3 M_T^2 f^2}
\int\!\!\frac{d^4k}{(2\pi)^4}\, y\,
\Bigg[
  \frac{(\overline{M}_T^2 - m_\phi^2)^2 (\Delta_T^2 - m_\phi^2)}
       {D_\phi\, D_T\, D_\Lambda^5}				\notag\\
& &
+ \frac{(k^2 + m_\phi^2)(2 \overline{M}_T^2 + \Delta_T^2 - m_\phi^2)
	-(\overline{M}_T^2 - 2\Delta_T^2) \overline{M}_T^2 - k^4}
       {D_T\, D_\Lambda^5}					\notag\\
& &
+ \frac{1}{D_\phi\, D_\Lambda^5}
  \Big(
    4 (p \cdot k)^2 + 3 m_\phi^4
    - (3 k^2 + 3 \overline{M}_T^2 + \Delta_T^2 - \overline{M}_T \Delta_T)
      m_\phi^2							\notag\\
& & \hspace*{2cm}
    - (4 k^2 - 6 m_\phi^2 + 2 \overline{M}_T^2)\, (p \cdot k)
    + k^2\, \overline M_T^2
    + \overline{M}_T^3\, \Delta_T
    + k^4
  \Big)								\notag\\
& &
- \frac{1}{D_\Lambda^5}
  \Big( 3\,M_T^2+ 5\,M^2+4\,M\, M_T-3 m_\phi^2- 6\,  p \cdot k \Big)
\Bigg]
\delta\left(y-\frac{k^+}{p^+}\right).
\label{eq:dfTKRdef}
\end{eqnarray}
With integration over $k^-$, the splitting function for the
nonlocal KR gauge link diagram can be simplified to a sum of
nonlocal off-shell and $\delta$-function contributions,
\begin{eqnarray}
\delta f^{(\rm KR)}_T(y)
&=& \frac{C_{T\phi}^2 \overline{M}_T^2}{(4\pi f)^2}
\Bigg[
  - 4\, \delta f^{\rm (off)}_T(y)
  + \frac{1}{18} \delta f^{(\delta)}_T(y)\,
  -\, \frac{\overline{M}^2 (\overline{M}_T^2 - m_\phi^2)}
           {6 M_T^2\, \overline{M}_T^2}\,
    \delta f^{(\delta)}_{\phi}(y)
\Bigg].
\label{eq:fdfTKR}
\end{eqnarray}
From Eqs.~(\ref{eq:dkaon}), (\ref{eq:fTphi-rbw}), (\ref{eq:fTKR})
and (\ref{eq:fdfTKR}), one can then explicitly verify that gauge
invariance for the decuplet baryon contributions is satisfied
through the relation
\begin{eqnarray}
f^{\rm (rbw)}_{\phi T}(y)
&=& f^{\rm (rbw)}_{T \phi}(y)
 +  f^{\rm (KR)}_T(y)
 +  \delta f^{\rm (KR)}_T(y).
\label{eq:g2}
\end{eqnarray}
This generalizes the result from Ref.~\cite{XGWangPRD} to nonlocal
interactions in the presence of vertex functions parametrizing the
extended nature of the proton.

\newpage
\subsection{Leading nonanalytic behavior}

Having derived the complete set of splitting functions for the one-loop
diagrams in Fig.~\ref{fig:loop8} for the dissociation of a proton to a
pseudoscalar meson ($\phi$) and an SU(3) octet ($B$) or decuplet ($T$)
baryon, in the rest of this section we discuss the characteristics of
each of the functions and illustrate their relative shapes and magnitudes
numerically.
The full set of functions includes 8 basis functions that are nonzero
in the local limit,
 $\{ f^{\rm (on)}_B$,
    $f^{\rm (off)}_B$,
    $f^{\rm (on)}_T$,
    $f^{\rm (on\, end)}_T$,
    $f^{\rm (off)}_T$,
    $f^{\rm (off\, end)}_T$,
    $f^{(\delta)}_T$,
    $f^{(\delta)}_\phi \}$,
and 4 nonlocal functions,
 $\{ \delta f^{\rm (off)}_B$,
    $\delta f^{\rm (off)}_T$,
    $\delta f^{(\delta)}_T$,
    $\delta f^{(\delta)}_\phi \}$,
that vanish for pointlike particles.
All of the diagrams in Fig.~\ref{fig:loop8} are then represented by
splitting functions that can be written as linear combinations of
these basis functions.

Before presenting the numerical results for the splitting functions
for the case of the covariant dipole form factor in Eq.~(\ref{eq:re}),
we first identify some features of the basis functions that do not
depend on details of the regularization method, but are entirely
determined by the infrared behavior of the chiral loops.
Namely, expanding the lowest moments $\langle\, f_i\, \rangle$ of
the basis splitting functions,
\begin{eqnarray}
\langle f_i \rangle
&=& \int_0^1 dy\, f_i(y),
\label{eq:intfy}
\end{eqnarray}
as a series in the pseudoscalar meson mass $m_\phi$, the coefficients
of terms that are nonanalytic (NA) in $m_\phi^2$ (either odd powers of
$m_\phi$ or logarithms of $m_\phi$) are determined by the low-energy
properties of the nucleon and do not depend on the ultraviolet behavior
of the functions \cite{TMS00, Chen01, Arndt02, Chen02, Detmold01}.
In particular, the moments of the on-shell and off-shell functions
  $f^{(\rm on)}_B$, $f^{(\rm off)}_B$,
  $f^{(\rm on)}_T$, $f^{(\rm off)}_T$
and the $\delta$-function terms
  $f^{(\delta)}_\phi$ and $f^{(\delta)}_T$
all receive NA contributions, while the purely nonlocal functions
and the decuplet end-point contributions $f^{(\rm on\, end)}_T$
and $f^{(\rm off\, end)}_T$ are entirely analytic.

For the octet intermediate states, we find the NA contribution
to the on-shell moment $\langle f^{\rm (on)}_B \rangle$ is given by
\begin{eqnarray}
\label{eq:fonLNA}
\overline{M}^2\,
\big\langle f^{(\rm on)}_B \big\rangle\Big|_{\rm NA}
= \left\{
\begin{array}{l}
  (4 m_\phi^2 - 6 \Delta^2) \log m_\phi^2
   + 6 R\, \Delta\, \log\dfrac{\Delta-R}{\Delta+R},
			\hspace*{1.9cm} \Delta > m_\phi, \\
							\\
  (4 m_\phi^2 - 6 \Delta^2) \log m_\phi^2
   + 6 \overline{R}\, \Delta
     \Big( \pi - 2 \arctan\dfrac{\Delta}{\overline{R}} \Big),
			\hspace*{0.7cm} \Delta < m_\phi,
\end{array}
  \right.
\end{eqnarray}
where
$R=\sqrt{\Delta^2 - m_\phi^2}$ and
$\overline{R}=\sqrt{m_\phi^2-\Delta^2}$.
This agrees with the result found in Ref.~\cite{XGWangPRD} for
strange octet contributions.
In particular, for the latter case, when $\Delta <  m_\phi$, the
mass difference $\Delta$ approaches zero first in the chiral limit,
$m_\phi \to 0$.  The resulting LNA term is then simply
	$4 m_\phi^2 \log m_\phi^2$,
consistent with Refs.~\cite{Arndt02, Chen01, Chen02, Detmold01,
Burkardt12, Salamu:2014pka}.
For the case $\Delta > m_\phi$, expanding $R$ as
	$R = \Delta - m_\phi^2/2 \Delta + {\cal O}(m_\phi^4)$
one finds that the $\Delta^2 \log m_\phi^2$ terms cancel, leaving
behind the same LNA behavior $\sim m_\phi^2 \log m_\phi^2$,
\begin{eqnarray}
\label{eq:fonLNAexp}
\overline{M}^2\,
\big\langle f^{(\rm on)}_B \big\rangle\Big|_{\rm LNA}
&=& (4 m_\phi^2 - 6 \Delta^2) \log m_\phi^2
   + 6 \Delta^2 \log m_\phi^2 -3 m_\phi^2 \log m_\phi^2  \notag\\
&=& m_\phi^2 \log m_\phi^2,
\hspace*{5.5cm} \Delta > m_\phi.
\end{eqnarray}
but with a coefficient that is now 4 times smaller than
for the $\Delta <  m_\phi$ case.

For the off-shell moment $\langle f^{\rm (off)}_B \rangle$,
the NA contribution is
\begin{equation}
\overline{M}^2\,
\big\langle f^{(\rm off)}_B \big\rangle\Big|_{\rm NA}
= \left\{
\begin{array}{l}
  - 2 m_\phi^2 \log m_\phi^2
    - \dfrac{2 R^3}{M_B} \log\dfrac{\Delta-R}{\Delta+R},
			\hspace*{2.5cm} \Delta > m_\phi,\\
							\\
  - 2 m_\phi^2 \log m_\phi^2
    + \dfrac{2 \overline{R}^3}{M_B}
      \Big( \pi - 2 \arctan\dfrac{\Delta}{\overline{R}} \Big),
			\hspace*{1.3cm} \Delta < m_\phi.
\end{array}
  \right.
\end{equation}
The LNA behavior of the moment, $\langle f^{(\delta)}_\phi \rangle$,
of the $\delta$-function term is
\begin{equation}
\overline{M}^2\,
\big\langle f^{(\delta)}_\phi \big\rangle\Big|_{\rm LNA}
= - m_\phi^2 \log m_\phi^2.
\end{equation}
These results generalize the LNA expressions given for hyperons
and kaons in Ref.~\cite{XGWangPRD}.

For the decuplet intermediate states, the NA term for the
on-shell moment $\langle f^{(\rm on)}_T \rangle$ is
\begin{eqnarray}
\overline{M}_T^2\,
\big\langle f^{(\rm on)}_T \big\rangle\Big|_{\rm NA}
= \left\{
\begin{array}{l}
  \dfrac{(8 m_\phi^2 - 12 \Delta_T^2)}{3} \log m_\phi^2
  + 4 R_T \Delta_T\, \log\dfrac{\Delta_T-R_T}{\Delta_T+R_T},
		\hspace*{1.3cm} \Delta_T > m_\phi,	\\
							\\
  \dfrac{(8 m_\phi^2 - 12 \Delta_T^2)}{3} \log m_\phi^2
  + 4 \overline{R}_T \Delta_T
    \Big( \pi - 2 \arctan\dfrac{\Delta_T}{\overline{R}_T} \Big),
		\hspace*{0.3cm} \Delta_T < m_\phi,
\end{array}
  \right.
\end{eqnarray}
where $R_T=\sqrt {\Delta_T^2-m_\phi ^2}$ and
$\overline{R}_T=\sqrt{m_\phi^2 - \Delta_T^2}$.
For the case $\Delta_T < m_\phi$, one finds in the $\Delta_T \to 0$
limit the LNA behavior $\tfrac{8}{3}\, m_\phi^2 \log m_\phi^2$.
For $\Delta_T > m_\phi$, one may again expand $R_T$ as
	$R_T = \Delta_T - m_\phi^2/2\Delta_T + {\cal O}(m_\phi^4)$,
and note that the LNA term remains $\sim m_\phi^2 \log m_\phi^2$ due to
a cancellation of the terms proportional to $\Delta_T^2 \log m_\phi^2$,
\begin{eqnarray}
\label{eq:fTonLNAexp}
\overline{M}_T^2\,
\big\langle f^{(\rm on)}_T \big\rangle\Big|_{\rm LNA}
&=& \dfrac{(8 m_\phi^2 - 12 \Delta_T^2)}{3} \log m_\phi^2
 +  4 \Delta_T^2 \log m_\phi^2 - 2 m_\phi^2 \log m_\phi^2     \notag\\
&=& \dfrac{2}{3} m_\phi^2 \log m_\phi^2,
\hspace*{5.5cm} \Delta_T > m_\phi.
\end{eqnarray}
In both cases, therefore, the LNA term is given by
$m_\phi^2 \log m_\phi^2$, although the the coefficient for
$\Delta_T > m_\phi$ is 4 times smaller than that for
$\Delta_T < m_\phi$ in the chiral limit.

The NA contribution to the moment of the decuplet off-shell
function $\langle f^{(\rm off)}_T \rangle$ is given by
\begin{equation}
\overline{M}_T^2\,
\big\langle f^{(\rm off)}_T \big\rangle\Big|_{\rm NA}
= \left\{
\begin{array}{l}
  \dfrac{2}{3} m_\phi^2 \log m_\phi^2
  + \dfrac{4 R_T^3}{3 M_T} \log\dfrac{\Delta_T-R_T}{\Delta_T+R_T},
		\hspace*{2.9cm} \Delta_T > m_\phi,	\\
							\\
  \dfrac{2}{3} m_\phi^2 \log m_\phi^2
  - \dfrac{4 \overline R_T^3}{3 M_T}
    \Big( \pi - 2 \arctan\dfrac{\Delta_T}{\overline{R}_T} \Big),
		\hspace*{1.9cm} \Delta_T < m_\phi.
\end{array}
  \right.
\end{equation}
The decuplet $\delta$-function moment does not have an LNA term,
but has contributions at higher order in $m_\pi$,
\begin{equation}
\overline{M}_T^2\,
\big\langle f^{(\delta)}_T \big\rangle\Big|_{\rm LNA}
= 0.
\end{equation}
The decuplet results for the total LNA behavior coincide with those
for the $\pi \Delta$ intermediate states in Ref.~\cite{Salamu:2014pka},
arising from the $f_T^{(\rm on)}$ and $f_\phi^{(\delta)}$ terms in
Eq.~(\ref{eq:dkaon}), if the $\pi N \Delta$ coupling constant
$g_{\pi N \Delta}$ in \cite{Salamu:2014pka} is related to the
meson--octet--decuplet coupling constant ${\cal C}$ in
Eq.~(\ref{eq:ch8}) by $g_{\pi N \Delta}^2 = {\cal C}^2/(2 f^2)$.

We stress that these results are completely general, depending
only on the infrared properties of pseudoscalar meson loops,
following directly from the symmetries of the chiral Lagrangian.
They are independent of short-distance contributions, which are
model dependent, and so provide us with a powerful tool that
can be used to verify whether any model is consistent with the
chiral symmetry properties of QCD.

\subsection{Phenomenology of meson--baryon splitting functions}

In this section we explore the features of the meson--baryon
splitting functions for the various octet and decuplet contributions
that are nonzero at $y > 0$, for a finite dipole cutoff parameter
$\Lambda$ in Eq.~(\ref{eq:re}).
For illustration, we consider the nucleon and lightest $\Lambda$
hyperon states for the octet baryons, and the $\Delta$ and $\Sigma^*$
for the decuplet states.
Unless otherwise indicated, we will use a typical value for the
cutoff mass of $\Lambda = 1$~GeV.

\begin{figure}[t]
\centering
\begin{tabular}{ccc}
\hspace{-0.3cm}{\epsfxsize=3.5in\epsfbox{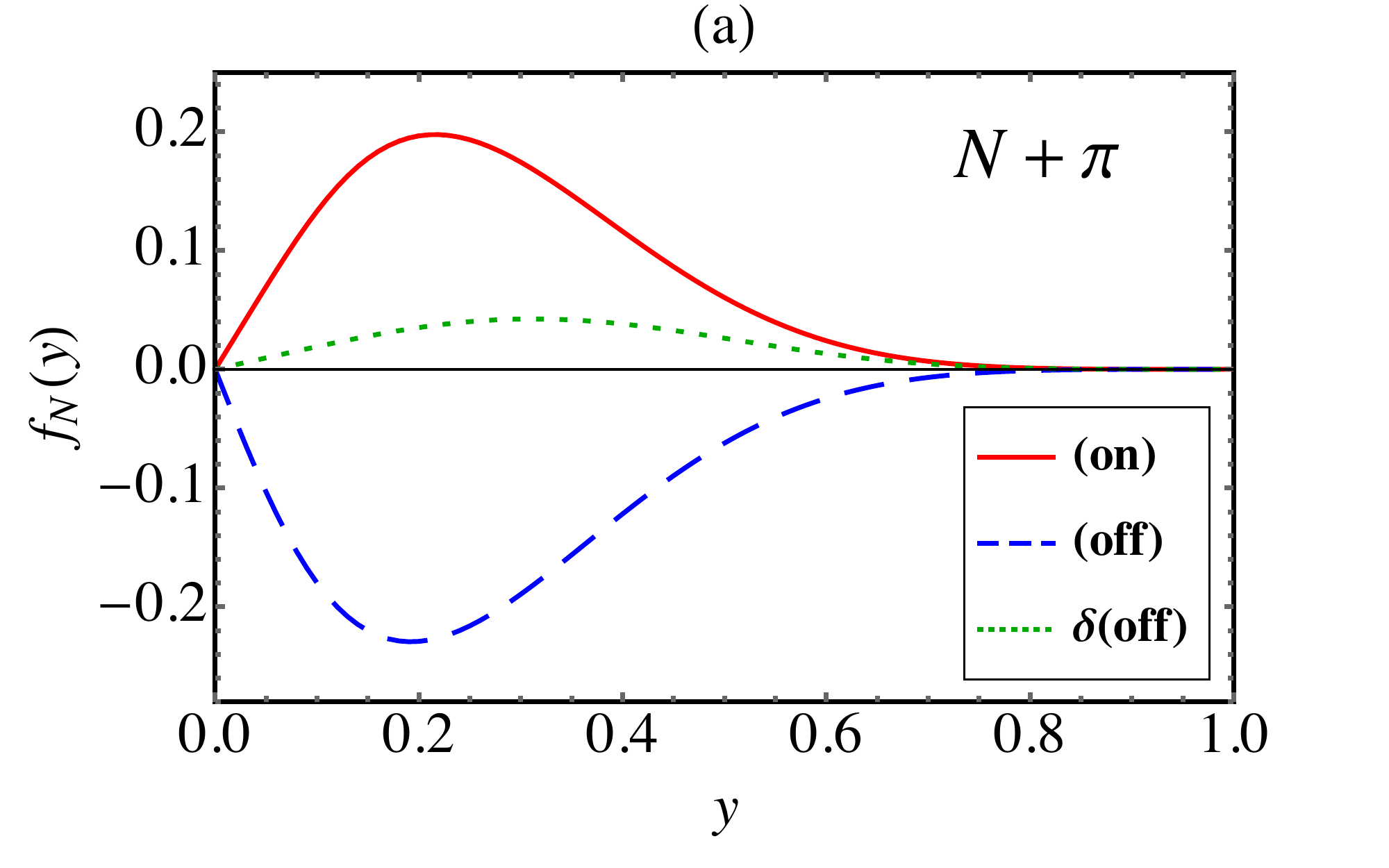}}&
\hspace{-0.5cm}{\epsfxsize=3.5in\epsfbox{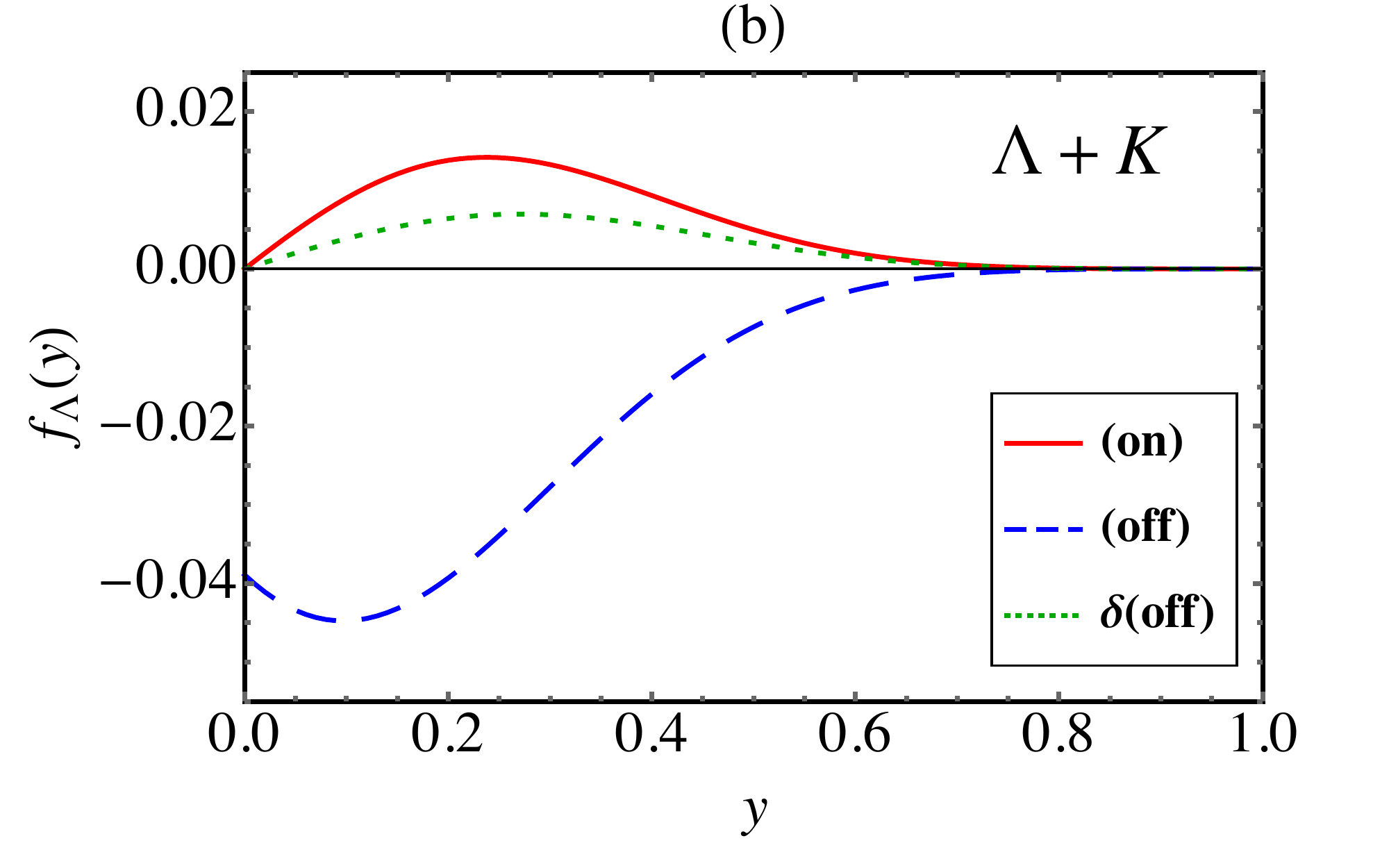}}& \\
\hspace{-0.6cm}{\epsfxsize=3.5in\epsfbox{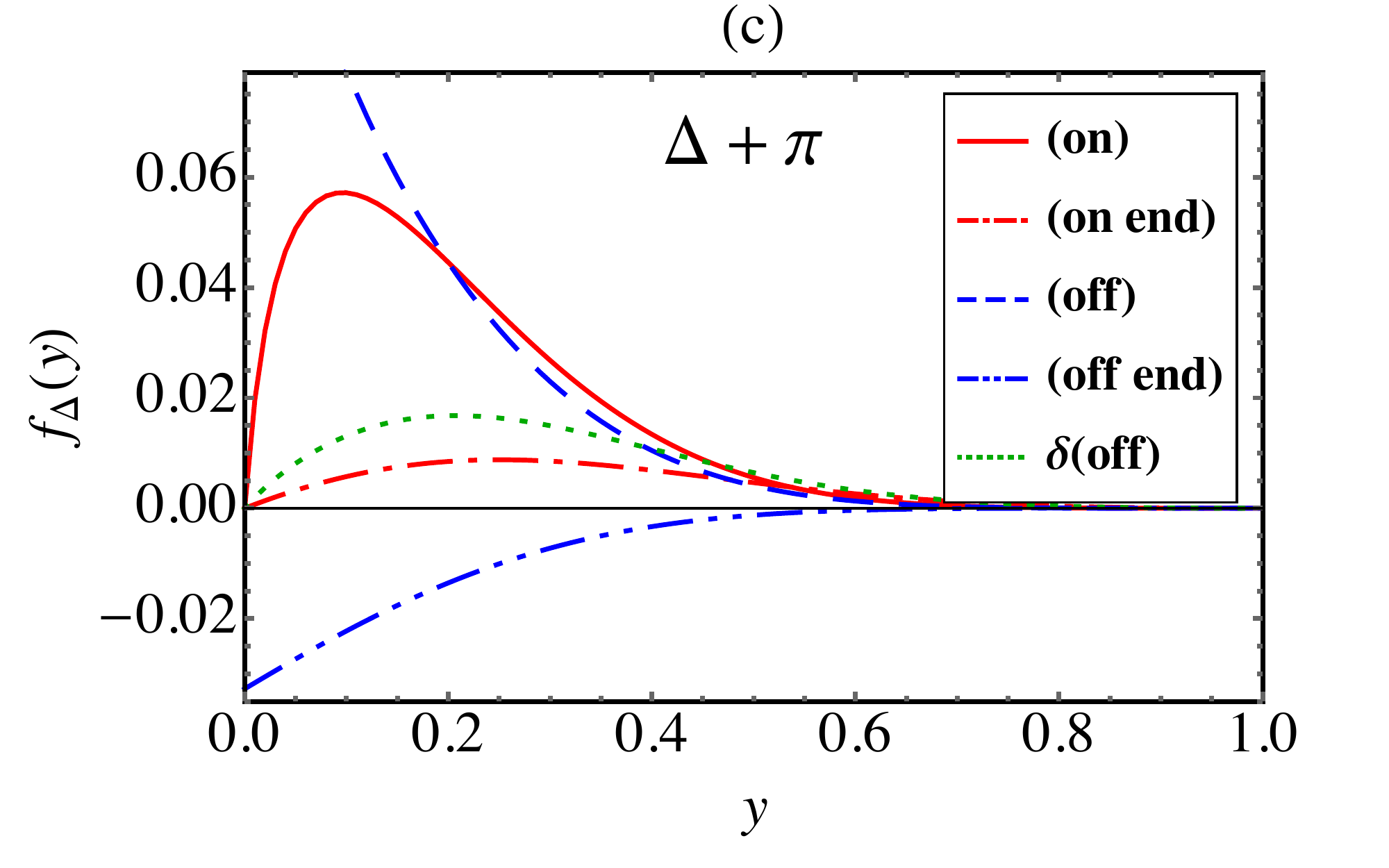}}&
\hspace{-0.5cm}{\epsfxsize=3.5in\epsfbox{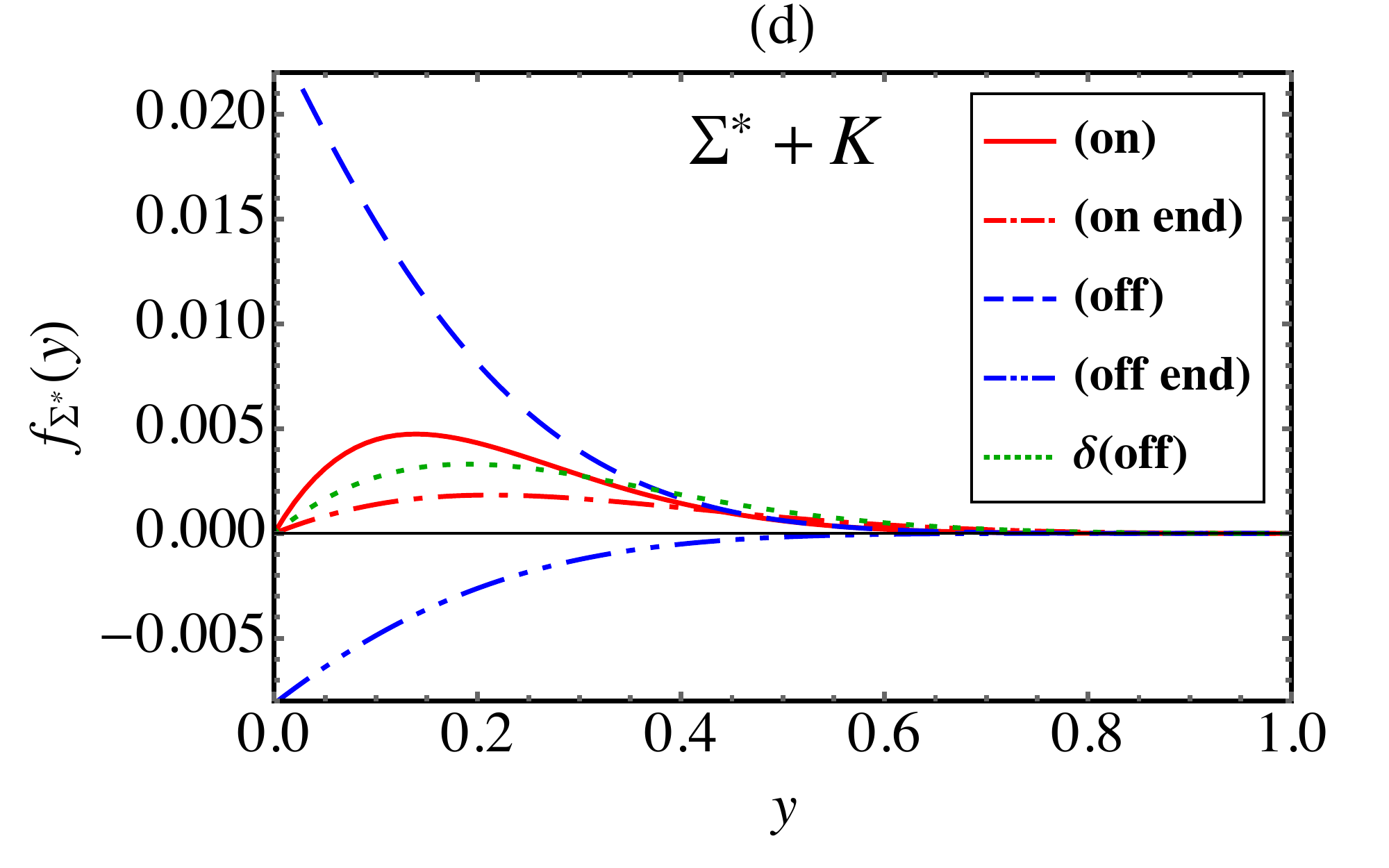}}&
\end{tabular}
\caption{Splitting functions versus meson momentum fraction $y$
	for the proton dissociations into
	(a)~$N + \pi$,
	(b)~$\Lambda + K$,
	(c)~$\Delta + \pi$, and
	(d)~$\Sigma^* + K$ state, for the
	on-shell $f^{\rm (on)}$ (red solid curves),
	off-shell $f^{\rm (off)}$ (blue dashed), and
	nonlocal off-shell $\delta f^{\rm (off)}$ (black dotted)
	contributions.
	For the decuplet $\Delta$ and $\Sigma^*$ states,
	additional contributions from
	on-shell end point $f^{\rm (on\, end)}$ (red dot-dashed) and
	off-shell end point $f^{\rm (off\, end)}$ (blue dot-dot-dashed)
	are included.
	All results correspond to the covariant dipole form factor
	in Eq.~(\ref{eq:re}) with cutoff mass $\Lambda=1$~GeV.}
\label{fig:fy}
\end{figure}

In Fig.~\ref{fig:fy} we show the basis splitting functions for
the on-shell $f^{\rm (on)}_{B,T}$, off-shell $f^{\rm (off)}_{B,T}$,
and nonlocal off-shell $\delta f^{\rm (off)}_{B,T}$ contributions,
as well as the on-shell and off-shell end point functions
$f^{\rm (on\, end)}_T$ and $f^{\rm (off\, end)}_T$ for the
decuplet $\Delta$ and $\Sigma^*$ states.
For all baryon intermediate states, the on-shell functions
$f^{\rm (on)}_{B,T}$ are positive at all $y$ values and peak
at around $y = 0.1-0.2$, depending on the mass of the baryon.
The main difference between the on-shell functions for the
different baryons is the magnitude: for the strange baryons
the functions are approximately an order of magnitude smaller
than for the non-strange.

The off-shell functions $f^{\rm (off)}_{B,T}$ for the octet baryons
are negative, with magnitude comparable to the on-shell functions.
For decuplet baryons, the off-shell functions increase as $y \to 0$,
and in fact dominate the small-$y$ region.
The nonlocal off-shell functions $f^{\rm (off)}_{B,T}$ have the same
sign as the on-shell contributions, but are somewhat smaller in
magnitude.
The additional on-shell and off-shell end point contributions
$f_T^{\rm (on\, end)}$ and $f^{\rm (off\, end)}$ for the decuplet
intermediate states in Eqs.~(\ref{eq:Tend}) and (\ref{eq:Tphi-offend})
are positive and negative, respectively, with the former vanishing
at $y=0$ and the latter increasing in magnitude as $y \to 0$.

\begin{figure}[t]
\centering
\begin{tabular}{ccc}
\hspace{-0.2cm}{\epsfxsize=3.7in\epsfbox{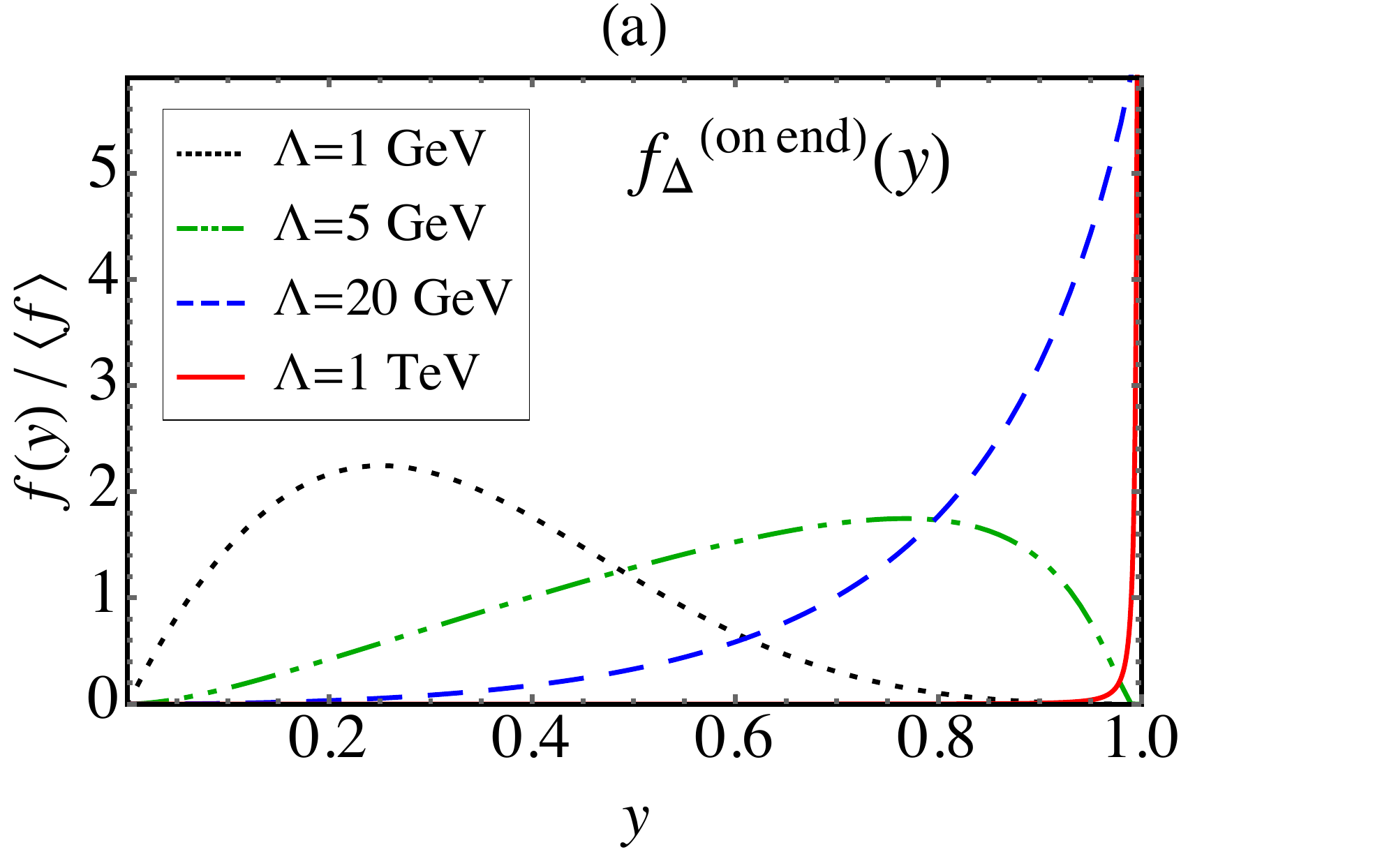}}&
\hspace{-1.2cm}{\epsfxsize=3.7in\epsfbox{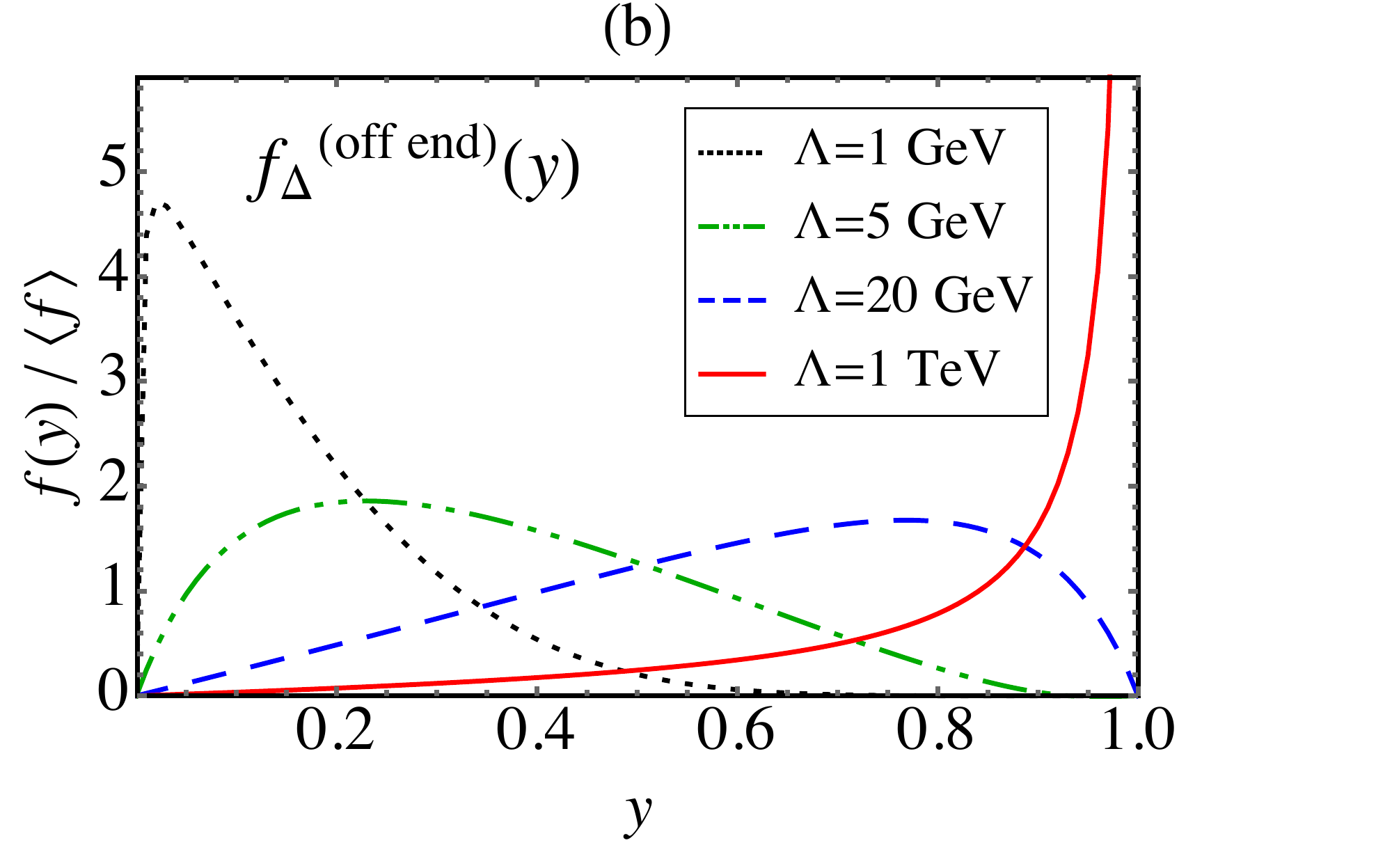}}& \\
\end{tabular}
\caption{Normalized splitting functions
	$f_i(y)/\langle f_i \rangle$ for the
	(a) on-shell end point and
	(b) off-shell end point contributions
	for the $\Delta + \pi$ intermediate state,
	for different values of the dipole cutoff mass
	$\Lambda$ (1~GeV to 1~TeV) and a fixed value of
	the constant $\Omega_0=100$~GeV$^2$.}
\label{fig:fyO}
\end{figure}

Interestingly, both the on-shell and off-shell end point functions
at $\Lambda=1$~GeV peak at rather small values of $y$, while formally
they become $\delta$-functions at $y=1$ for $\Lambda \to \infty$.
The dramatic change in the shape of the end point functions with
increasing $\Lambda$ is illustrated in Fig.~\ref{fig:fyO}, which
shows the on-shell and off-shell end point terms as a function
of $y$ for a range of $\Lambda$ values from 1~GeV to 1~TeV.
Of course, in practical calculations relevant for phenomenological
applications, the relevant values of $\Lambda$ would typically be
of the order of hadronic scales, $\sim 1$~GeV; the results for the
larger $\Lambda$ values shown in Fig.~\ref{fig:fyO} are simply to track
numerically the evolution of the nonlocal results to the local limit.

Note that the derivation of the local limit of the end point splitting
functions, as in Eq.~(\ref{eq:ghj}), includes the $D_0$ term.
There, it was assumed that the constant $\Omega_0$ in $D_0$ is
very large, although in the local limit it also satisfies
$\Omega_0 \ll \Lambda^2$ [see Eq.~(\ref{eq:HGHG})].
In order to observe the $D_0$ contribution to Eq.~(\ref{eq:interim})
in practice, we fix the parameter $\Omega_0$ to a very large value,
$\Omega_0 = 100$~GeV$^2$.
As shown in Fig.~\ref{fig:fyO}, when $\Lambda$ is small, the
contribution of $D_0$ is negligible, and the on-shell and off-shell
end point distributions coincide with those in Fig.~\ref{fig:fy}(c)
for $\Lambda = 1$~GeV.
(The end point functions decrease in magnitude at $y < 1$ with
increasing $\Lambda$, so for clarity these are normalized by their
integrals, $\langle f_i \rangle$, over all $y$.  This then renders the
ratio for the off-shell end point function in Fig.~\ref{fig:fyO}(b)
positive, whereas the unnormalized distribution in Fig.~\ref{fig:fy}(c)
is negative.)
The $D_0$ term can therefore be dropped when considering the
contribution of the nonlocal end point functions for finite values
of~$\Lambda$.
On the other hand, Fig.~\ref{fig:fyO} clearly indicates that as
$\Lambda \to \infty$ the peaks of the end point functions migrate
to higher values of $y$, approaching a shape that resembles a
$\delta$-function, $\delta(1-y)$, in the local limit.

\begin{figure}[t]
\centering
\begin{tabular}{ccc}
\hspace{-0.6cm}{\epsfxsize=3.5in\epsfbox{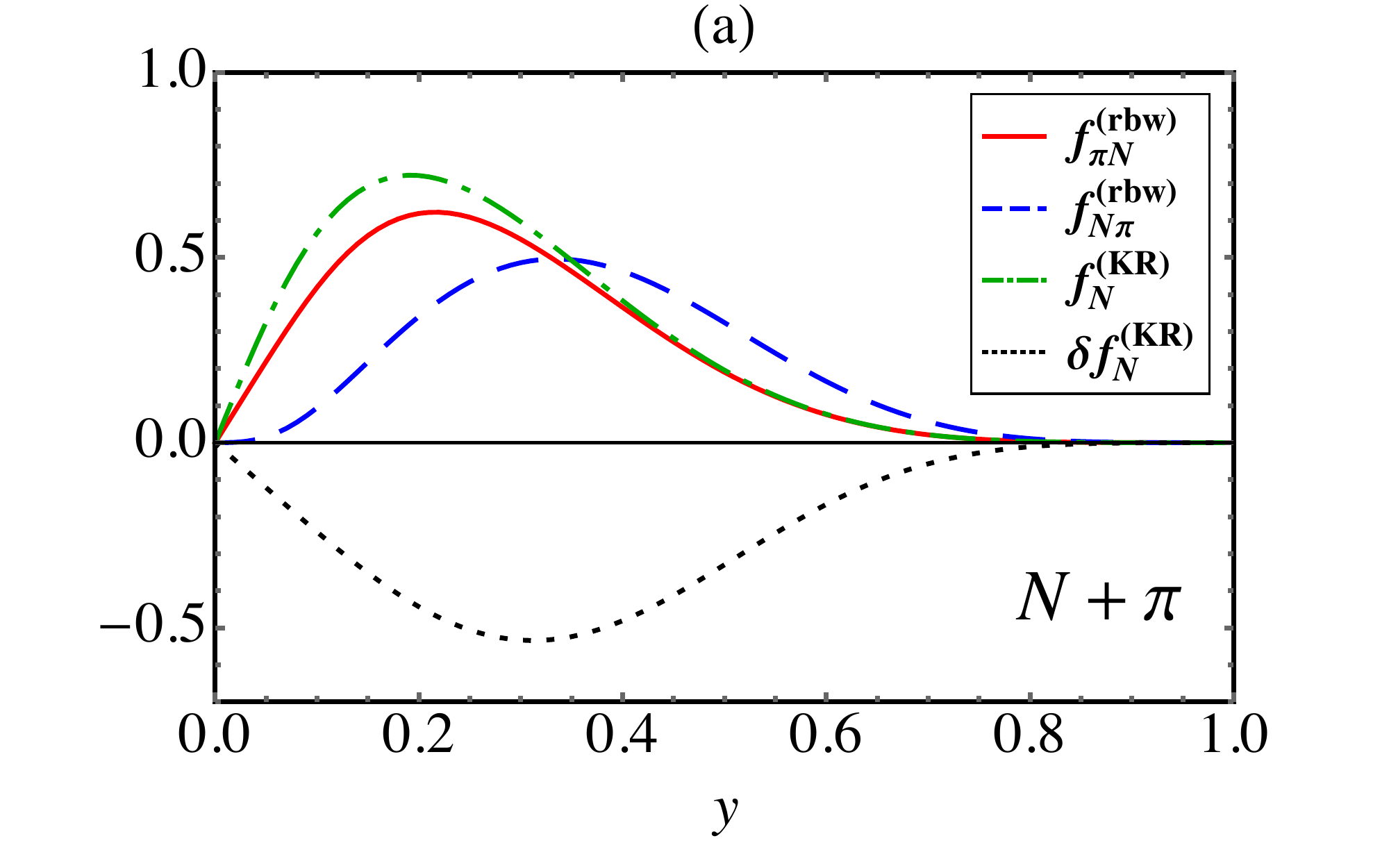}}&
\hspace{-0.6cm}{\epsfxsize=3.5in\epsfbox{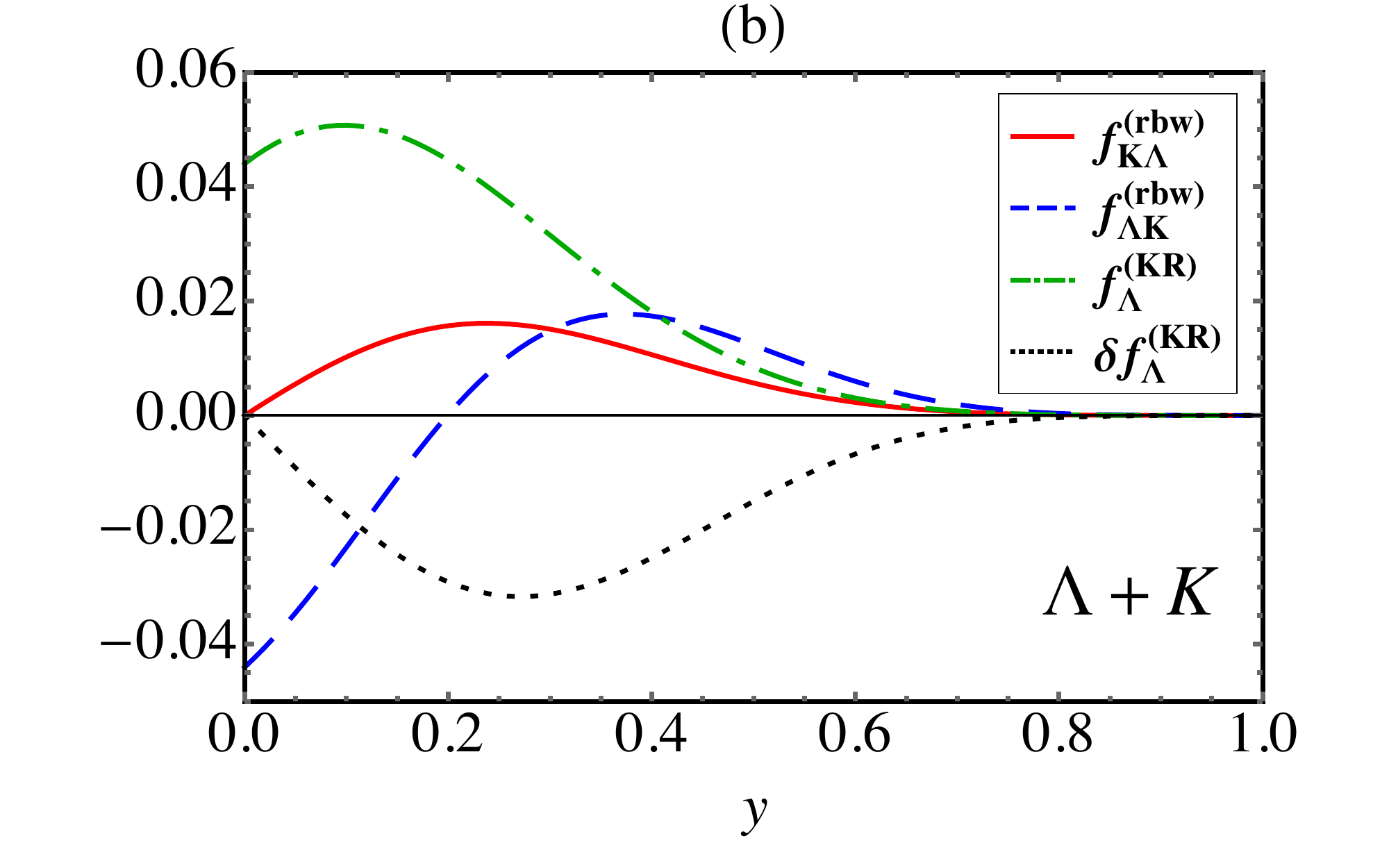}} &\\
\hspace{-0.6cm}{\epsfxsize=3.5in\epsfbox{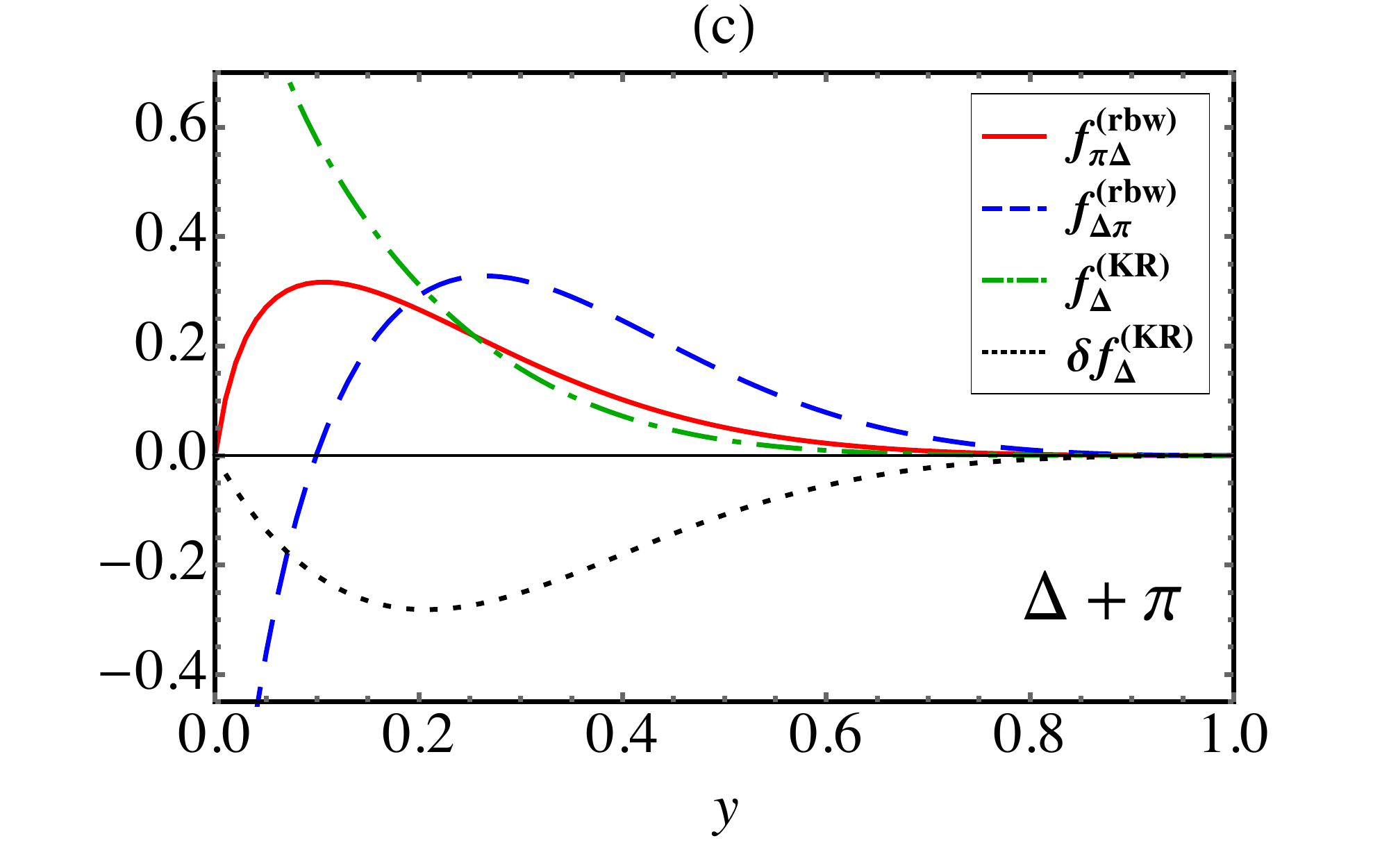}}&
\hspace{-0.6cm}{\epsfxsize=3.5in\epsfbox{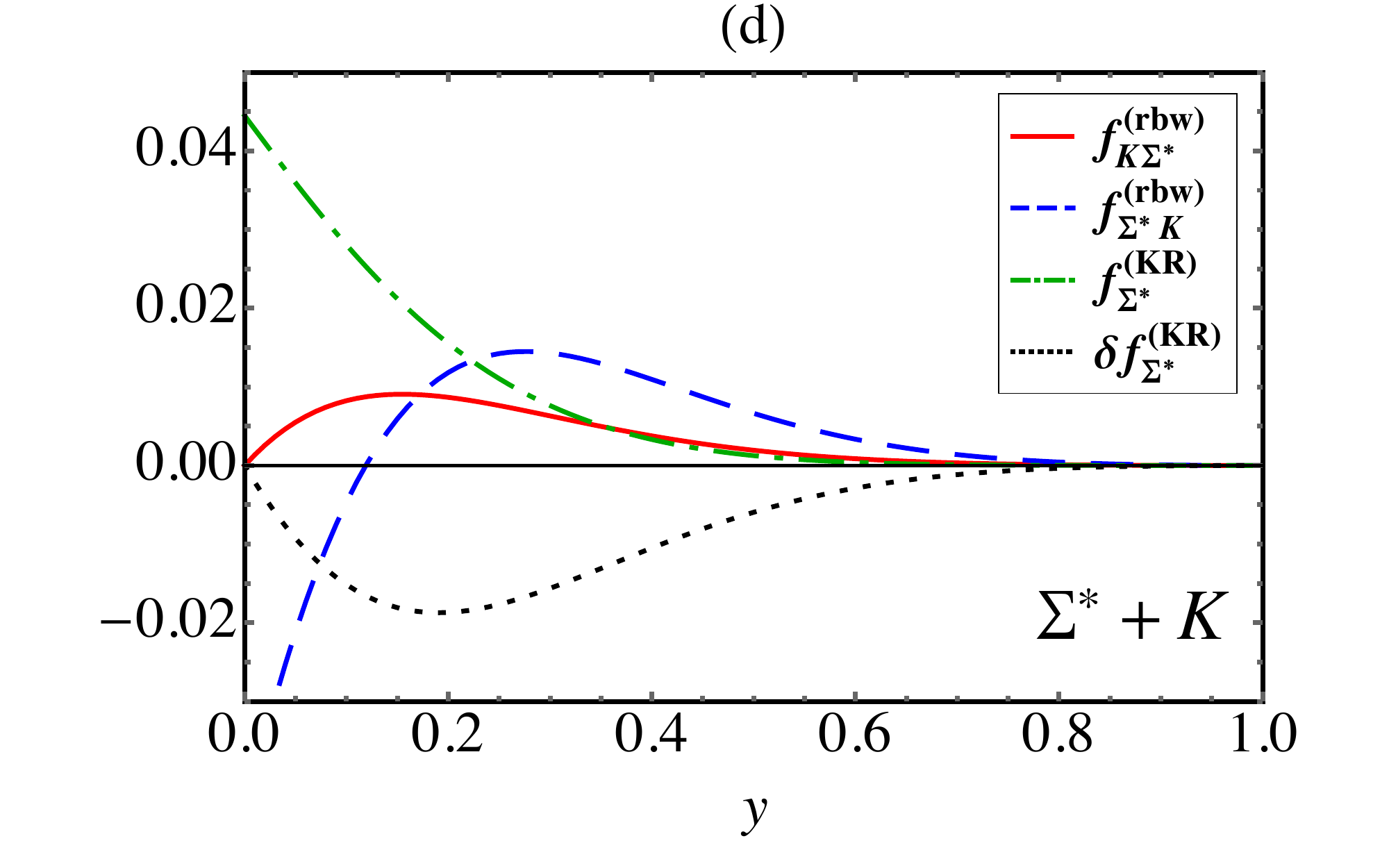}}&
\end{tabular}
\caption{Splitting functions versus $y$ for proton dissociations
	into various meson--baryon intermediate states as in
	Fig.~\ref{fig:fy}, but for the total contributions to the
	  meson-coupling rainbow diagrams in
	Fig.~\ref{fig:loop8}(a) and (h) (red solid curves),
	  baryon-coupling rainbow diagrams in
	Fig.~\ref{fig:loop8}(b) and (i) (blue dashed),
	  KR diagrams in
	Fig.~\ref{fig:loop8}(c) and (j) (green dot-dashed),
	  and nonlocal KR diagrams in
	Fig.~\ref{fig:loop8}(d) and (k) (black dotted).
	Contributions from the tadpole and bubble diagrams in
	Fig.~\ref{fig:loop8}(e)--(g) at $y=0$ are not shown here.}
\label{fig:fydiag}
\end{figure}

The combinations of the various basis functions corresponding to the
rainbow and KR diagrams in Fig.~\ref{fig:loop8} are illustrated in
Fig.~\ref{fig:fydiag} for the same intermediate states as in
Fig.~\ref{fig:fy}.
Again the main difference between the nonstrange and strange baryon
contributions is the magnitude of the functions, with the strange
being an order of magnitude or more suppressed.
The total meson-coupling rainbow functions,
  $f^{\rm (rbw)}_{\phi B}$ and $f^{\rm (rbw)}_{\phi T}$,
generally have very similar shape to the corresponding on-shell
functions in Fig.~\ref{fig:fy}.
The baryon-coupling rainbow functions,
  $f^{\rm (rbw)}_{B\, \phi}$ and $f^{\rm (rbw)}_{T\, \phi}$,
have similar magnitude and are generally positive at intermediate $y$,
but become more negative as $y \to 0$.
The latter behavior is canceled by the KR functions
  $f_{B,T}^{\rm (KR)}$
at small $y$, especially for the decuplet contributions,
such that the sum of the baryon-coupling rainbow and KR diagrams
satisfies Eqs.~(\ref{eq:g0}) and (\ref{eq:g2}).
The nonlocal KR functions,
  $\delta f_{B,T}^{\rm (KR)}$,
at nonzero $y$ values are proportional to $-4$ times the nonlocal
off-shell functions [Eqs.~(\ref{eq:f-link-Bphi}) and (\ref{eq:fdfTKR})],
and hence are negative at $y > 0$.
Some degree of cancelation therefore takes place between the local
$f_{B,T}^{\rm (KR)}$ and nonlocal $\delta f_{B,T}^{\rm (KR)}$
functions at intermediate and large values of $y$.

\begin{figure}[t]
\centering
\begin{tabular}{ccc}
\hspace{-0.3cm}{\epsfxsize=3.5in\epsfbox{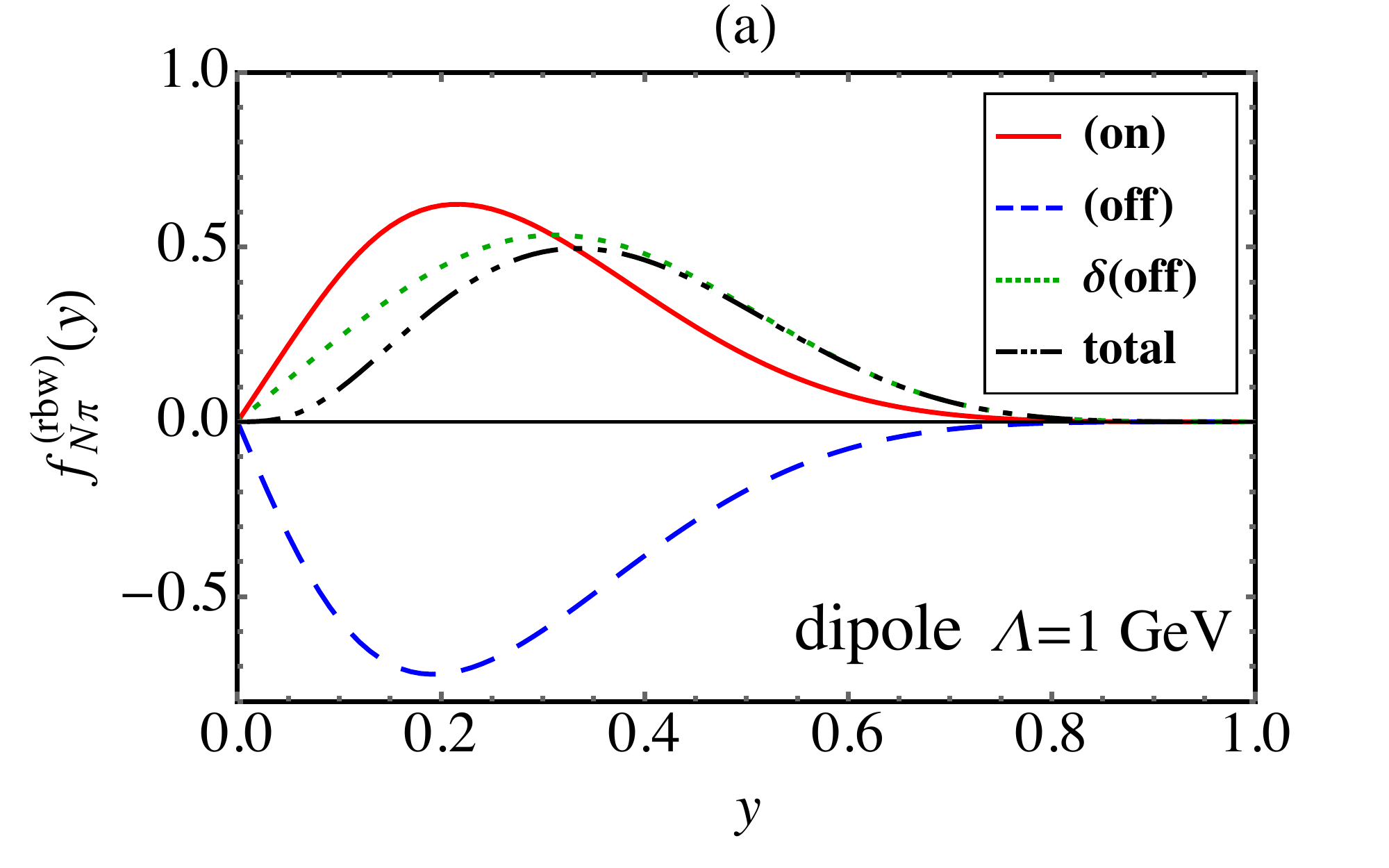}}&
\hspace{-0.5cm}{\epsfxsize=3.5in\epsfbox{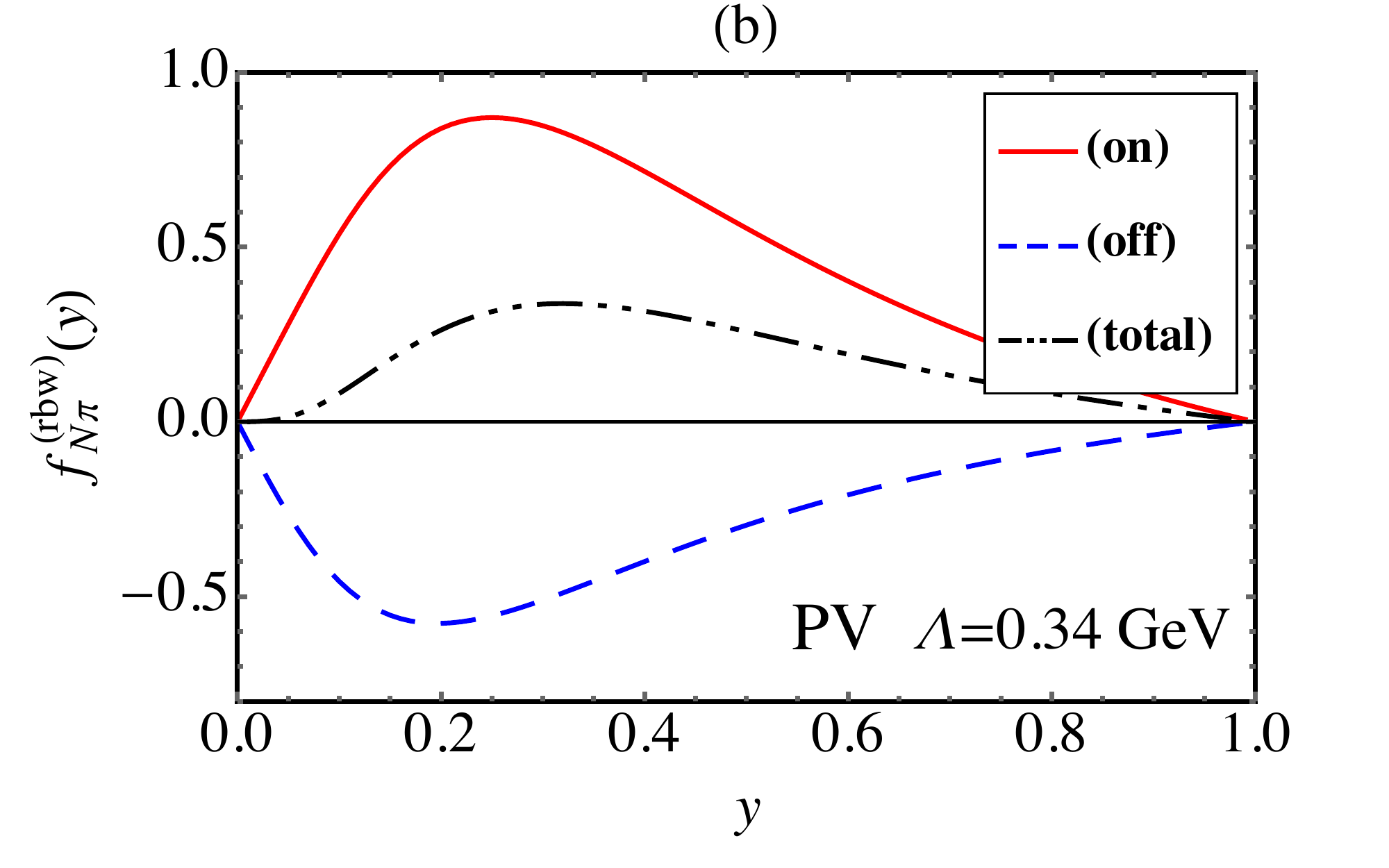}}&
\end{tabular}
\caption{Decomposition of the splitting function for the
	nucleon-coupling rainbow digram in Fig.~\ref{fig:loop8}(b) for
	(a) the nonlocal chiral theory with dipole regulator, and
	(b) the local chiral theory with a symmetry preserving
	Pauli-Villars regulator.
	The value of the Pauli-Villars mass parameters $\Lambda$ is
	determined by normalizing to the momentum carried by the
	interacting nucleon,
	$\langle y \rangle = \int_0^1 dy\, y\, f(y)$,
	for the dipole regulator with $\Lambda=1$~GeV.}
\label{fig:fy2}
\end{figure}

The pattern of cancelations between the various contributions from
the basis functions to particular diagrams in Fig.~\ref{fig:loop8}
is further explored in Fig.~\ref{fig:fy2}, which shows the
decomposition of the splitting function for the nucleon-coupling
rainbow diagram, $f_{N\pi}^{(\rm rbw)}$.
For the case of the covariant dipole form factor with $\Lambda=1$~GeV,
Fig.~\ref{fig:fy2}(a), one observes very strong cancelation between
the positive on-shell and negative off-shell contributions, with the
total closely resembling the purely nonlocal off-shell function
$\delta f^{(\rm off)}$.
At first sight this may be perplexing, if one interprets the result to
suggest that the total nucleon-coupling rainbow function may be very
small in the pointlike limit, where $\delta f^{(\rm off)}$ vanishes.
In practice, however, the on-shell and off-shell functions vary
differently with $\Lambda$, so that the degree of cancelation
depends on the cutoff.

This is illustrated in Fig.~\ref{fig:fy2}(b), which shows the
decomposition of $f_{N\pi}^{(\rm rbw)}$ for the case of a local
theory with a Pauli-Villars regulator, which preserves the necessary
symmetries of the theory \cite{XGWangPLB, XGWangPRD}.
In this case there is no nonlocal contribution, and the total
is given by the sum of the on-shell and off-shell terms.
For the on-shell splitting function $f_N^{(\rm on)}$ the
Pauli-Villars regulating function takes the form
\begin{equation}
\widetilde{F}_{\rm PV}^{(\rm on)}(k)
= 1 - \frac{D_{\phi B}^2}{D_{\Lambda_{\rm PV}}^2},
\label{eq:F_PVon}
\end{equation}
while for the off-shell splitting function $f_N^{(\rm off)}$
the regulator is given by
\begin{equation}
\widetilde{F}_{\rm PV}^{(\rm off)}(k)
= 1 - \frac{D_{\phi B}}{D_{\Lambda_{\rm PV}}}.
\label{eq:F_PVoff}
\end{equation}
In order to compare the shapes more directly, we choose the
Pauli-Villars regulator to give the same total momentum
$\langle y \rangle = \int_0^1 dy\, y\, f(y)$ carried by the
interacting nucleon in $f_{N\pi}^{(\rm rbw)}$, which yields
$\Lambda_{\rm PV} = 0.34$~GeV.
These have similar general features as the functions for the nonlocal
theory with covariant dipole regulator, with the small differences
in magnitude for the on-shell and off-shell contributions for the
dipole and Pauli-Villars regulators allowing a sizeable nonzero total
to remain.

\begin{figure}
\centering
\begin{tabular}{ccc}
\hspace{-0.4cm}{\epsfxsize=3.5in\epsfbox{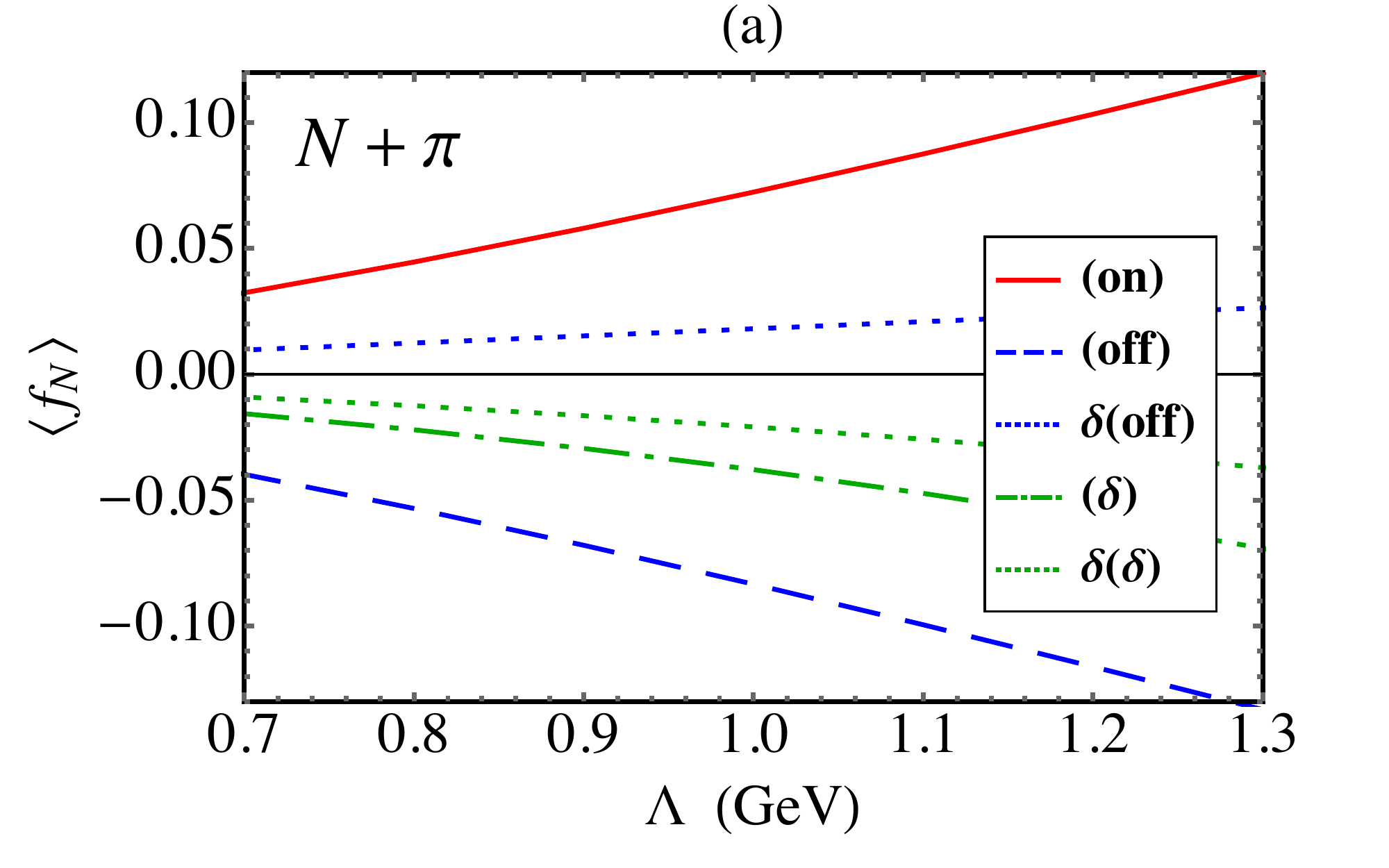}}&
\hspace{-0.8cm}{\epsfxsize=3.5in\epsfbox{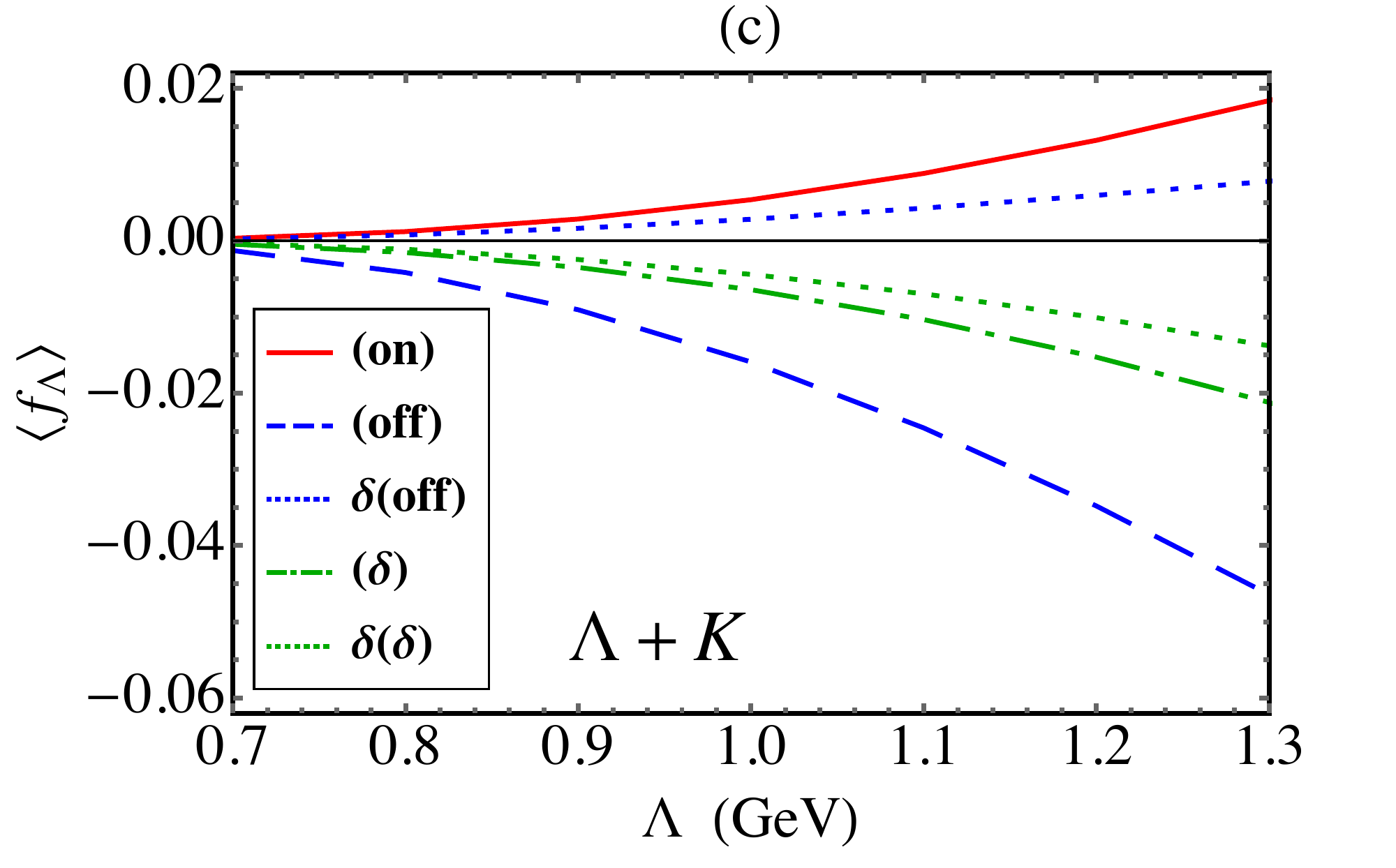}}&\\
\hspace{-0.4cm}{\epsfxsize=3.5in\epsfbox{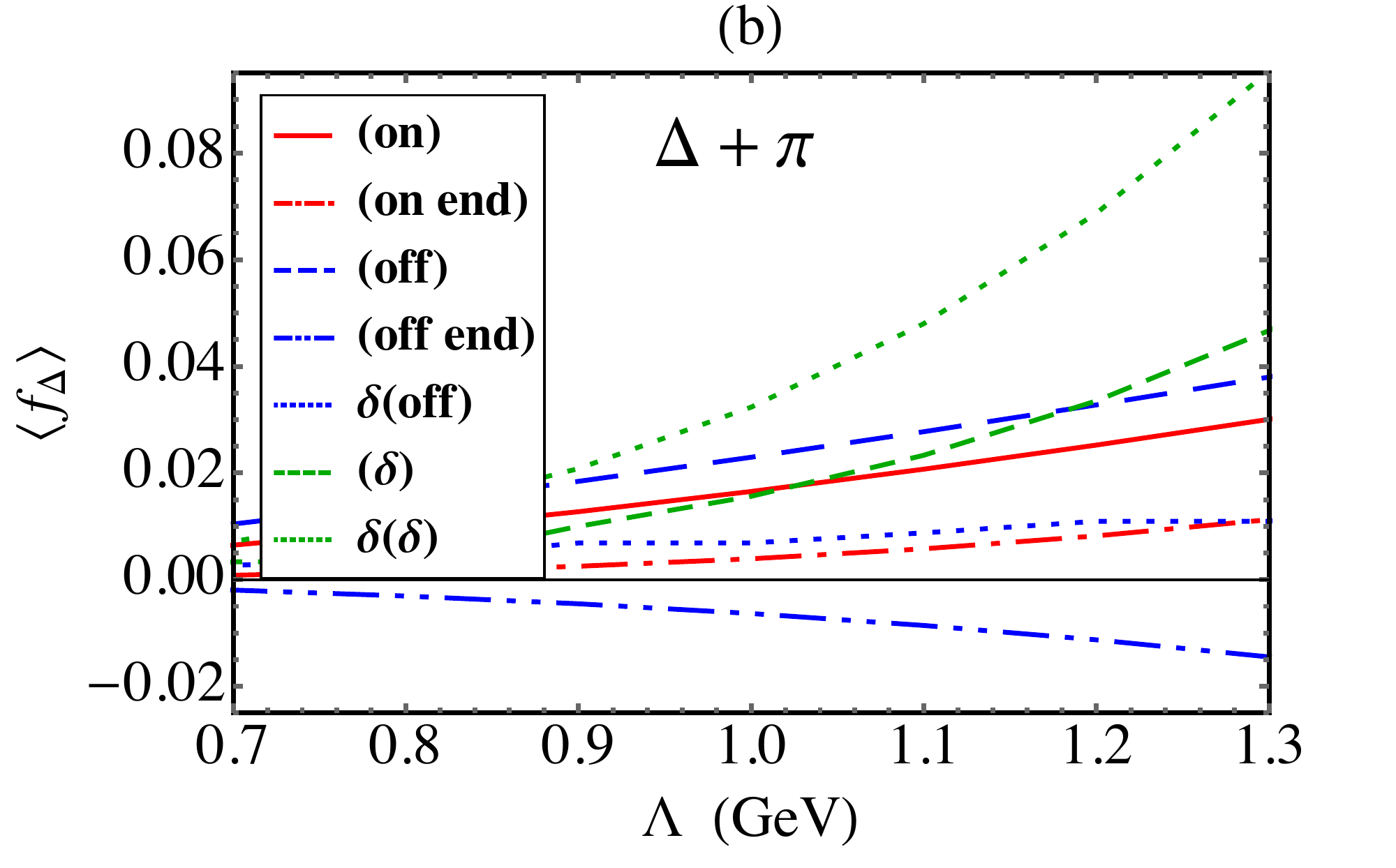}}&
\hspace{-0.4cm}{\epsfxsize=3.5in\epsfbox{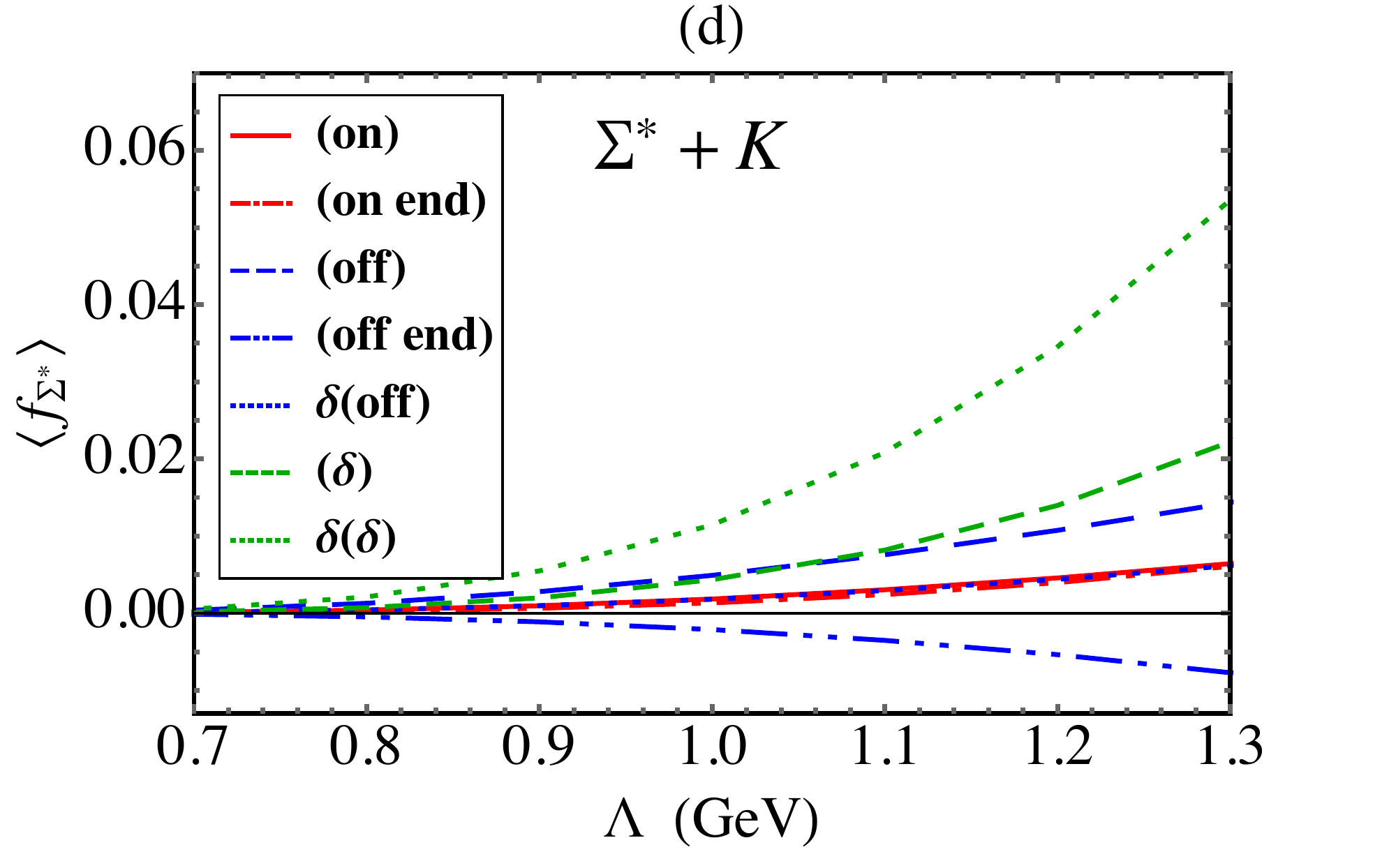}}&
\end{tabular}
\caption{Integrals of splitting functions $\langle f \rangle$
	versus $\Lambda$, for
	(a) $N + \pi$,
	(b) $\Lambda + K$,
	(c) $\Delta + \pi$ and
	(d) $\Sigma^* + K$ intermediates states, for the
	on-shell (red solid curves),
	off-shell (blue dashed),
	nonlocal off-shell (blue dotted),
	local $\delta$-function (green dot-dashed), and
	nonlocal $\delta$-function (green dotted)
	contributions.
	The decuplet states include additional contributions from
	on-shell end point (red dot-dot-dashed) and
        off-shell end point (blue dot-dot-dashed) terms.
        All results correspond to the covariant dipole form factor
        in Eq.~(\ref{eq:re}).}
\label{fig:intfy}
\end{figure}

While the contributions of the various splitting functions at $y > 0$
are illustrated in Figs.~\ref{fig:fy} and \ref{fig:fydiag}, the relative
importance of the $\delta$-functions terms at $y = 0$ is demonstrated
in Fig.~\ref{fig:intfy} by the integrated values of the basis functions,
$\langle f \rangle$ as a function of the covariant dipole form factor
cutoff mass $\Lambda$.
As expected, the magnitude of each of the integrated functions increases
with $\Lambda$, as more short-distance contributions are included.
For the nominal $\Lambda=1$~GeV used in Figs.~\ref{fig:fy} and
\ref{fig:fydiag} the $\pi N$ intermediate states dominate, with the
hyperon and decuplet contributions an order of magnitude smaller.
The picture changes for larger cutoff values, and for
$\Lambda \gtrsim 1.2$~GeV some of the $\pi \Delta$ contributions
become as large as the $\pi N$.
Of course, the validity of a one-loop calculation for larger cutoffs
is more questionable, as contributions from higher-order terms become
increasinbly more important.
Interestingly, for the octet baryons, the on-shell and nonlocal
off-shell contributions are positive, while the local off-shell
and both the (local and nonlocal) $\delta$-function contributions
are negative.
In contrast, for the decuplet states, all contributions are positive,
with the exception of the off-shell end point terms, as already
indicated in Fig.~\ref{fig:fy}.

\newpage
\section{Conclusion}
\label{sec.summary}

In this paper we have for the first time used a nonlocal covariant
formulation of SU(3) chiral effective theory to construct the
framework necessary for systematically computing the contributions
from pseudoscalar meson loops to parton distributions in the nucleon.
The main result of the present work has been the derivation from the
nonlocal theory of the lowest order proton $\to$ meson $+$ baryon
splitting functions arising from transitions of the initial state
to intermediate states involving octet and decuplet baryons,
as well as those involving contact interactions at zero momentum.

Since the contributions from the loop diagrams are ultraviolet
divergent, care must be taken to ensure that the integrals are
regularized in a way that preserves the underlying symmetries of
the effective theory, such as gauge invariance, Lorentz invariance,
and chiral symmetry.
A common approach adopted in the literature involves the use of
local interactions with regulators that explicitly depend on the
3-momentum of the meson.  While this does take into account the
extended nature of hadrons and renders finite results, this approach
is in practice {\it ad hoc} and destroys the local gauge and Lorentz
invariance of the theory.

The virtue of the nonlocal formulation, on the other hand, is that
it allows the use of a 4-dimensional regulator while preserving
the gauge and Lorentz symmetries.
In this case the regulator is generated directly from the nonlocal
Lagrangian, and gives rise to additional diagrams that appear from
the expansion of the gauge link [see Fig.~\ref{fig:loop8}(d), (g)
and (k)].

To illustrate the characteristic features of the new nonlocal splitting
functions, we have used a simple dipole function for the 4-dimensional
regulator.
The approach is analogous to a resummation of chiral perturbation
theory using dimensional regularization, which is known to provide
better convergence at larger momenta, at the expense of losing the
power counting of the traditional chiral perturbation theory.
Our results reveal some novel patters of cancelations among the
local and nonlocal functions in the rainbow and Kroll-Ruderman diagrams,
and illustrate the importance of nonlocal contributions for finite
values of the regulator mass $\Lambda$.
For the decuplet intermediate states, our analysis is able to study
numerically the transition from the case of a finite $\Lambda$ to
the pointlike limit, which is realized most dramatically for the
on-shell and off-shell end point contributions to the baryon-coupling
rainbow and Kroll-Ruderman diagrams.
We verify explicitly that in the $\Lambda \to \infty$ limit the
nonlocal generalization does indeed reproduce the results of the
local theory.

The results derived here will serve as a basis for future applications
of the formalism to computing meson loop contributions to parton
distributions in the nucleon.
Within the effective theory, these can be computed by matching
twist-two quark level and effective hadronic level operators,
which leads to a convolution representation for the PDFs,
\begin{eqnarray}
q(x) &=& \sum_j \int_0^1 \frac{dy}{y}
	 f_j(y)\, q_j^v\Big( \frac{x}{y} \Big),
\label{eq:conv}
\end{eqnarray}
where $f_j(y)$ are the meson--baryon splitting functions, and
$q_j^v$ is the valence distribution for the quark flavor $q$
in the hadronic configuration $j$.
In a forthcoming paper \cite{nonlocal-II}, we will use this formalism
to study flavor asymmetries in the nucleon generated through meson
loops, such as in the light antiquark sea ($\bar{d}-\bar{u}$) or for
strange quarks ($s-\bar{s}$), consistently within the 4-dimensional
chiral effective theory framework.

\section*{Acknowledgments }

We thank Xuangong Wang for helpful discussions.
This work was supported by
NSFC under Grant No. 11475186 and 11747094, CRC 110 by DFG and NSFC;
the DOE Contract No.~DE-AC05-06OR23177, under which Jefferson
  Science Associates, LLC operates Jefferson Lab;
DOE Contract No.~DE-FG02-03ER41260;
the Australian Research Council through the ARC Centre of
  Excellence for Particle Physics at the Terascale (CE110001104);
and an ARC Discovery Project DP151103101.


\end{document}